\documentclass[ showpacs, preprintnumbers, amsmath, amssymb,twocolumn]{revtex4}

\usepackage{graphicx}
\usepackage{dcolumn}
\usepackage{bm}
\usepackage{subfigure}

\begin{document}

\preprint{...}

\title{Entangled multi-knot lattice model of anyon current}

\author{Tieyan Si $^*$}

\affiliation{ Physics department, School of Sciences,and Key Laboratory of Microsystems and Microstructures Manufacturing-Ministry of Education, Harbin Institute of Technology, Yi Kuang street 2, Harbin 150080, China }

\date{\today}

\begin{abstract}

We proposed an entangled multi-knot lattice model to explore the exotic statistics of anyon. This knot lattice model bears abelian and non-abelian anyons as well as integral and fractional filling states that is similar to quantum Hall system. The fusion rules of anyon are explicitly demonstrated by braiding on crossing states of the multi-knot lattice. The statistical character of anyon is quantified by topological linking number of multi-knot link. Long-range coupling interaction is a fundamental character of this knot lattice model. The short range coupling models, such as Ising model, fermion paring model, Kitaev honeycomb lattice model and so on, appears as the short range coupling case of the knot lattice model. We introduced link lattice pattern as geometric representation of the eigenstate of quantum many body model to explore the topological nature of quantum eigen-states. For example, a convection flow loop is introduced into the well-know BCS fermions pairing model to show the Pseudo-gap state in unconventional super-conducting state. The integral and fractional filling numbers in quantum Hall system is directly quantized by topological linking number. The quantum phase transition between different quantum states in quantum spin model is also directly quantified by the change of topological linking number, which revealed topological character of phase transition. This multi-knot lattice has a promising physical implementation by circularized photons in optical firbre network. It may also provide another different path to topological quantum computation.

\end{abstract}

\pacs{05.30.Pr,73.43.-f,03.65.Vf,02.10.Kn}

\maketitle

\section{Introduction}

Anyon in quantum many body system is exotic quasiparticle that bear special statistical character beyond fermion and boson. It is the elementary unit for constructing fault-tolerant quantum computation \cite{Kitaev1}\cite{Nayak}. Anyon can exist as Majorana fermion or vortex core in topological superconductor, which is classified into the widespread existence of topological matter in recent years \cite{kou}. However a solid experimental manipulation of non-abelian anyon remains hard challenge so far. One promising candidate for experimental implementation of anyon is electron gas in strong magnetic field \cite{Read}. Exchanging two abelian anyons generates an arbitrary phase upon the wave function \cite{Frank}, $ \phi(r_1, r_2) = e^{i\theta}\phi(r_2, r_1)$, where $\theta\in[0, 2\pi]$. The statistical phase can be controlled by the enclosed magnetic flux within the exchanging path loop. The interference fringes between Laughlin quasiparticles provided a controversial experimental signal of non-abelian statistics \cite{Camino} that is predicted by fractional quantum Hall theory \cite{Read}\cite{Wen}. While the experimental operation of Ising anyons suggested by conformal field theory of critical two-dimensional Ising model \cite{Georgiev} is still a difficult challenge for experiment \cite{Stern}.

The plaquette vortex excitation in Kitaev's toric code model \cite{Kitaev1}and honeycomb model \cite{Kitaev2} is typical anyon in quantum lattice models, so does Wen's toric code model and topological color code model on two dimensional lattice \cite{Bombin}. Anyons also exists as fractional quasi-excitation states in one-dimensional optical superlattice \cite{SChen}\cite{HGLuo}\cite{HGLuo2}. In the toric code model, the plaquette excitation and vertices excitation form a dual pair of abelian anyons. The exchanging of the two anyons is performed on a loop of lattice squares. Non-abelian anyon exists in the gapless phase of Kitaev honeycomb model \cite{Kitaev2}. Both the toric code model and honeycomb model are difficult to find a solid material correspondence in reality, even though the physic of Kitave honeycomb model shows relevance to certain transition metal compounds, such as $(Na,Li)_{2}IrO_3$ iridates and $RuCl_3$ \cite{Hermanns}.

Here we proposed another different approach to anyons: periodically entangled knot current of hopping particles on lattice. The over-crossing point of many entangled knots are placed on periodical lattice. The over-crossing states are mapped into spin state. In this model, anyons exist as running particles in these entangled wires. For the anyon knot model on square lattice, anyons are conventional positrons(or electrons) and magnetic monopoles. For anyon knot model on honeycomb lattice, there are three color anyons: red, blue and yellow anyon as we named. These anyon knot model have exact correspondence with two dimensional Ising model, Kitaev honeycomb model as well as Heisenberg model. Each eigenstate of the anyon knot model corresponds to a knot configuration. Each knot configuration bear a topological invariant Jones polynomial, which are related to the non-abelian Chern-Simons field theory \cite{Witten}
\begin{eqnarray}\label{nonAbeCS}
\mathcal{L}={\int}_{M}Tr(\epsilon^{ijk}A_{i}(\partial_{j}A_{k}-\partial_{k}A_{j})+\frac{2}{3}A_{i}[A_{j},A_{k}]).
\end{eqnarray}
Abelian Chern-simons field theory suggest that entangled many knot are classified by linking number, self-linking number as well as writhing number \cite{Duan}. The fusion rules of anyons has explicit demonstration in this anyon knot model with the assistance of braiding operations. Unlike the braiding operation along the time lines, the braiding operation on knot lattice was implemented on spatial lattice.

The paper is organized into two sections. In the first section, we explore the anyon states and fusion rules on square knot lattice model with long range coupling interaction, which incorporate the two-state and three-state block spin to demonstrate integral and fractional filling quantum Hall state, long range hopping model, spin Hall model, and convective fermion current pairing model. In the second section, we constructed knot lattice on honeycomb lattice to study non-abelian anyons in quantum states in various long range coupling quantum models. In the approximation of nearest neighboring coupling, this knot lattice model reveals the topological order in the conventional transverse Ising model, Kitaev honeycomb model, Haldane model and topological flat band model. The topological change from one ordered state to another is distinguished by the variation of topological linking number.

\section{ Entangled Anyons current knot on square lattice}

\subsection{Dual anyon pairs in long-range coupling knot lattice model of two-state spins}

\begin{figure}
\begin{center}
\includegraphics[width=0.42\textwidth]{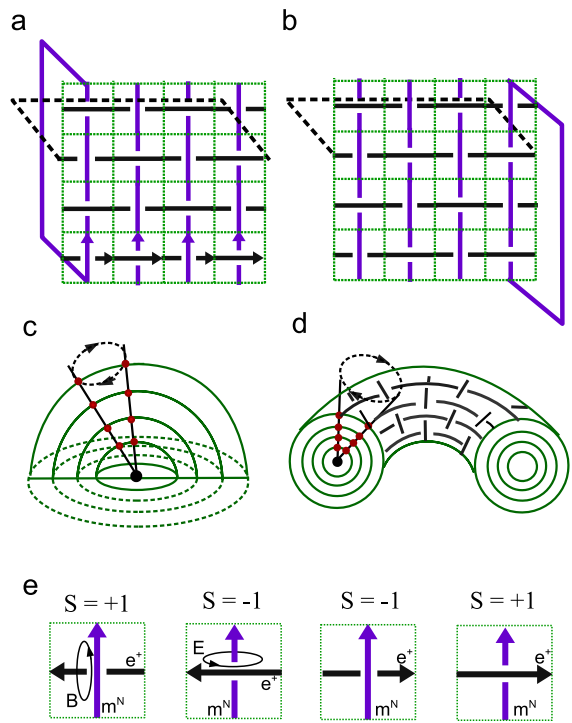}
\caption{\label{spin12sphere} (a) The lattice of over crossing points with spherical periodic boundary condition. (b) The lattice of over crossing points with torus periodic boundary condition. (c) Multilayer sphere of many layers of entangled knots. Each layer corresponds to a square lattice above. (d) Multilayer torus of many layers of entangled knots. (e) The mapping from over crossing states of electric current and magnetic current to the two states of Ising spin.}
\end{center}
\vspace{-0.5cm}
\end{figure}

 A knot is a closed loop in three dimensional manifold that can map into an unit circle. Many entangled knots together define a link. We take M knot current that project horizontal current and N knot current that project vertical current, and entangle them to project a two dimensional $M\times{N}$ lattice of over crossing points (or under crossing). If both the horizontal knots and vertical knots bends upward (downward), the base manifold of the over crossing lattice is a sphere in thermal dynamic limit (M,N $\rightarrow\infty$) (Fig. \ref{spin12sphere} (a)(c)). If the horizontal knots bend upward and the vertical knots bend downward, the knot lattice is equivalent to a torus (Fig. \ref{spin12sphere} (b)(d)).

The horizontal knot (black lines in Fig. \ref{spin12sphere} (a) (b)) could be implemented by electrical conductor wires, such as super-conducting wires. Only electron or positron runs in the horizontal knot. While the vertical knot (purple lines in Fig. \ref{spin12sphere} (a) (b)) are currents for running magnetic monopoles with positive or negative magnetic charges. According to the electromagnetic induction effect, a running positron induce a circular magnetic field around the electric current. There exist an electromagnetic interaction between positrons and magnetic monopoles at each over crossing point. If the magnetic current lies in the same direction as the induced magnetic field by the electric current according to the right hand rule, then the energy of system would increase by one unit. This is the case for Fig. \ref{spin12sphere} (e)(S=+1), where the upward magnetic current is above the left-moving electric current. On the contrary case (Fig. \ref{spin12sphere} (e)(S=-1)), both the magnetic current and electric current are slowed down, the energy of the system drops one unit. Each over crossing point can be mapped into an effective Ising spin with two states, $|\uparrow\rangle$ and $|\downarrow\rangle$. Under the action of effective Hamiltonian $H_{z} = S^{z}$, the eigenvalues with respect to these two spin states are $S = \pm 1$ (Fig. \ref{spin12sphere} (e)).

The electromagnetic induction effect also introduced the long range coupling interaction between different crossing sites.Because if any one of the local crossing sites along the loop is cut, it would induce a global magnetic field that acts on all of the rest crossing sites. The magnetic monopole current generates a global electric field in the direction parallel to electric current. The same phenomena occurs for the electric current. Whenever an electric loop is cut into segment, the total magnetic flux loses one flux. All of the positrons along this electric current become static and lose its interaction with magnetic monopole at the crossing point. Thus there exist a topological correlation between Ising spin in each loop. Therefore we introduce long range coupling between spins along the same loop,
\begin{eqnarray}
H_{LIsing}=\sum_{i,j,m,n}^{M,N} J_{ij} (J_{m} S^{z}_{ij}S^{z}_{i+m\textbf{e}_{x},j} + J_{n} S^{z}_{ij}S^{z}_{i,j+n\textbf{e}_{y}}),
\end{eqnarray}
where $S^{z} = \pm1.$ The correlation length is proportional to the length of the loop. This long range coupling Ising model assigns different flipping probability on crossing state from that assigned by the nearest neighboring coupling Ising model. Non-abelian anyons exist in this long range coupling Ising-like knot model.

Every knot lattice configuration can be classified by a topological number called linking number, which is defined as the total number of positive crossing minus the total number of negative crossing, $L_{link} = ({N_{+}-N_{-}})/{2}$. This linking number is equivalent to the total magnetization of spins in magnetic system,
\begin{eqnarray}\label{link}
L_{link} = M = \sum_{{i}}^{N}S^{z}_{i}.
\end{eqnarray}
Thus total magnetization is a topological invariant for one knot lattice configuration in this lattice model. Since every current segment is confined on local lattice site, here Reidmeister move in knot theory is strictly confined within one lattice site. Total magnetization is not a knot variant, because different knot configuration may share the same linking number. In physics language, different degenerated spin configurations may have the same magnetization value. Suppose the over-crossing state has probability to flip from $+1$ to $-1$ (or vice versa) under random cutting and reunite. A temperature $T$ can be defined as a number that is positively correlated to this flipping probability. Then different knot lattice configurations have different existence probability with respect to the total energy and temperature. We assume that the knot lattice with lower total energy has higher probability to exist at a fixed temperature, i.e., obeying the Maxwell distribution. Then this probability weight of a certain knot lattice configuration $A$ follows the same rule as spins in statistical mechanics,
\begin{eqnarray}\label{partition}
p=\frac{1}{Z}e^{-H(A)/k_{b}T},\;\;\; Z = \sum_{A} e^{-H(A)/k_{b}T},
\end{eqnarray}
where $Z$ is the partition function which summarized the total probability of all possible configurations. As we know, each spin state flipping indicates a topological change of knot lattice, the entanglement between two knot either increase or decrease by 1. Since partition function is the summation of all possible knot lattice configurations, it is a topological invariant under arbitrary flipping operations. When the knot lattice configuration is exposed to an external electric field (or magnetic field) that is perpendicular to the knot lattice plane, the electric current of positrons (or magnetic current of monopoles) likens to stay above (or below). The probability of a certain knot configuration is determined by its linking number, which is equivalent to the effective Hamiltonian
\begin{eqnarray}\label{Hz}
H_z = - \sum_{i}h_{i} S^{z}_{i},
\end{eqnarray}
here $h_{i}$ represent the strength of external field. Obviously the ground state knot configuration corresponds to the highest linking number, $L = M{\times}N$. With the torus boundary condition, all the magnetic currents are above the electric currents at every over-crossing sites. The magnetic monopole generates electric field to propel the positrons in electric currents that pass through the inner zone enclosed by the magnetic loops. These magnetic loops cannot separate from electric loops without cutting. For the spherical boundary condition, it is equivalent to making a copy of lattice and connect it with the original lattice on the boundary point by point, but magnetic current is below the electric current in the copied lattice. This case is equivalent to that of torus boundary condition but with a larger linking number $L = 2 M{\times}N$. The eigenvalue corresponds to the Hamiltonian $H_z$ is $E_{i} = - h L_i$, where $L_i$ represents linking number with respect to the $i$th excited states. There are $(MN-1)$-fold degenerated first excited state with respect to $L_1 = (MN-2)$. The average Linking number reads,
\begin{eqnarray}\label{averLink}
\langle{L_{link}}\rangle = \frac{\sum_{A}[(\sum_{i} S^{z}_{i})e^{- E(A)/k_{b}T}]}{Z} =\frac{1}{k_{b}T}\frac{\partial}{\partial{h}}{{\ln}[Z(h)]},
\end{eqnarray}
where $Z(h)$ is the partition function for the free spin Hamiltonian Eq. (\ref{Hz}) with homogeneous external field $h$.

For a classical knot lattice of elastic wires, we introduce the coupling interaction between two nearest neighboring crossing points. Two neighboring sites with opposite crossing states generate a kink between them, which increased the elastic energy by one unit. While neighbors with the same over-crossing states bear smooth crossover between them, that decreased the elastic energy by one step. Thus the total energy of the knot lattice is
\begin{eqnarray}\label{Hising}
H_{Ising}=\sum_{\langle{ij}\rangle} J_{z} S^{z}_{i}S^{z}_{j},\;\;\;S^{z}= \pm1,
\end{eqnarray}
where $J_{z}<0$. Usually the coupling strength $J_{z}$ between two spins, $J_{z} = J_{z}(i,j)$, decreases if the distance between two spins ($S^{z}_{i}$ and $S^{z}_{j}$) increases. Conventional Hamiltonian is not a topological invariant which is invariant under continuous transformation of knot lattice. However for this knot lattice model, the coupling strength between two spins within the same loop is induced by electromagnetic wave which fills in the conducting channel at the speed of light. The coupling strength $J_{z}(i,j)$ is independent of the separation distance between two spins. In this case, the knot Hamiltonian is still a topological invariant, so does the partition function as well as other thermaldynamic observable. In the ground state of this ferromagnetically coupled system, all over-crossings are oriented in the same say. The transition from disordered orientation to this uniform ordering occurs at critical temperature $T_c$ \cite{Yang}. Since the topological linking number changes during this transition, we can also call this transition a topological phase transition.

The ground state of ferromagnetic Ising model has two fold degeneracy, i.e., all spins either point up or down, $|g\rangle = |\uparrow\uparrow\uparrow\cdots\uparrow\uparrow\uparrow\rangle
+|\downarrow\downarrow\downarrow\cdots\downarrow\downarrow\downarrow\rangle$. For this knot lattice model, the ground state can be represented by two layers of multi-knot sphere or multi-knot lattice (Fig. \ref{spin12sphere} (c)(d)). The first excited states is generated by flipping one spin, thus $|e\rangle_{1} = |\downarrow\uparrow\uparrow\cdots\uparrow\rangle+|\uparrow\downarrow\uparrow\cdots\uparrow\rangle+\cdots
+|\uparrow\downarrow\downarrow\cdots\downarrow\rangle+|\downarrow\uparrow\downarrow\cdots\downarrow\rangle+\cdots$.
There must be $2MN$ layers of knot lattice in total to represent the first excited state (Fig. \ref{spin12sphere} (c)(d)). Different knot configurations of eigenstate can transform into each other by braiding anyons, i.e., the positron and magnetic monopole.

\begin{figure}
\begin{center}
\includegraphics[width=0.4\textwidth]{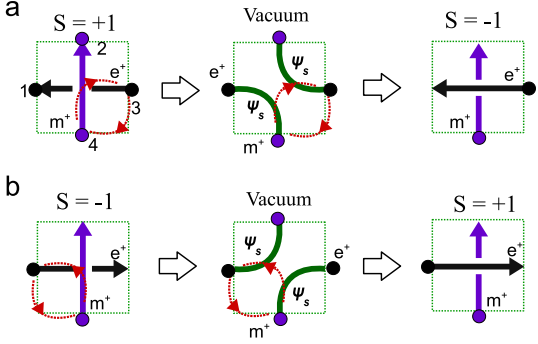}
\caption{\label{spin12} (a) Flipping $|\uparrow\rangle$ to $|\downarrow\rangle$ by braiding the positron $(e^+)$ and magnetic monopole $(m^+)$ twice in clockwise way. (b) Flipping $|\downarrow\rangle$ to $|\uparrow\rangle$ by braiding  $e^+$ and $m^+$ in counterclockwise way.}
\end{center}
\vspace{-0.5cm}
\end{figure}

The positron and magnetic monopole in this knot lattice model are dual anyon to each other. Each spin state is a collective wave function of two open fermionic strings of electric current or magnetic current. Since the two open strings are controlled by the four ending points at the middle point of each edge, the spin state here are also collective wave function of four anyons on the interface (Fig. \ref{spin12} (a) (b)). The $|\uparrow\rangle$ state can transform into $|\downarrow\rangle$ by braiding the positron $(e^+)$ No. 3 and magnetic monopole $(m^+)$ No. 4 twice in clockwise direction (Fig. \ref{spin12} (a)),
\begin{eqnarray}\label{bradoper}
|\downarrow\rangle = (R_{34\circlearrowright}^{m^+e^+})^{2}|\uparrow\rangle.
\end{eqnarray}
Since the spin state after braiding gains a phase factor $ e^{i\pi}$, the statistical phase factor for $e^+$ and $m^+$ is $R_{34\circlearrowright}^{m^+e^+} = e^{i\pi/2}$. Thus positron and magnetic monopole are dual anyons. Follow the same process, braiding $e^+$ and $m^+$ in counterclockwise direction leads to the same statistical phase (Fig. \ref{spin12} (b)), i.e., $R_{14\circlearrowleft}^{m^+e^+} = e^{i\pi/2}$. This braiding operation generates an intermediate vacuum state. The two strings (the green arc in Fig. \ref{spin12} (a)(b)) in this state are forbidden to touch each other, that is why they are fermionic strings $\psi_{s}$. Since monopole only runs in vertical current, while positron only run in the horizontal current, the two fermionic strings are effective converter that transform a monopole into a positron, or vice versa. The vacuum state physically implemented the fusion rules of anyons (Fig. \ref{spin12} (a)(b)),
\begin{eqnarray}\label{abefusion}
e \times \psi_{s} = m, \;\;\; e \times m = \psi_{s}, \;\;\;  m \times \psi_{s} = e,
\end{eqnarray}
and the trivial fusion rules: $e \times e = I, m \times m = I, \psi_{s} \times \psi_{s} = I$. However this braiding operation is only focused on one spin in one layer, it is not the eigenstate of Ising model.

\begin{figure}
\begin{center}
\includegraphics[width=0.40\textwidth]{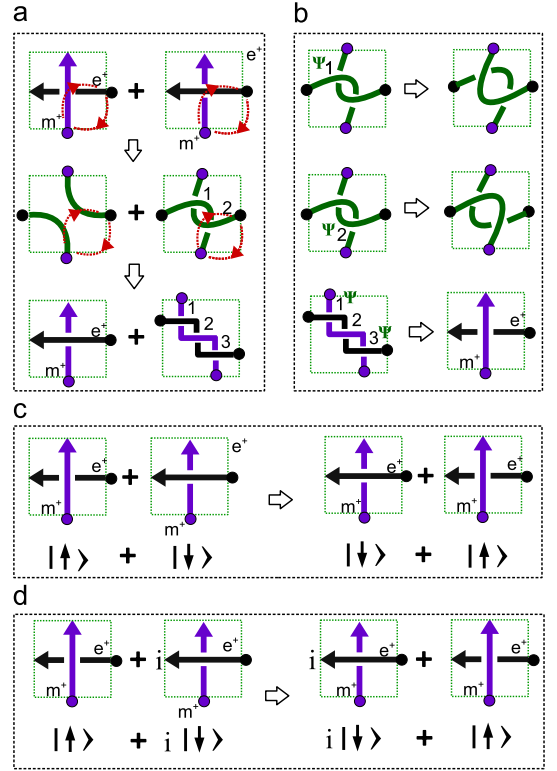}
\caption{\label{spinx} (a) braiding two anyons of knot state $(|\uparrow\rangle+|\downarrow\rangle)$ twice in clockwise direction. (b) Performing a Majorana fermion operator on the redundant crossing points. (c) Mapping the eigenstate $S^x$ to its eigenstate by braiding. (d) Mapping of the eigenstate $S^y$ to its eigenstate by braiding.}
\end{center}
\vspace{-0.5cm}
\end{figure}

\begin{figure}
\begin{center}
\includegraphics[width=0.45\textwidth]{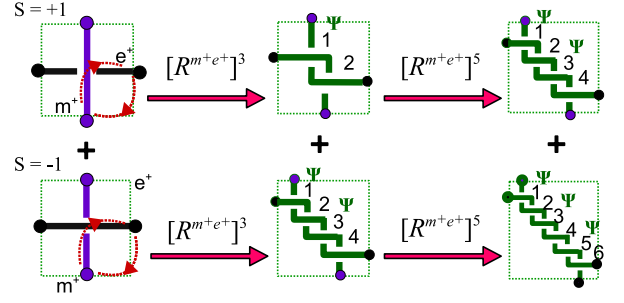}
\caption{\label{braid3} The knot configuration after braiding anyons for odd number of times. Here three and five times of braiding are showed as examples.}
\end{center}
\vspace{-0.5cm}
\end{figure}

Braiding anyons in the eigenstate of Ising model leads to more complicate statistical behavior beyond an abeian phase factor. We consider a synchronous braiding of two anyons in the bilayer knot configuration of Ising ground state (Fig. \ref{spin12sphere} (c)(d)). The two layers of knot lattice keep conformal invariance, thus one can draw two lines out of their common center to locate the two anyons at the same projected position on the base manifold. Suppose the upper layer represent spin up $|\uparrow\uparrow\uparrow\cdots\uparrow\uparrow\uparrow\rangle$, while the bottom layer represent spin down $|\downarrow\downarrow\downarrow\cdots\downarrow\downarrow\downarrow\rangle$. If we flip one spin at the same site of the two layers, it would generate a quasiparticle in the first excited state. This spin flipping can be implemented by braiding operation. At a given lattice site, we braid anyons of $(|\uparrow\rangle+|\downarrow\rangle)$ in the two layers synchronously (Fig. \ref{spinx} (a)). Upon one braiding in clockwise direction, the upper layer spin transforms into vacuum state, while the fermionic string in the bottom layer becomes nontrivially entangled with one crossing. While the sum of these two knot configurations are not the eigenstate of the system. Furthermore, one more braiding brings the spin up state in the upper layer to a spin down state, while the spin in the bottom layer now becomes entangled electric current and magnetic current with two crossings on one site. The sum of the two layer states is still not eigenstate of the system. That means these states are not physically accessible. In order to find the right knot configuration for the eigenstate, we have to introduce a Majorana fermionic operation ($\psi$) on the internal crossings within one site (Fig. \ref{spinx} (a)). The output of this Majorana fermion is to flip the crossing state, performing the same action as $S^{x}$ which is formulated into Jordan-Wigner transformation,
\begin{eqnarray}\label{spincmap}
c_{i} &=& \left(\prod_{j<i} S_{j}^{z}\right)S_{i}^{+},\;\;\;\;
c_{i}^{\dag} = \left(\prod_{j<i} S_{j}^{z}\right) S_{i}^{-},\nonumber\\
\psi_{i} &=& (c_{i}^{\dag}+ c_{i})/2.
\end{eqnarray}
Here the raising operator and lowering operator have the familiar form, $S^{+} = (S^{x} +i S^{y})/2$,
$S^{-} = (S^{x} - i S^{y})/2$. Here $c$ is conventional annihilation fermion, obviously $\psi$ is Majorana fermion, $\psi^{\dag} = \psi$. In fact, the spin-string operator representation of fermions has a geometric implementation in this knot lattice. First cutting each isolated horizontal loops on a chosen edge and connecting one ending point at the cutting edge of one loop with the ending point of another loop, then the two loops fuse into one. Repeating this operation unites  all the horizontal loops into one global loop. Thus Jordan-Wigner transformation has a natural implementation in this knot lattice model. We require that the flipping operation of spin operator $S^{x}$ only acts on spin states, i.e., $S^{x}|\uparrow\rangle=|\downarrow\rangle$ and $S^{x}|\downarrow\rangle=|\uparrow\rangle$, to avoid its undefined operation on the crossing of two fermionic strings. Here the Majorana operator $\psi$ acts on both the crossing of electric/magnetic current and two fermionic strings. After the first braiding operation $R_{\circlearrowright}^{m^+e^+}$, there are two crossings appeared in the bottom layer, in order to bring it back to vacuum state, the Majorana operator could act on either the first crossing or the second crossing to disentangle the two fermionic strings, then perform a Reidemeister move \cite{Wu} to reach the exact vacuum state as the upper layer (Fig. \ref{spinx} (b)). For the $|\downarrow\rangle$ acted by braiding operator $R_{\circlearrowright}^{m^+e^+}$ twice, two more Majorana fermion operators ($\psi_{1}\psi_{3}$) have to be performed to map it back to eigenspace on different crossing points (Fig. \ref{spinx} (b)). Thus the magnetic monopole and positron obey non-abelian fusion rules in the ground state of Ising model,
\cite{Nayak}
\begin{eqnarray}\label{nonfusion}
 e \times m = I + \psi, \;\psi \times \psi = I, \; e \times \psi = m, \;m \times \psi = e.
\end{eqnarray}
More over, the statistical factor of braiding two Ising anyons twice at one lattice site is no longer $e^{i\pi}$, it reads now
\begin{eqnarray}\label{braidings2}
[R_{\circlearrowright_{i}}^{m^+e^+}]^2
\left(
  \begin{array}{c}
    |\uparrow\rangle_{i} \\
    |\downarrow\rangle_{i} \\
  \end{array}
\right)=
\left(
  \begin{array}{cc}
    e^{i\pi} &  0\\
    0 & \psi_{1}\psi_{3} \\
  \end{array}
\right)_{i}
  \left(
  \begin{array}{c}
    |\uparrow\rangle_{i} \\
    |\downarrow\rangle_{i} \\
  \end{array}
\right).
\end{eqnarray}
Here $i$ represent the $i$th lattice site. The final state after this operation is the first excited state of ferromagnetic Ising model. The quasiparticle (or kink excitation around a lattice site) on the upper layer is a vacuum-like excitation, while the quasiparticle in the bottom layer is a Majorana fermion pair excitations. If the braiding operation was performed in counterclockwise direction, the two types of quasiparticles simply exchange their layer levels. Note that fermionic strings and unpaired Majorana fermion only exist for odd number of times of braiding (Fig. \ref{braid3}). For instance, three clockwise braiding on the eigenstate, $(|\uparrow\rangle+|\downarrow\rangle)$, generates one Majorana fermion and one Majorana fermion pair,
\begin{eqnarray}\label{braidings3}
[R_{\circlearrowright_{i}}^{m^+e^+}]^3
\left(
  \begin{array}{c}
    |\uparrow\rangle_{i} \\
    |\downarrow\rangle_{i} \\
  \end{array}
\right)=
\left(
  \begin{array}{cc}
    \psi_{1} &  0\\
    0 & \psi_{1}\psi_{3} \\
  \end{array}
\right)_{i}
  \left(
  \begin{array}{c}
    |\uparrow\rangle_{i} \\
    |\downarrow\rangle_{i} \\
  \end{array}
\right).
\end{eqnarray}

Furthermore, five clockwise braiding generate one Majorana fermion pair $\psi_{1}\psi_{3}$ and a triplet cluster of Majorana fermions $\psi_{1}\psi_{3}\psi_{5}$ (Fig. \ref{braid3}). Thus the magnetic monopole and positron fused into a pair of Majorana fermion on the $S=+1$ sector, and fused into three Majorana fermions on the $S=-1$ sector. When the Majorana fermion operators are acted on the even number of crossing sites, the overlapped vacuum state would flip a sign. The fusion rule for magnetic monopole and positron passing through the two fermionic strings which is braided for (2n+1) times is following,
\begin{eqnarray}\label{fusionodd}
&& e \times m = \psi_{1}\psi_{3}\psi_{5}...\psi_{2n-3} + \psi_{1} \psi_{3}\psi_{5}...\psi_{2n-1}\, \;\;\nonumber\\
&& e \times \prod_{n=1}^{M}\psi_{2n-3} = m, \;\;m \times \prod_{n=1}^{M}\psi_{2n-3} = e.
\end{eqnarray}
This fusion rule is a natural output of multi-knot lattice model, but almost invisible in the conventional Ising model.

\subsection{ Dual anyon pair in knot lattice model of block spin-1}

The knot lattice model for two-state spin, $S = \pm1$, does not admit fermionic string state as its eigen-knot state. In fact, the vacuum state generated by braiding operation can be naturally defined as spin zero state (Fig. \ref{spin12} (a)(b)), i.e., $S^{z} = 0$. The complete Hamiltonian of the knot lattice model includes the long range coupling along the loop current,
\begin{eqnarray}
H_{LS1}=\sum_{i,j,m,n} J_{ij} (J_{m} S^{z}_{ij}S^{z}_{i+m\textbf{e}_{x},j} + J_{n} S^{z}_{ij}S^{z}_{i,j+n\textbf{e}_{y}}),
\end{eqnarray}
Where $S^{z} = (+1,0,-1)$. If only the nearest neighboring coupling terms are included, this knot lattice model reduced to the conventional Ising model of spin-1 spins,
\begin{eqnarray}\label{Hspin1}
H_{S1}=\sum_{\langle{ij}\rangle} J_{z} S^{z}_{i}S^{z}_{j}.
\end{eqnarray}
Block spin-1 Ising model is not exactly solved so far. For ferromagnetic coupling $J_{z}<0$, the minimal energy state is magnetically ordered state with spins pointing up collectively or pointing down collectively. However the Hilbert space is highly enlarged due to the two degenerated spin-zero states. Upon one braiding operation of $R_{\circlearrowright}^{m^+e^+}$, the vacuum state and Majorana fermion state can coexist in the Hilbert sub-space with a zero eigen-energy. More braiding operations generate more Majorana fermions in other excited states. The knot configuration of the zero-energy states is made of many entangled loops with different sizes.

\begin{figure}
\begin{center}
\includegraphics[width=0.45\textwidth]{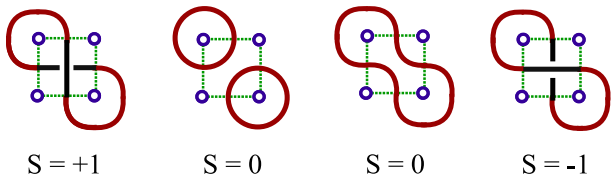}
\caption{\label{skein} The four knot states of single spin operator $S_{i}^{z}$. Obviously the two $S_{i}^{z}=0$ states are not topologically equivalent.}
\end{center}
\vspace{-0.5cm}
\end{figure}

Even though the lattice model is confined in two dimensions, the knot configuration is in fact in three dimensional space. Topological quantum field theory offers a method to calculate Jones polynomial and partition function of these links \cite{Witten}. For a chosen common lattice site, $i$, of the multi-layer knot lattice, the knot configuration of the rest lattice sites (except the lattice site $i$) is first fixed. For each fixed spin state, for instance $S_{i}^{z} = 0$, the partition function (or Feynmann path integral) of this layer could be computed as $Z(S_{i}^{z} = 0)_{0}$, here $0$ denotes ground state. The partition function of $Z(S_{i}^{z} = +1)_{0}$ and  $Z(S_{i}^{z} = -1)_{0}$ is obtained following the same procedure. Repeating the same computation on all of the other knot lattice layers of eigen-states, it leads to the partition function, $Z(S_{i}^{z} = 0) = \sum_{r = 0}^{M} Z(S_{i}^{z} = 0)_{r}$, $r$ represents eigen-energy levels. The partition function of the three states satisfy the familiar linear relation in topological quantum field theory and knot theory \cite{Witten},
\begin{eqnarray}\label{jones}
{\alpha}Z (S_{i}^{z} = +1) + {\beta}Z (S_{i}^{z} = 0) +{\gamma} Z (S_{i}^{z} = -1)=0.
\end{eqnarray}
The three coefficients $(\alpha,\beta,\gamma)$ in this Skein relation is computable for an explicit knot lattice state. Since partition function is a topological invariant, the linear combination of them is also topological invariant. Here the partition function depends on spin coupling strength and temperature. This topological relation is solid for finite system. It is convenient for quantum gate manipulations even though it is far from thermal dynamic limit. The single spin Hamiltonian, $H=hS_{i}^{z}$, has four knot states: one over-crossing, two zero-crossings and one under-crossing (Fig. \ref{skein}). The corresponding partition functions with respect to the four knot states are,
\begin{eqnarray}\label{Z101}
&&Z(S_{i}^{z} = +1) = e^{-\frac{h}{K_{b}T}},\;\; Z(S_{i}^{z} =0) = 1, \nonumber\\
&&Z(S_{i}^{z} = -1) = e^{-\frac{-h}{K_{b}T}}.
\end{eqnarray}
The two degenerated zero states are not distinguishable by partition function. The coefficients of the knot invariant Jones polynomial reads,
\begin{eqnarray}\label{Zjones}
\alpha = 1/t,\;\; \gamma = -t,\;\; \beta = \sqrt{t}-1/\sqrt{t};\;\;t = e^{-2\frac{h}{K_{b}T}}.
\end{eqnarray}
Here the abstract variable $t$ is a familiar Boltzmann factor in physics. Repeating this Skein recursion relation across the whole knot lattice generate a global knot invariant polynomial. Then transition between different crossing states is manipulated by braid group. Braiding operation can be performed upon ground state, which has the same bilayer knot configuration as two state Ising model. The braiding generates one vacuum and one Majorana state, both the two spins of the two layers at the $i$ lattice flipped to $S_{i}^{z}=0$ state. Note the zero crossing state has two fold degeneracy. The Jones polynomial now depends on external field strength and temperature. Combining the average linking number Eq. (\ref{averLink}) with the Skein relation equation (\ref{jones}) generates the average linking number for vacuum state $S_{i}^{z}=0$,
\begin{eqnarray}\label{links0}
\langle{L_{link}}{(S(0))}\rangle =\frac{1}{k_{b}T}\frac{\partial{{\ln}[\frac{-\alpha {Z(S(+1))}-\gamma{Z(S(-1))}}{\beta}]}}{\partial{h}}.
\end{eqnarray}
This equation offers one method for computing the average linking number in spin zero state as well as that of spin up state and spin down state. It is also useful for exploring topological phase transitions in quantum system of block spin-1 particles.

\begin{figure}
\begin{center}
\includegraphics[width=0.45\textwidth]{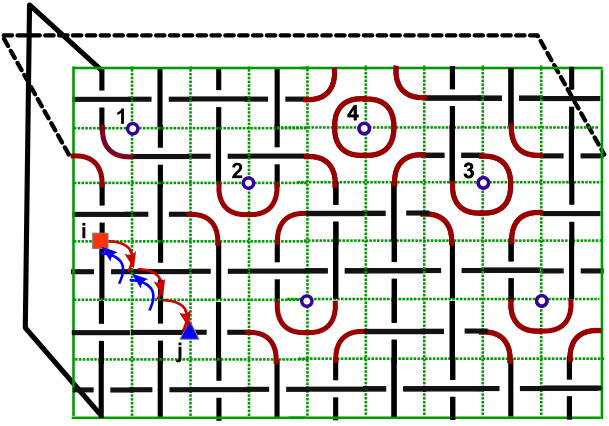}
\caption{\label{Hall} Electron gas runs through a spherical lattice of knot. A magnetic monopole is placed at center of this spherical lattice to exert an external magnetic field. The blue circle indicates magnetic flux. The blue triangle and red square indicates the far separated initial and final anyon correspondingly. The red arrow sequence bring the initial anyon to the final location. The blue arrow sequence brings the initial anyon back to its original location.}
\end{center}
\vspace{-0.5cm}
\end{figure}

\subsection{Electronic anyon states with integral and fractional filling factors in block spin-1 knot lattice model}

The knot lattice exposed to a homogeneous external magnetic field demonstrates the same integral and fractional quantum Hall effect. Replacing the magnetic monopole and positron in the knot lattice (Fig. \ref{spin12sphere}) with electrons is a natural implementation of two dimensional electron gas. If a magnetic monopole with magnetic charge $Q_{m}$ is placed at the center of the spherical lattice to exert a magnetic field perpendicular to the knot lattice plane (Fig. \ref{Hall}), this electron gas knot lattice shows similar quantum Hall effect with quantized Hall resistance. Hall voltage is defined on the four virtual edges in Fig. \ref{spin12sphere}. The Hall resistance tensor increases as the magnetic field strength increases \cite{Murthy}. A serial plateaus show up for certain magnetic field strength \cite{Murthy},
\begin{equation}\label{Rxy}
R_{xy} = \frac{h}{e^2}\frac{1}{\nu}, \;\;\nu = \frac{\phi_{0}\rho}{B},
\end{equation}
$\nu$ is the filling factor which counts how many electrons are filled into each magnetic flux quanta. $(\nu = 1,2,...,n)$ corresponds to integral quantum Hall effect. $\nu = n/(2pn\pm1)$ leads to fractional quantum Hall effect \cite{Murthy}. If the electronic current in this knot lattice either oriented along horizontal direction or in vertical direction (Fig. \ref{spin12sphere}), the off-diagonal terms of Hall resistance vanished, because electrons have no chance to bend their velocity from X- to Y-direction. Such kind of knot lattice state only exist for a zero magnetic field, i.e., the external magnetic charge is $Q_{m}=0$. A non-zero magnetic field bends the electric current. A finite value of Hall resistance $R_{xy}$ exist for a knot lattice state with zero-crossing states, i.e., the vacuum knot configuration in Fig. \ref{spin12} and red arcs showed in Fig. \ref{Hall}, which are called turning arcs in the following. These turning arcs only appear for odd number of braiding operations (Fig. \ref{braid3}). The output effect of the braiding operator of Eq. (\ref{braidings3}) is physically equivalent to an effective magnetic field operator,
\begin{equation}\label{RtoB}
[R_{\circlearrowright_{i}}^{m^+e^+}] = \hat{B}_{i}.
\end{equation}
Wherever one unit of magnetic field $\hat{B}$ is applied, the two anyons at the ending points of the electric current are exchanged in clockwise direction. The inverse operator of $\hat{B}$, i.e., $\hat{B}^{-1}$, braids the two anyons in counterclockwise direction (Fig. \ref{spin12}). The three lowest Landau levels of electron gas in strong magnetic field can be equivalently mapped into the three quantum states of block spin-1. Only odd number of times of braiding by magnetic field operators upon the state $|+1\rangle$ generates turning arcs in vacuum state (Fig. \ref{spin12}, Fig. \ref{spinx} and Fig. \ref{braid3}). The number of crossing points of these entangled turning arcs within one unit cell increase by one for each braiding operation. In order to unknot a pair of entangled arcs with $2n+1$ crossing points back to trivial vacuum state, there must be $n$ Majorana fermion operators acting on the crossing points alternatively. This geometric operation is summarized as following algebra,
\begin{eqnarray}\label{Bodd}
&&\hat{B}^{2n+1}|+1\rangle = \prod_{i=1}^{n}\psi_{i}|0\rangle,\;\;\hat{B}^{2n-1}|-1\rangle = \prod_{i=1}^{n}\psi_{i}|0\rangle,\nonumber\\
&&\hat{B}^{2n}|0\rangle = \prod_{i=1}^{n}\psi_{i}|0\rangle,\;\;[\hat{B}^{-1}]^{2n} |0\rangle = \prod_{i=1}^{n}\psi_{i}|0\rangle,\nonumber\\
&&[\hat{B}^{-1}]^{2n+1} |-1\rangle = \prod_{i=1}^{n}\psi_{i}|0\rangle.\nonumber\\
&&[\hat{B}^{-1}]^{2n-1} |+1\rangle = \prod_{i=1}^{n}\psi_{i}|0\rangle.
\end{eqnarray}
These Majorana fermion operators can be effectively implemented by electrons filled into a magnetic flux bundle which is denoted by the number of operations of Magnetic field operator. Since the magnetic field is homogeneously distributed, all lattice sites are braided simultaneously. The filling factor atone lattice site is the same as other lattice sites. Then we arrived at composite-fermions attached by magnetic field. The filling factor for these quasi-particles excited out of vacuum state is the same as that for fractional quantum Hall system \cite{Murthy},
\begin{equation}\label{filling}
\nu =\frac{L_{link}}{N{(B)}} =\frac{N(\psi)}{N{(B)}}= \frac{n}{2n\pm1};\;\;\nu = \frac{1}{2}.
\end{equation}
$L_{link}$ is linking number. $N(\psi)$ and $ N(B)$ are the number of Majorana fermion operators and the number of magnetic field operators correspondingly. This is a reasonable physical result, because the two entangled electric currents are equivalent to two solenoids which generates magnetic field passing through their interior. These highly entangled knot vacuum is a different implementation of composite fermions \cite{jain}.

\begin{figure}
\begin{center}
\includegraphics[width=0.42\textwidth]{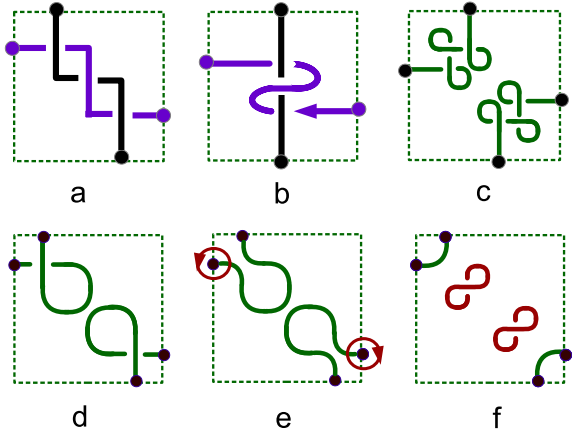}
\caption{\label{selflink} (a) Exchange two anyons within $|\pm1\rangle$ states with three crossings. (b) The equivalent solenoid helix scheme for the same braiding operation with respect to (a). (c) The odd number of many writhing loops generated on square lattice. (d) One writhing loop generated on one arc. (e) The twisted free arcs generated by one braiding. (f) The output of one braiding on the crossing point of writhing loop leads to one untwisted free arc and one free loop.}
\end{center}
\vspace{-0.5cm}
\end{figure}

As for the state transitions from $|+1\rangle$ to $|-1\rangle$ (or vice versa) by a number of braiding operations of magnetic field operator, it is equivalent to injecting one electric current through the interior center line of the solenoid of the other electric current (Fig. \ref{selflink} (a)(b)). Magnetic flux exists by pairs in this case. This state transition can exist at some lattice sites and can be generated from vacuum states which corresponds to resistance plateaus similar to quantum Hall effect. $2n$ numbers of braiding operations on $|+1\rangle$ generates $n$ Majorana fermions as its eigen-excitation,
\begin{eqnarray}\label{B101}
\hat{B}^{2}|+1\rangle &=& |-1\rangle,\;\;\hat{B}\hat{B}^{-1} = 1. \nonumber\\
\hat{B}^{4}|+1\rangle &=& \psi_{2}|-1\rangle = \psi_{1}\psi_{3}|+1\rangle,\nonumber\\
\hat{B}^{6}|+1\rangle &=& \psi_{2}\psi_{4}|-1\rangle = \psi_{1}\psi_{3}\psi_{5}|+1\rangle,\nonumber\\
\hat{B}^{2n}|+1\rangle &=& \prod_{i=1}^{n-1}\psi_{2i}|-1\rangle = \prod_{i=1}^{n}\psi_{2i-1}|+1\rangle,\nonumber\\
\hat{B}^{2n}|-1\rangle &=& \prod_{i=1}^{n}\psi_{2i}|-1\rangle = \prod_{i=1}^{n+1}\psi_{2i-1}|+1\rangle,\nonumber\\
(\hat{B}^{-1})^{2n}|-1\rangle &=& \prod_{i=1}^{n}\psi_{2i-1}|-1\rangle = \prod_{i=1}^{n-1}\psi_{2i} |+1\rangle,
\end{eqnarray}
so does the $|-1\rangle$ state. The filling factor for the quasi-excitations from $|+1\rangle$ to $|-1\rangle$ (or vice versa) obeys the following filling factors,
\begin{eqnarray}\label{filling2}
\nu = \frac{L_{link}}{N{(B)}} = \frac{N(\psi)}{N{(B)}}= \frac{n\pm1}{2n};\;\;\nu = \frac{1}{2}.
\end{eqnarray}
The filling factor in the eigen-energy level of $|+1\rangle$ or $|-1\rangle$ state is $\nu = n/2n = 1/2$. One unpaired Majorana fermion maps a $|+1\rangle$ to $|-1\rangle$ (or vice versa). If the total number of Majorana fermions approaches to infinity, the fractional filling factor also reaches this half-filling state. However this unpaired Majorana fermion always exist due to topological braiding.

The fractional filling factor above only appears for a local knot without self-linking. According to
White formula \cite{Pohl}, the self-linking number is the sum of twisting number ($T_{t}$) and writhing number $W_{w}$, i.e., $S_{link}=W_{w}+T_{t}$. The writhing number counts the number of loops made by the loop itself (Fig. \ref{selflink} (c)), while the twisting number counts the number of twisting around its circular column section center (assuming the wire here has finite thickness). The physical implementation of this twisting wire is an electric current generated by a running electron which carries a rotating spin. The twisting number and writhing number can transform into each other, but the sum of the two keeps a conserved self-linking number. Suppose the electric current are twisted during the braiding operation, then this twist would induce the writhing of current and form loops. For instance, there exists a writhing loop on the block square covering four unit squares around local site No. 3 in Fig. \ref{Hall}. Suppose there are two of such kind of writhing loops in a 3 $\times$3 block square covering 9 unit squares (Fig. \ref{selflink} (d)). Applying one braiding on the local crossing point unknots the writhing loop into two different types of free arcs: one is twisted free arc (Fig. \ref{selflink} (d)), the other is untwisted arc together with a free writhing loop (Fig. \ref{selflink} (e)). For the first case, $W_{w}=0$, $T_{t}=2$, the self-linking number is $S_{link}=2$. For the second case, $W_{w}=2$, $T_{t}=0$, then $S_{link}=2$. Thus each twist or each writhing loop is equivalent to one fermion. For a turning arc with many writhing loops or twisting units, it was born together with many fermions. One braiding by magnetic flux operator is filled with many Majorana fermions (Fig. \ref{selflink} (c),(d)). The filling factor for this case demonstrates similar filling factors like integral and half integral quantum Hall effect,
\begin{eqnarray}\label{filling3}
\nu = \frac{({L_{link}}+W_{w}+T_{t})}{N{(B)}} = \frac{n}{2},\;\;n=1,2,3,....
\end{eqnarray}
The linking number ${L_{link}} = {N_{i}(\psi)}$ counts the number of Majorana fermion operators to unknot entangled vacuum strings. ${N_{i}{(B)}}$ counts the total number of magnetic field operators. $W_{w}$ counts the number of writhing loops. $T_{t}$ counts the number of twisting. For example, there are two fermions encoded by the writhing loops in Fig. \ref{selflink} (d). After two braiding operations $B^{2}$, it turns into a vacuum state with two crossing points plus the two writhing loops. Then one Majorana fermions must be added to unknot this state, i.e., ${N_{i}(\psi)}=1$. In total, there are three free fermions in two magnetic flux, thus the filling factor is $\nu = 3/2$. The writhing number $W_{w}$ is an independent topological number for counting fermions, $W_{w}=0,1,2,\cdots.$ In mind of the additional twisting numbers on the ending points, the self-linking number could be even or odd. If the total number of fermions is odd, it reaches a half-integral filling factor, $\nu = n/2$. Otherwise, the filling factor is integral $\nu =n$. This integral or half-integral filling factor only exist at the eigen-energy level of vacuum state, $|+1\rangle$ and $|-1\rangle$. The quasi-particle excited out of these eigen-states obeys a similar fractional filling factor equations (\ref{filling}) (\ref{filling2}) by replacing $N(\psi)$ with $({N(\psi)}+W_{w}+T_{t})$. The braiding algebra is independent of space scale, thus it is quite robust under renormalization. For example, if the two free turning arcs in the vacuum state generate writhing loops in the same way, then the total writhing number is an even serial, $W_{w}=0,2,4,6,\cdots.$ If the twisting number is zero, the even writhing number leads to the half filling serial for filling factors in eigenenergy levels, $\nu = ({2n+1})/{2}$. For the quasi-particle or quasi-holes excited out of vacuum, even writhing number generates $\nu = ({n+W_{w}})/({2n\pm1})$. For instance, $n=1$ leads to the serial, $\nu = ({n+W_{w}})/(2n+1) = 1/3,1,5/3,7/3,3,\cdots$. $n=2$ leads to the serial, $\nu = ({n+W_{w}})/(2n-1) = 2/3, 4/3, 8/3, 10/3,4,\cdots$. Odd number of writhing loops only exist for the case that the two turning arc generates different number of writhing loops. The linking number of crossing current within one unit square could be equivalently viewed as the writhing number of a larger current crossing covering many unit squares. This equivalence induced one-to-one mapping between integral filling factor and fractional filling factors. This knot lattice model offers a topological explanation to Jain's composite fermion theory.

If the external magnetic field distribution is not homogeneous but has a fluctuating strength distribution within the scale of a few lattice sites, the filling factor would be more continuous. Even though it is quite hard to confine a magnetic field in a nanoscale circle, inhomogeneous doping of magnetic particle still provides a possible way. In this case, the crossing states in different block squares would be acted by different times of braiding operators at the same time and generate Majorana fermions at different locations simultaneously. In that case, the renormalized filling factor is computed by direct summations,
\begin{eqnarray}\label{filling4}
\nu = \frac{\sum_{i}({L_{link}}+W_{w}+T_{t})}{\sum_{i}N{(B)}}.
\end{eqnarray}
The linking number is ${L_{link}} = {N_{i}(\psi)}$. These filling factors are mostly likely showed as plateau in Hall resistance, but fits in the continuous straight lines. The writhing free loop in the counting above is equivalent to free fermion. While free loop without self-crossing is simply vacuum. It takes four fermionic turning arcs to form a vacuum loop. Thus the vacuum loop behaves as boson. Even number of electrons are trapped in vacuum loop and can not move freely in the whole space. If the whole lattice is covered by free vacuum loops without any exception point, the total linking number of this insulating state is zero. The Hall resistance of this insulating phase depend the oddness or eveness of the length of the virtual edge. If the total number of columns is even, then the local in- and out-current pair cancelled each other to reach a zero Hall current. If the total number of columns is odd, there always exists one unpaired in- or out-current on the edge. Thus the Hall conductance has an odd-even dependence on the finite size of lattice.

The quantum Hall effect for the knot lattice model of electrons is intrinsically originated from the Chern-Simons field theory: the non-abelian Chern-Simons action is a topological invariant of many entangled knots \cite{Witten}, while the abelian Chern-Simons theory determines the topologically quantized Hall resistance in fractional quantum Hall effect \cite{Zhang}. Here the Hamiltonian of electrons moving along the tangential vector of the horizontal loop or vertical loop can be formulated as the same for quantum Hall systems \cite{Murthy},
\begin{eqnarray}\label{hallhami}
H_{qh} &=& \sum_{i}\hbar\omega\left[(\textbf{P}_{i,x}+a_{cs,ix})^2+(\textbf{P}_{i,y}+a_{cs,iy})^2\right]+V,\nonumber\\
\textbf{P}_{i,x} &=& -i\partial_{i,x}+\textbf{A}_{i,x},\;\;\;\textbf{P}_{i,y}=-i\partial_{i,y}+\textbf{A}_{i,y},
\end{eqnarray}
here the potential term $V = V_{i}+V_{ij}+V(S)$. $V_{i}={e^2}/{d\epsilon}$ is the on-site repulsive interaction at each over crossing point with $d$ the distance between the upper current and lower current. $V_{ij}$ is column interaction between electrons. Since the on-site distance between electron is much smaller than the distance between electron at different lattice site, then $V_{i}{\gg}V_{ij}$. The on-site repulsion prevents two electric currents from touching each other, which results in knot current lattice. V(S) indicates the interaction between the twisting spin 1 of composite electron and magnetic field, which encoded the topological filling factors of quantum Hall effect. The gauge field vectors, $\textbf{A}_{i,x}$ and $\textbf{A}_{i,y}$, is induced by magnetic field and obeys symmetric gauge ($\textbf{A}_{i,x} = eBy/2$, $\textbf{A}_{i,y}=eBx/2$). Since electron are still moving in continuous channels, the continuous Hamiltonian theory for fractional quantum Hall effect also work here. Note here the on-site repulsion potential $V_{i}$ has periodical distribution. $a_{cs}$ is the Chern-Simons gauge field. 2p flux quanta is attached to electron under Chern-Simons transformation. In the knot lattice model, each ending point is attached by a flux quanta assigned by Chern-Simons field. The collective wave function of electron currents in knot lattice can be described by an extended Laughlin wave function,
\begin{eqnarray}\label{laughlin}
\Psi_{2n+1} &=& \prod_{a}\prod_{i<j} (z_{i,a}-z_{j,b})^{2n+1}(z_{i,a}-z_{i,b})^{2n+1} f(z),\nonumber\\
f(z) &=& \exp\left(-\sum_{i,a} |z_{i,a}|^{2}/4l^{2}\right).
\end{eqnarray}
$z_{i,a}$ represent the coordinate of the four ending points at the middle point of the edge of unit square, which is the ending point of two crossing strings. Here $a=|0\rangle$ represent the vacuum state. $b=|-1\rangle$ or $b=|+1\rangle$ represent the spin-up state or spin-down state. This Laughlin wave function indicates that two fermionic strings in the same unit square cell obeys fractional statistics. The composite fermions in different unit square cell also obey fractional statistics. For instance, suppose the ending point at the $i$th unit cell (the blue square in Fig. \ref{Hall}) exchanges its position with one point in the $j$th unit cell (the red triangle in Fig. \ref{Hall}). It takes three braiding to bring blue square to the position of red triangle, but takes only two inverse braiding to bring the red triangle to the home of blue square (Fig. \ref{Hall}). As a result, there is only one braiding survived at the $j$th unit cell. Thus braiding two fermions in different unit cell obeys the same statistics as that within one unit cell.

The collective wave function of other filling factors can be constructed by Jain's composite fermions theory \cite{jain}. Since these electrons are always running in closed loop which bears non-zero vorticity. Each loop of electric current also generate a magnetic field. The abelian Chern-Simons action counts the total helicity of these entangled knots. When the knot configuration fluctuates from one pattern to another, some knot inevitably will be cut and reunite, this induces some opposite magnetic fields against the external magnetic field due to the Lenz's law in electromagnetism theory. Thus fractional quantum Hall effect can still exist in this knot lattice model even if there is no external magnetic field. In that case, the total phase flux in different sublattices must cancel each other.

\subsection{ Anyons in long range hopping insulator model and quantum spin Hall model on square knot lattice}

The presence of magnetic field in square knot lattice breaks time reversal symmetry. Without external magnetic field but introducing spin-orbital coupling into the Hamiltonian results in quantum anomalous Hall effect \cite{Zhang}. The knot configurations provide a geometric representation of the fermion filling states of anomalous Hall Hamiltonian. On the square knot lattice Fig. (\ref{spin12sphere}), we define the crossing state that blue wire is below the black wire as the zero filling state of fermionic operator, i.e., $|0\rangle=|S=-1\rangle$. The output of annihilation fermion operator $c_{i}$ on $|0\rangle$ is zero. The creation fermion operator, $c^{\dag}_{i}$, generates one fermion out of zero filling state, $c^{\dag}_{i}|0\rangle=|1\rangle$. The one fermions sate is defined as the opposite crossing state of $|0\rangle$, i.e., $|1\rangle=|S=+1\rangle$ (Fig. (\ref{spin12sphere}) (e)). For spinless particle, Pauli principal forbids the existence of two fermions at the same site, thus $c^{\dag}_{i}c^{\dag}_{i}|1\rangle=0$. The annihilation operator brings the one fermion state to vacuum, $c_{i}|1\rangle=|0\rangle$. If the fermions bear intrinsic spin state, each current at the crossing point has to be oriented as the four crossing states (Fig. (\ref{spin12sphere}) (e)) to match the coupling between different spin states. We first focus on spinless fermions in the following. One long range fermion hopping model on the knot lattice model in Fig. (\ref{spin12sphere}) (a)(b) can be constructed as
\begin{eqnarray}\label{Lahallhami}
H_{Lhop}&=&\sum_{i,l,n}\frac{1}{2}\left[(c^{\dag}_{i}S_{z}c_{i+l\textbf{e}_x}+c^{\dag}_{i}S_{z}c_{i+n\textbf{e}_y})\right]\nonumber\\
&-&\sum_{i,l,n}\frac{e^{i\frac{\pi}{2}}}{2}\left[(c^{\dag}_{i}S_{x}c_{i+l\textbf{e}_x}+c^{\dag}_{i}S_{y}c_{i+n\textbf{e}_y})\right]\nonumber\\
&+&2mc^{\dag}_{i}S_{z}c_{i}+h.c.
\end{eqnarray}
The spin 1/2 operators ($S_{x}$, $S_{y}$,$S_{z}$) are global spin operator which couples to the horizontal (or vertical) fermion current. The Fourier transformation of fermion operators on this knot lattice is naturally anisotropic,
\begin{eqnarray}\label{cxyfourier}
c^{\dag}_{k_{x}}&=&\frac{1}{\sqrt{M}}\sum_{l}e^{-ik_{x}{l}\textbf{e}_x}c^{\dag}_{j+l\textbf{e}_x},\nonumber\\
c_{k_{y}}&=&\frac{1}{\sqrt{N}}\sum_{n}e^{ik_{y}{n}\textbf{e}_y}c_{j+n\textbf{e}_y}.
\end{eqnarray}
This two-band topological insulator model reduces to diagonal Hamiltonian in momentum space,
\begin{eqnarray}\label{ahallmomen}
&&H_{ah}=\sum_{k}[\sin(k_{x})c^{\dag}_{k_{x}}c_{k_{x}}S_{x}+\sin(k_{y})c^{\dag}_{k_{y}}c_{k_{y}}S_{y}\nonumber\\
&&+[mc^{\dag}_{k}c_{k}+\cos(k_{x})c^{\dag}_{k_{x}}c_{k_{x}}+\cos(k_{y})c^{\dag}_{k_{y}}c_{k_{y}}]S_{z}].
\end{eqnarray}
For the nearest neighboring coupling case, this knot lattice model naturally reduces to a conventional topological insulator model \cite{qiprb},
\begin{eqnarray}\label{ahallhami}
H_{ah}&=&\sum_{i}\frac{1}{2}\left[(c^{\dag}_{i}S_{z}c_{i+\textbf{e}_x}+c^{\dag}_{i}S_{z}c_{i+\textbf{e}_y}
+2mc^{\dag}_{i}S_{z}c_{i})\right]\nonumber\\
&-&\sum_{i}\frac{e^{i\frac{\pi}{2}}}{2}\left[(c^{\dag}_{i}S_{x}c_{i+\textbf{e}_x}+c^{\dag}_{i}S_{y}c_{i+\textbf{e}_y})\right]+h.c..
\end{eqnarray}

Non-zero Chern numbers exist for the energy spectrum with respect to different polarization degrees \cite{qiprb}. The Chern number in momentum space is not solely determined by the real space topology, since the same knot configuration acted by different Hamiltonian maps out different energy spectrum. Different Hamiltonian organizes the knot lattice layers in different way. However topological physics in real space area still encoded in momentum space. The linking number is defined as total number of positive crossing minus the total number of negative crossings. While here the negative crossing is represented by zero filling state, $n_{i} = c^{\dag}_{i}c_{i}=0$. The positive crossing is counted by the one fermions filling state, $n_{i} = 1$. Thus the Linking number in this fermion-spin coupling model is in fact the total number of fermions,
\begin{eqnarray}\label{ahalllink}
L_{link} = N_{c} = \sum_{i} c^{\dag}_{i}c_{i}.
\end{eqnarray}
The total number of fermions is a topological number in this knot square lattice. The diagonal electronic conductance of this knot lattice model is quantized by the number of unbroken channels in X-direction or Y-direction. The global spin component $S_{x}$ is coupled to running fermions in X-loops $\sum_{k}\sin(k_{x})n_{k_{x}}$. $S_{y}$ is coupled to running fermions in Y-loops $\sum_{k}\sin(k_{y})n_{k_{y}}$. However the $S_{z}$ component is not only coupled to on-site occupation, but also coupled to the fermion current in X- and Y-loops. As along as the fermion numbers in X- or Y-loops are not zero, they will contribute to the $S_{z}$ component.

The real space Hamiltonian Equation (\ref{ahallhami}) assigned a spin $S_{z}$ component on each crossing point. The loop currents along X-direction carry $S_{x}$. The loop currents along Y-direction carry $S_{y}$. The evolution of each spin component is governed by Heisenberg equation,
\begin{eqnarray}\label{szheisen}
\partial_{t}S_{z}=\frac{1}{\hbar}[\sum_{k_{x}}\sin(k_{x})n_{k_{x}}S_{y}-\sum_{k_{y}}\sin(k_{y})n_{k_{y}}S_{x}].
\end{eqnarray}
In the continuum limit, the right hand side of Eq. (\ref{szheisen}) is equivalent to Rashba spin-orbital coupling. For a constant polarization $\partial_{t}S_{z}=0$, then $\sum_{k_{x}}\sin(k_{x})n_{k_{x}}{\langle}S_{y}{\rangle}=\sum_{k_{y}}\sin(k_{y})n_{k_{y}}{\langle}S_{x}{\rangle}$. In the classical representation of spin, $S_{x}$ can be expressed as the projection of a total spin,
\begin{eqnarray}\label{sxsycos}
{\langle}S_{x}{\rangle}=S\cos(\theta),\;\;{\langle}S_{y}{\rangle}=S\sin(\theta).
\end{eqnarray}
Then the stable configuration of spin components obeys equation,
\begin{eqnarray}\label{sxsycos==}
\sum_{k_{x}}\sin(k_{x})n_{k_{x}}\sin(\theta)=\sum_{k_{y}}\sin(k_{y})n_{k_{y}}\cos(\theta).
\end{eqnarray}
The orientation of spin in plane is labeled by the projection angle $\theta$. Obviously ${\langle}S_{y}{\rangle}$ increase when ${\langle}S_{x}{\rangle}$ decreases. In order to fulfill the balance Eq. (\ref{sxsycos==}), the total number of fermions in X-loops has to be reduced, in the meantime, the total number of fermions in the Y-loop must increase. Since the total fermions number is a conserved number, there must exist turning arcs in the knot square lattice to fuse the X-loop into Y-loop, driving the fermions from X-loop into Y-loop. In this sense, the output effect of spin-orbital coupling is equivalent to an external magnetic field. Since the global spin plays the same action at every lattice site, the turning arc shows up around every lattice site.

The Hall conductance of this two band model is quantized by the first Chern number in momentum space \cite{qiprb}. As all know, the energy function derived in real space model is exactly the same as its equivalent model in momentum space, even though sometimes it is quite difficult to get the formulation of energy spectrum in real space. Fourier transformation does not change the intrinsic topology of the energy manifold, except a coordination transformation from space index into wave vector index. The momentum space is the reciprocal space of real space. However the wave vector in thermal dynamics limit turns into a continuum variable. While the space index for the Hamiltonian in real space is still discrete. If we consider a finite system with finite particle number and lattice size. A knot in real space can still map into a knot in momentum space, maybe it is expressed into different geometry, but its topology should remain the same. As an explicit example to support above conjecture, we study an extreme simple case that both the fermion occupation in X-loop and Y-loop are single occupied, $n_{k_{x}}=n_{k_{y}}=1$. For a general consideration, the lattice constant $a_{x}$ in X-loop is controlled at a different value from that of Y-loops, i.e., $a_{x}\neq{a_{y}}$. A vector in real space, $R = n_{x}a_{x}+n_{y}a_{y}$, is dual vector of the reciprocal vector in momentum space, $K = k_{x}b_{x}+k_{y}b_{y}$, here ($b_{x}=2\pi/a_{x}, b_{y}=2\pi/a_{y}$). The two dual vectors obey the unit relation, $a_{i}\cdot{b_{j}}=2\pi\delta_{ij}$. The coupling current of the three spin components are listed as following:
\begin{eqnarray}\label{sxsysz3}
{\langle}S_{x}{\rangle}&:&I_{x}=\sum_{k_{x}}\sin(k_{x}),\nonumber\\
{\langle}S_{y}{\rangle}&:&I_{y}=\sum_{k_{y}}\sin(k_{y}),\nonumber\\
{\langle}S_{z}{\rangle}&:&I_{z}=\sum_{k}[m+\cos(k_{x})+\cos(k_{y})].
\end{eqnarray}
In order to encode the lattice constants into the equations above, we reformulate the wave vector components as
\begin{eqnarray}\label{kxkylamda}
k_{x}= 2\pi\omega_{x}k,\;\;k_{y}= 2\pi\omega_{y}k,
\end{eqnarray}
where $\omega$ is the spatial oscillation frequency of the current defined by ($\omega_{x} = 1/a_{x}$,$\omega_{y} = 1/a_{y}$), which counts how many unit lattice length is covered by one wavelength. For a general consideration, a phase factor $\phi_{i}$ is introduced into each current, then the three coupling fermion currents read,
\begin{eqnarray}\label{IxIyIz3}
I_{x}&=&\sum_{k}\sin(2\pi\omega_{x}k+\phi_{x}),\nonumber\\
I_{y}&=&\sum_{k}\sin(2\pi\omega_{y}k+\phi_{y}),\nonumber\\
I_{z}&=&\sum_{k}[m+\cos(2\pi\omega_{x}k+\phi_{x})\nonumber\\
&+&\cos(2\pi\omega_{y}k+\phi_{y})].
\end{eqnarray}
The three current functions above define a Fourier knot in
momentum space \cite{Soret}. One familiar example of Fourier knot
for physicist is Lissajous knots \cite{Bogle}, which is usually
visualized by oscilloscope. Inputting two sinusoidal electric
currents into the vertical and horizontal channels at the same
time, the oscilloscope displays different closed loops with
respect different to different frequency ratios.

\begin{figure}
\begin{center}
\includegraphics[width=0.47\textwidth]{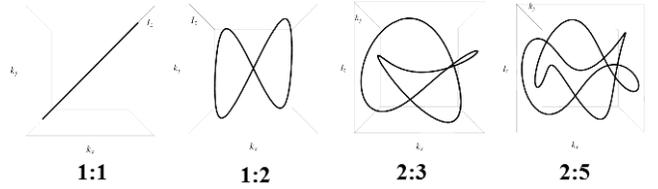}
\caption{\label{knot} The Lissajous knots in momentum space for original fermion current $I_{i}$ with different frequency ratios $\omega_{x}:\omega_{y}$.}
\end{center}
\vspace{-0.5cm}
\end{figure}

The fourier knot is not always a closed knot. Closed knot only appears if the wavelength ratio ($\omega_{x}:\omega_{y}$) is a rational number. For instance, if ($\omega_{x}:\omega_{y}=1:1$, $\phi_{x}=\phi_{x}=0$), the Lissajous curve is a circle and project a straight line in $k_{x}-k_{y}$ plane. The knot for $\omega_{x}:\omega_{y}=1:2$ is wave circle in three dimensions but project a $\infty$ shape in $k_{x}-k_{y}$ plane (Fig. (\ref{knot})). For higher number of ratios, the Lissajous knot demonstrate more fluctuations (Fig. (\ref{knot})). The Fourier knot configuration of the original fermion current $I_{i}$ is independent of the value of polarization $m$. Different $m$ simply shift the whole knot upward or downward. However this is not the case for a normalized fermion current.

\begin{figure}
\begin{center}
\includegraphics[width=0.45\textwidth]{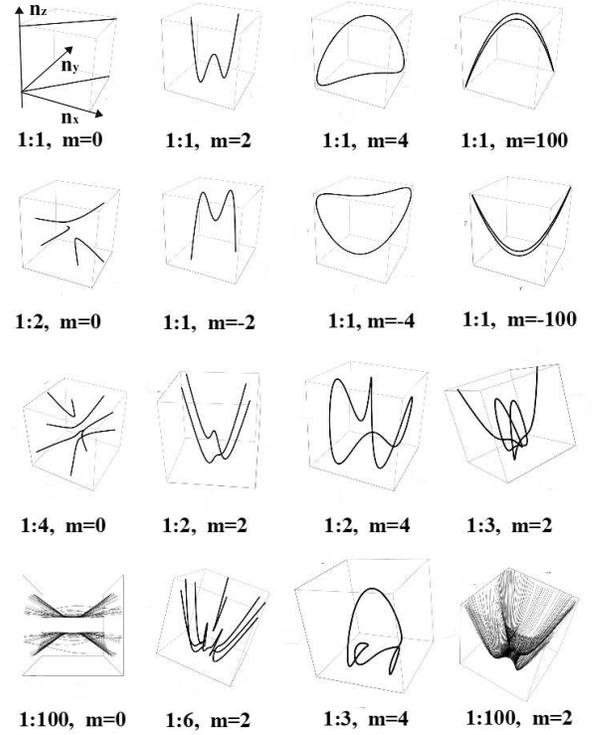}
\caption{\label{knot2} The Lissajous knots in momentum space for normalized fermion current $n_{i}$ with different frequency ratios $\omega_{x}:\omega_{y}$ and magnetization degree $m$.}
\end{center}
\vspace{-0.5cm}
\end{figure}

The topological Chern number is defined by normalized fermion current in momentum space \cite{qiprb}\cite{Duan}. Here we choose the similar formulation of three fermion current as Ref \cite{qiprb} for simplicity but with adjustable frequency, ($I_{x}=\sin(\omega_{x}k)$, $I_{y}=\sin(\omega_{y}k)$, $I_{z}=[m+\cos(\omega_{x}k)+\cos(\omega_{y}k)]$), then use the total energy spectrum, $E(k)=\sqrt{I^2_{x}+I^2_{y}+I^2_{z}}$, to normalize the fermion currents,
\begin{eqnarray}\label{nxnynz}
n_{x}&=&{\sin(\omega_{x}k)}/{E}, \;\;\;\; n_{y}=\sin(\omega_{y}k)/{E},\nonumber\\
n_{z}&=&[m+\cos(\omega_{x}k)+\cos(\omega_{y}k)]/{E}.
\end{eqnarray}
This normal fermion current also demonstrates three dimensional knot in momentum space. We first study the special case, $\omega_{x}:\omega_{y}=1:1$. The knot configuration shows different geometry with respect different polarization degree $m$. For m = 0, it is two parallel lines instead of a closed loop. For $0< m <3.5$, the normal current is an upward parabola with double wells instead of a closed loop. For $3.5 < m$, the normal current is a closed loop, which approaches to a downward parabola shape for $m\rightarrow\infty$. For $-3.5 < m < 0$, the normal current is an downward parabola with double wells. For $ m < -3.5$, the normal current is also a closed loop but approaches to an upward parabola (Fig. \ref{knot2}).

\begin{figure}
\begin{center}
\includegraphics[width=0.48\textwidth]{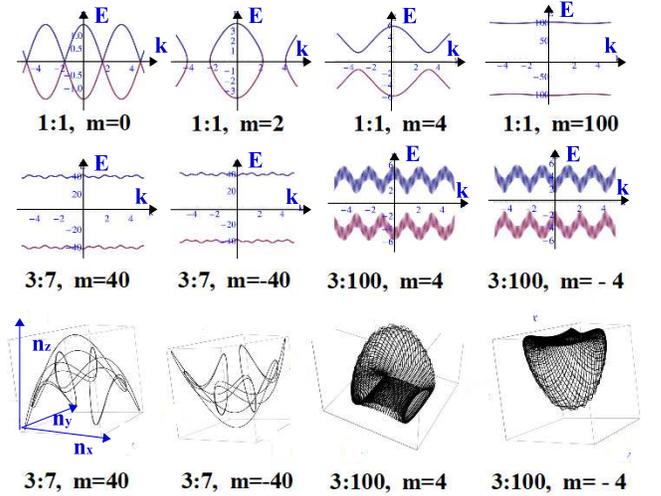}
\caption{\label{knot22} The energy spectrum $E(k)$ with respect to different frequency ratios and magnetization value $m$. The bottom row shows the Lissajous knots of normalized fermion current $n_{i}$ for large frequency ratios $\omega_{x}:\omega_{y}$, i.e., ($3:7, 3:100$).}
\end{center}
\vspace{-0.5cm}
\end{figure}

The normal fermion current for other frequency ratios shows more fluctuating knots in momentum space. Different value of magnetization $m$ classified the knots or unknots into different zones. For m = 0, there only exist parabola curves instead of closed loops. For most cases, the total number of these parabola curves equals to the sum of the two frequency number. As showed in Fig. (\ref{knot2}), there are 2 branches for $\omega_{x}:\omega_{y}=1:1$, three branches for $1:2$ and five branches for $1:4$. However, this rule does not hold for all cases, there are only two branches for $1:3$, 6 branches for $3:5$, and 10 branches for $5:7$, but only 12 branches for $7:9$. The physics reason for this serial is not fully understand yet. However, the output of extremely high frequency ratios are the same, it all leads to two separated band with different edge branches (Fig. (\ref{knot2}) 1:100, m=0).

The Lissajous curves for a magnetization $m<3.5$ are always a collection of parabola curves for arbitrary frequency ratio $\omega_{x}:\omega_{y}$. For example, the output curve of $\omega_{x}:\omega_{y}=1:2$ with $m=2$ is two upward parabola curves with double wells, which turn into downward parabola curves for $m=-2$ (Fig. (\ref{knot2})). While the ratio $1:3$ with $m=2$ generates one parabola curve with three wraps (\ref{knot2}). $\omega_{x}:\omega_{y}=1:6$ results in four deformed upward parabola curves (Fig. (\ref{knot2})). The high frequency ratio with $m=2$ finally converges to a cup-like network with four touching point on the bottom (Fig. (\ref{knot2}), (1:100, m=2)).

Closed Lissajous knot only exist for $m>3.5$. The $1:1$ case is a loop without self-writhing loops. While $1:2$ and $1:3$ generate a loops with two writhing loops and three writhing loops correspondingly (Fig. (\ref{knot2})). For high magnetization $m=40$, the Lissajous knot form a ten-like knot with many writing loops (Fig. (\ref{knot22}), 3:7, m=40), which turns into a tent-like network cage, as showed by the case of $3:100, m=4$ (Fig. (\ref{knot22})). These Lissajous knots approaches to tent-like parabola network for high magnetization $m\rightarrow\infty$, one example is showed in Fig. \ref{knot2}, 1:1, m=100.

The topological quantum field theory of non-abelian Chern-Simons action provides a topological invariant for many entangled knots. While Fourier knot is only a special case of general link with many entangle knot. Thus we could introduce the Chern-Simons action \cite{Witten}\cite{Duan} to quantify the Fourier knots here. Another topological invariant for the collection of all of these knots is partition function. The partition function for knots in real space share the same formulation as that for the knot in momentum space. These topological invariants are global topological invariant, sometimes the local topological invariant for a special eigenstate is really relevant to experiment measurement. For instance, the ground state and the first excited state is the most concern for physicists. For any knot configuration of ground state, Euler number is always a topological invariant. The Euler number of a closed curve which is homotopic to circle is always zero. The Euler characteristic number for a continuous closed curve can be computed by Morse theorem. For a given knot in momentum space, there always exist some critical points at which fermion current satisfies $\partial_{k_{s}}n_{s}=0$. For instance, the $\infty$ shaped knot in $k_{x}-k_{y}$ plane (Fig. (\ref{knot})) has 6 critical points. The local curve of two critical points on the upper boundary is approximated by $n(k_{x})=- a k_{x}^2$, and $n(k_{x})= a k_{x}^2$ is for two points on the bottom boundary. The critical point on the left boundary is $n(k_{y})= b k_{y}^2$, and $n(k_{y})= - b k_{y}^2$ for the right one. Then the Euler characteristic number is computed by the morse theorem \cite{nash},
\begin{eqnarray}\label{eulermorse}
\chi=\sum_{q=0}^{M}(-1)^{q}c_{q},
\end{eqnarray}
$q$ is an index counting the independent directions in which the current decreases. For the $\infty$ shaped knot, there are two points with $q=1$ above, two points with $q=0$ on the bottom, one $q=0$ for the left boundary and one $q=1$ for the right boundary. $c_{q}$ counts the total number of points with index $q$. The Euler characteristic number of this $\infty$ shaped knot is $\chi= (-1)^{0}3+(-1)^{1}3 = 0$. For the Lissajous knot, the number of critical points of the horizontal critical point to that of the vertical point is directly readable by the frequency ration on oscilloscope, $c^{x}_{q}:c^{y}_{q} = \omega_{x}:\omega_{y}$. For the magnetization $m>3.5$ or $m<-3.5$, the fermion currents are closed curves, thus the Euler number is zero.

The fermions current for the other two cases is not homotopic to knot anymore. In that case, Euler-Poincare equation is more effective for computing the topological numbers \cite{nash},
\begin{eqnarray}\label{eulerpointcaree}
\chi=\sum_{q=0}^{M}(-1)^{q}b_{q},
\end{eqnarray}
Here $b_{q}$ is the Betti number, which counts the number of the $q$ dimensional simplex. For instance, the $0$ dimensional simplex is a point. $1$ dimensional simplex is a line segment. $2$ dimensional simplex is a 2D surface. For the zero magnetization case, $m=0$, the fermion current in different frequency ration are composed of curves, In that case, these curve approaches to parabola curve for an infinite wave vector. For this knot lattice system, the momentum wave vectors has a cut-off at the unit lattice space, $k_{x}=1/a_{x}$ and $k_{y}=1/a_{y}$. In that case, there always exist two ending points ($b_{0}=2$) and one line ($b_{1}=1$) for each branch, thus the Euler number is $\chi=(-1)^{0}2+(-1)^{1}1=1$. This Euler number has the same value for the parameter range, $0< m <3.5$ or $-3.5 < m < 0$, but the fermion current for this case has only one band. This is because a finite magnetization $m$ breaks time reversal symmetry. Similarly, the fermion current for $m>3.5$ or $m<-3.5$ also has only one band, as shown by the tent-like cage ($3:100, m=4$ in (Fig. (\ref{knot22}))). The Euler number of a two dimensional simplex is equivalent to the first Chern number on manifold in its continuum limit. However the Euler number above are computed on one dimensional curves instead of two dimensional surface.

The energy spectrum shows closed knots in momentum space only exist for a gapped two band model (Fig. (\ref{knot22})). If the two bands form a periodic closed spectrum loops, its corresponding current knot in momentum space are collections of parabola with double wells. The two bands intersecting each other for zero magnetization $m=0$. Its corresponding current knot is a pair of separated flat branches. For a larger magnetization $m=40$, the two band in spectrum becomes almost flat band with fine wavy structure which is induced by the interference between the waves with two frequencies, 3 and 7. A reversed magnetization induced a phase flip of the spectrum wave. In the meantime, the corresponding knot in momentum space switched to opposite direction. For a larger frequency ratio, $3:100$, the spectrum wave becomes a modulated composite wave with many fine wave in each macroscopic wave section (Fig. (\ref{knot22})), the corresponding knot in momentum space is a tent-like network (Fig. (\ref{knot22})). For a finite lattice, these exist a gapless edge state on the boundary.

\begin{figure}
\begin{center}
\includegraphics[width=0.35\textwidth]{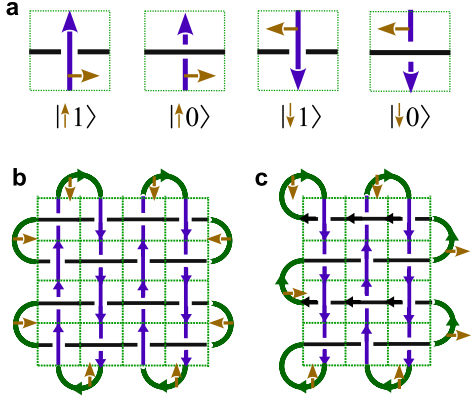}
\caption{\label{spinhall} (a) The four crossing states with respect to different spin states. The red arrow indicates spin. (b) The spin current on the edge with even number of Y-channels. (c) The spin current on the edge with odd number of Y-channels.}
\end{center}
\vspace{-0.5cm}
\end{figure}

Topological insulator model only consider the coupling between the orbital of spinless fermion and a global spin operator \cite{qiprb}. Thus the knot square lattice implementation of topological insulator model only incorporate undirected fermion current in X- and Y-loop. $|0\rangle$ state was defined as vertical current above horizontal current, while the opposite setup defines $|1\rangle$. The spin-up and spin-down fermion have a natural implementation by directed fermion currents in the knot lattice. For charged spin fermions in the loop channel, an electric field confined in the x-y plane could induce quantum spin Hall effect \cite{Bernevig}. Here electric field is oriented along the y-axis to drive electrons running in the Y-loops. Then the spin up state $|\uparrow0\rangle$ is defined as a state that the positive Y-current is above the X-current, $|\uparrow0\rangle$ corresponds to that the Y-current is below the X-current (Fig. \ref{spinhall} (a)). The negative Y-current defines the spin-down states correspondingly (Fig. \ref{spinhall}). The direction of spin is perpendicular to the electric current and electric field, following the equation $J^{i}_{j} = \sigma_{s}\epsilon_{ijk}E_{k}$. The action of spin fermion operators obey the following rules, $c^{\dag}_{{\uparrow}}|\uparrow0\rangle=|\uparrow1\rangle$, $c^{\dag}_{{\uparrow}}|\uparrow1\rangle=0$, $c_{{\uparrow}}|\uparrow1\rangle=|\uparrow0\rangle$, $c_{{\uparrow}}|\uparrow0\rangle=0$. The operation of spin-down fermion operator follows similar rules. The effective Hamiltonian for this spin Hall system is composed of two topological insulator model Eq. (\ref{ahallhami}), but incorporate an opposite Y-current,
\begin{eqnarray}\label{spinhallhami}
H_{s}(k)=\left(
        \begin{array}{cc}
          H_{\uparrow}(k) & 0 \\
          0 & H^{\ast}_{\downarrow}(-k)\\
        \end{array}
      \right).
\end{eqnarray}
Here the Hamiltonian $H_{s}(k)$ share the same formulation as topological insulator model Hamiltonian Eq. (\ref{ahallmomen}) in momentum space. This Hamiltonian reduces to the effective Hamiltonian of quantum spin Hall insulator near the $\Gamma$ point \cite{Konig}. The spin Hall current also turns from Y-loop into X-loop due to spin-orbital coupling interaction (Fig. \ref{spinhall} (a)). The gapless edge current runs along the edge without dissipation. The spin current on the upper edge carries opposite spin and runs in opposite direction to that of the spin current on the bottom edge (Fig. \ref{spinhall} (b)), so does the left edge and right edge. However, if the number of Y-loops is an odd number, the spin current on the left edge runs in the same direction as the spin current on the right edge that carries the same oriented spins (Fig. \ref{spinhall} (b)).

\subsection{ The long-range fermion pairing model on knot square lattice }

\begin{figure}
\begin{center}
\includegraphics[width=0.35\textwidth]{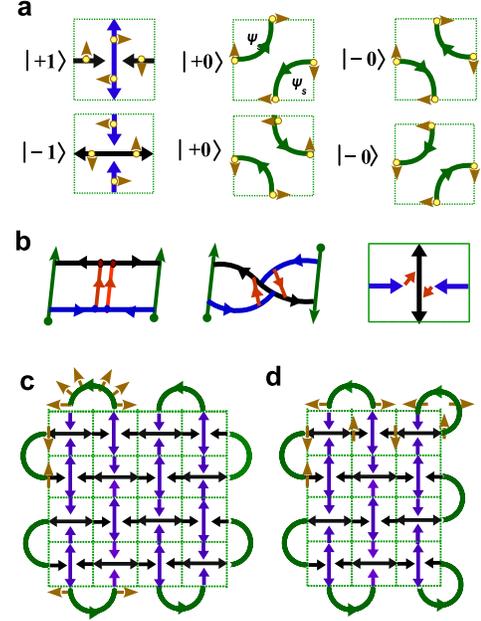}
\caption{\label{BCSpairing} (a) The two different fermion pairings with respect to two crossing states, and the corresponding four vacuum states. (b) The crossing currents as a projection of the two edges of Mobius strip. (c) The spin current on the edge of pairing square lattice with even number of Y-channels. (d) The spin current on the edge of pairing square lattice with odd number of Y-channels.}
\end{center}
\vspace{-0.5cm}
\end{figure}

The anyon current model above only considered continuous current without any local confliction. If we consider the conflicting patterns of directed anyon current at each local crossing point, this knot lattice model demonstrates a topological fermion pairing phenomena. For the nearest neighboring hopping case, the knot lattice model reproduces similar momentum pairing Hamiltonian with respect to the well-known Bardeen-Cooper-Schrieffer(BCS) fermion pairing model for conventional superconductor \cite{Tsuei}. Electrons with opposite spin and momentum are coupled into pairs as the main
carrier of super-conductor current. If we only consider the pairing pattern between the nearest neighboring crossing sites, the self-consistent construction
of fermions pairing in knot square lattice has only two possible
configurations (Fig. \ref{BCSpairing} (a)). An alternative
distribution of the $|+1\rangle$ and $|-1\rangle$ crossing states
construct one stable fermions pairing state. Flipping $|+1\rangle$
to $|-1\rangle$ (or vice versa) on the whole lattice is another
equivalent paring state (Fig. \ref{BCSpairing} (a)). Thus the
fermion paring on knot square lattice has two fold degeneracy. The
fermion pairing Hamiltonian in the two dimensional bulk area reads,
\begin{eqnarray}\label{pairhami}
H^{xy}_{bulk}&=&-V\sum_{{ij}}[c^{\dag}_{i-1,j\uparrow}c^{\dag}_{i-1,j\downarrow}c_{i-1,j-1\uparrow}c_{i-1,j+1\downarrow}
\nonumber\\
&+&c^{\dag}_{i,j-1\uparrow}c^{\dag}_{i,j+1\downarrow}c_{i,j\uparrow}c_{i,j\downarrow}
+c^{\dag}_{i,j\uparrow}c^{\dag}_{i,j\downarrow}c_{i-1,j\downarrow}c_{i+1,j\uparrow}\nonumber\\
&+&c^{\dag}_{i-2,j\downarrow}c^{\dag}_{i,j\uparrow}c_{i-1,j\uparrow}c_{i-1,j\downarrow}]+\sum_{{ij,s}}\epsilon_{ij}c^{\dag}_{ij,s}c_{ij,s}.
\end{eqnarray}
The Fourier transformation of fermions have separate wave vectors
in X-loops and Y-loops correspondingly,
$c_{i,s}=\frac{1}{\sqrt{N}}\sum_{k}e^{ik_{x}x_{i}}c_{k_{x},s}$,
and
$c_{i,s}=\frac{1}{\sqrt{N}}\sum_{k}e^{ik_{y}y_{i}}c_{k_{y},s}$.
Substituting this Fourier transformation into the Hamiltonian Eq.
(\ref{pairhami}) in real space leads to
\begin{eqnarray}\label{Hxybulk}
H^{xy}_{bulk}(k)=\sum_{k}\epsilon_{k}c^{\dag}_{k}c_{k}-\sum_{k,k'}V c^{\dag}_{k\uparrow} c^{\dag}_{-k\downarrow} c_{k'\downarrow} c_{-k'\uparrow}.\;\;
\end{eqnarray}
This Hamiltonian bears the same structure as the BCS Hamiltonian
in momentum space. Following the usual mean field approach, we
define the same energy gap function for exciting a Cooper pairing,
$\Delta=\sum_{k}V\langle{c_{k}c_{-k}}\rangle.$ Usually this energy
gap is a complex function, $\Delta=\Delta_{1}+i\Delta_{2}$,
$\Delta^{\ast}=\Delta_{1}-i\Delta_{2}$. The bulk pairing
Hamiltonian can be formulated as a fermion spinor
$\psi_{k}^{\dag}=[c_{k}^{\dag},c_{-k}]^{T}$ coupled to a
pseudo-spin vector, $\vec{\sigma}$,
\begin{eqnarray}\label{pseudospinfermi}
H^{xy}_{bulk}(k)=\psi_{k}^{\dag} [\epsilon_{k}\textbf{I} -
\Delta_{1}S_{x}+\Delta_{2}S_{y}]\psi_{k}.
\end{eqnarray}
Here $\textbf{I}$ is a 2 by 2 unit matrix. $S_{x}$ and $S_{y}$ are
conventional Pauli matrices. For different pairing states \cite{Tsuei}, this
paring Hamiltonian defines different Fourier knot in momentum. For instance, the p-wave pairing gap function defines a typical Fourier knot \cite{Soret},
\begin{eqnarray}\label{pairIxIyIz}
I_{x}&=&\Delta_{1}=\sum_{k}\sin(2\pi\omega_{x}k+\phi_{x}),\nonumber\\
I_{y}&=&\Delta_{2}=\sum_{k}\sin(2\pi\omega_{y}k+\phi_{y}),
\end{eqnarray}
However here there is no $\sigma_{z}$ component. Thus the pairing
gap function defines two dimensional Lissajous curves in momentum
space. The fermion pairing state here is in fact single plaquette
state in each unit square. It is an antiferromagnetic order state
for the block Ising spin $|+1\rangle$ and $|-1\rangle$ (Fig.
\ref{BCSpairing} (a)) in real space. The self-consistent pairing
state is the two fold degenerated ground state of the Ising
Hamiltonian for coupling block spin, $H_{pair} = J
S^{z}_{i}S^{z}_{j}$, with $J>0$. The eigenstate of $S^{z}_{i}$ is
the block spin states showed in Fig. \ref{BCSpairing} (a). Two
neighboring unit squares with same block spin can not match each
other self-consistently. Two fermions with opposite spins would
collide each other on the square boundary where no channel exist
for them to continue the current without turning back. While
inside each unit square, the two incoming fermions of X-loop could
tunnel up into the Y-channel and split up to fit into the
continuous current loops without frustration. This convective
fermion current actually keeps the total number of fermions
conserved (Fig. \ref{BCSpairing} (b)). The tunneling current from bottom current to upper current is also a fermion pairing Hamiltonian,
\begin{eqnarray}\label{pairhamiZ}
H_{bulk}^{z}=-V\sum_{{i,j}}[c^{\dag}_{i,j,u\uparrow}c^{\dag}_{i,j,u\downarrow}c_{i,j,b\uparrow}c_{i,j,b\downarrow}-h.c.],
\end{eqnarray}
Here the bottom current (upper current) is denoted by $b$ ($u$). A normal state is not fermion pairing state, the local crossing state maybe randomly distributed over the whole lattice. Then we have to use multi-layer knot lattice to represent the superposition of quantum states. For a periodically located multi-layer knot lattice, the paring Hamiltonian $H_{bulk}^{z}$ maps into four fermion interaction in momentum space by Fourier transformation. Because the fermion pair only moves upward along Z-axis, the time reversal symmetry is broken. The tunneling Hamiltonian under mean field approximations reads,
\begin{eqnarray}\label{pairhamiZk}
H_{bulk}^{z}(k) = -\sum_{{k}}[n_{\downarrow}(k')c^{\dag}_{k,\uparrow}c_{k,\uparrow}+n_{\uparrow}(k')c^{\dag}_{k,\downarrow}c_{k,\downarrow}-h.c.]\nonumber\\
\end{eqnarray}
Here $\Delta{N}(k_{z}) = N_{\downarrow}(k') - N_{\downarrow}(-k')$,
(with $N_{\downarrow}(k')= V\sum{\langle}c^{\dag}_{k',u\downarrow}c_{k',b\downarrow}\rangle$),
denotes the average occupation number difference between positive wave vector and negative wave vector. Because only the fermion moving to positive Z-axis reduce energy, an oppositely moving fermion would increase the total energy. In order to match the formulation of Hamiltonian Eq. (\ref{Hxybulk}), we replace $k$ with $-k$ in fermions operator $c_{k,\downarrow}$. Since the spin-up and spin-down move together upward as a pair, $n_{\uparrow}(k')$ should have the same occupation as $n_{\downarrow}(k')$. The Hamiltonian for Z-current reduces to a brief formulation,
\begin{eqnarray}\label{pairhamiSZ}
H_{bulk}^{z}(k) = \sum_{{k}}\psi_{k}^{\dag} [\Delta{N}(k_{z}) S_{z}]\psi_{k}.
\end{eqnarray}
$S_{z}$ is the Pauli matrices. Then the total Hamiltonian of fermion pairing is a three dimensional spin coupled to fermions, $H_{bulk}(k)=H_{bulk}^{xy}(k)+H_{bulk}^{z}(k)$,
\begin{eqnarray}\label{pseudospinfermi+SZ}
H_{bulk}(k)=\psi_{k}^{\dag} [\epsilon_{k}\textbf{I} -
\Delta_{1}S_{x}+\Delta_{2}S_{y}+\Delta{N}(k) S_{z}]\psi_{k}.
\end{eqnarray}
The gap function and the occupation number together defines a Fourier knot in three dimensional momentum space, $\vec{I}=[I_{x}=\Delta_{1},I_{y}=\Delta_{2},I_{z}=\Delta{N}]$. The energy spectrum of this mean-field Hamiltonian is $E(k)=\epsilon_{k}\pm\sqrt{\Delta{N}^2+|\Delta|^2}$. Following Duan's topological current theory of magnetic monopole \cite{duanmonopole}, a topological particle sits at the singular point of the normalized energy current, ${n_{a}}=I_{a}/\sqrt{\Delta{N}^2+|\Delta|^2}$, $(a = x,y,x)$. The topological current of magnetic monopole is
\begin{eqnarray}
C_{hern}=\frac{1}{2\pi}\int{dk^3}\epsilon_{ijl}\epsilon^{abc}\partial_{k_{i}}n^{a}\partial_{k_{j}}n^{b}\partial_{k_{l}}n^{c}.
\end{eqnarray}
Here $\epsilon_{ijl}$ is Levicivita symbol, $\epsilon_{ijl} = -\epsilon_{jil}$. Non-trivial topological particle exist in the gapless mode, i.e.,
$\sqrt{\Delta{N}^2+|\Delta|^2}=0$. In this fermion pairing model, the gapless points are located along a centerline passing through the center of a string of vortex in momentum space. However this topological vortex is also visible in real space on knot square lattice. Complete vortex only exist in the bulk area, while half vortex and a quarter vortex could exist both in bulk and edge.

The fermion current of unpaired fermions running on one edge is
always in the opposite direction to its opposite edge. The edge current in this
fermion pairing model does not show similar odd-even effect as that in spin Hall current. For even number of
Y-loops, the spin of fermion flips a sign when the Y-current
enters its neighboring opposite Y-current (Fig. \ref{BCSpairing}
(b)), but keeps the same orientation of electric field,
$\vec{E}/|E| = (\vec{J}\times{\vec{s}})/|J||s|$, here $\vec{J}$ is
the fermion current vector, $\vec{s}$ is the spin vector. While
for an odd number of Y-loops, the Y-current fuses into X-current
by flipping the orientation of spins in a certain way so that the
electric orientation also flips (Fig. \ref{BCSpairing} (c)), i.e.,
$\vec{E}\rightarrow-\vec{E}$. The effective Hamiltonian of the
edge current for both the two cases reads,
\begin{eqnarray}
H_{edge}=t_{ij}\sum_{\langle{ij}\rangle}(c^{\dag}_{i-e_{x}\downarrow}c_{i\uparrow}-c^{\dag}_{i+e_{x}\downarrow}c_{i\uparrow}).
\end{eqnarray}
The equivalent Hamiltonian in momentum space is $H_{edge}=2
t_{ij}\sin(k_{x})c^{\dag}_{k_{x}\downarrow}c_{k_{x}\uparrow}$,
which reduces to a gapless dispersion near $k_{x}\rightarrow{0}$.
The edge current on the upper boundary moves to the opposite
direction of the bottom current, so does the left edge current and
right edge current. This phenomena holds both for the even number
and odd number of Y-loops (Fig. \ref{BCSpairing} (b) (c)).

\begin{figure}[htbp]
%\centering
%\par
%\begin{center}
$
\begin{array}{c@{\hspace{0.06in}}c}
%\multicolumn{1}{1}{\mbox{}} & \multicolumn{1}{1}{\mbox{}} \\
\includegraphics[width=0.25\textwidth]{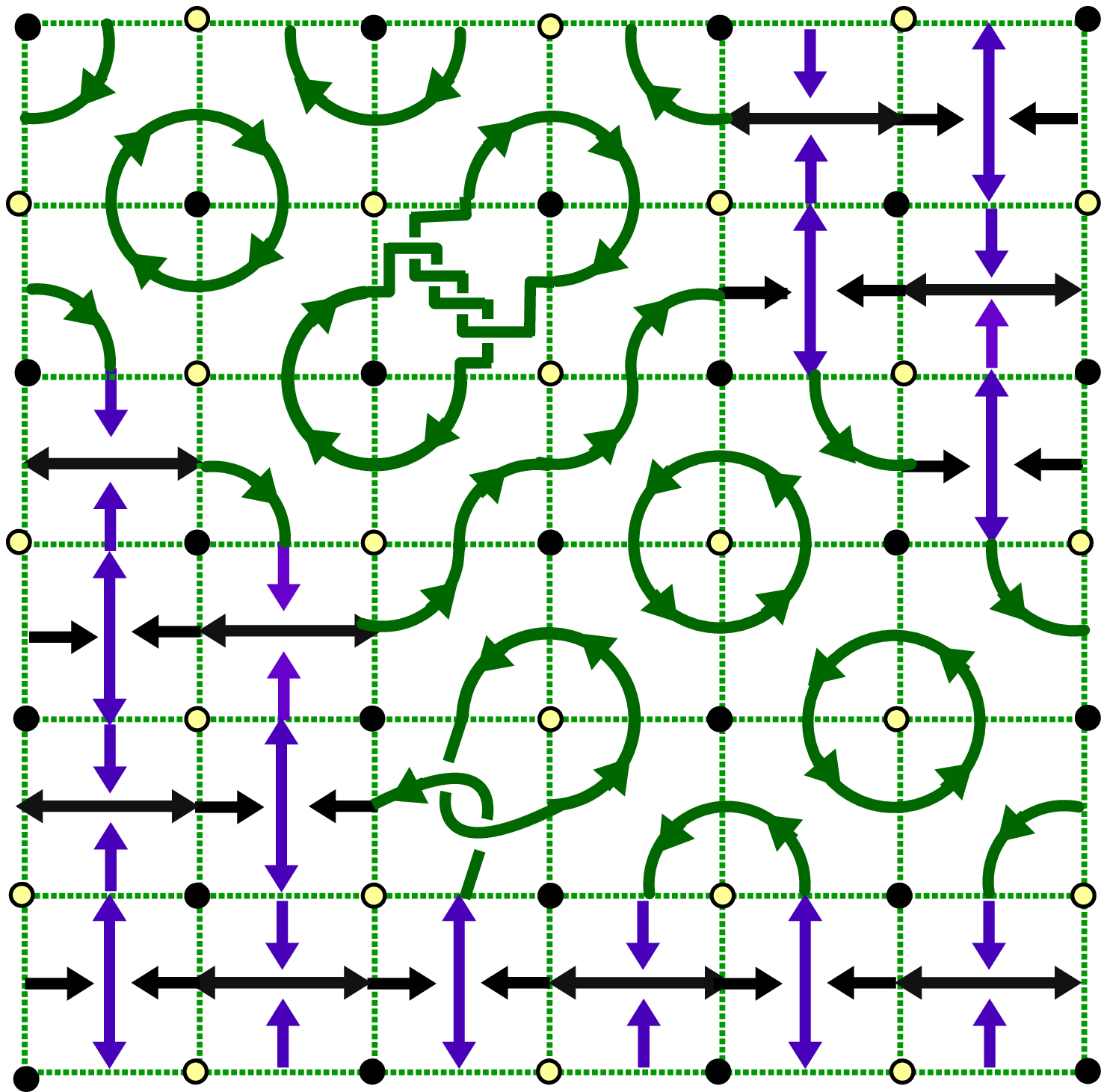}&\includegraphics[width=0.20\textwidth]{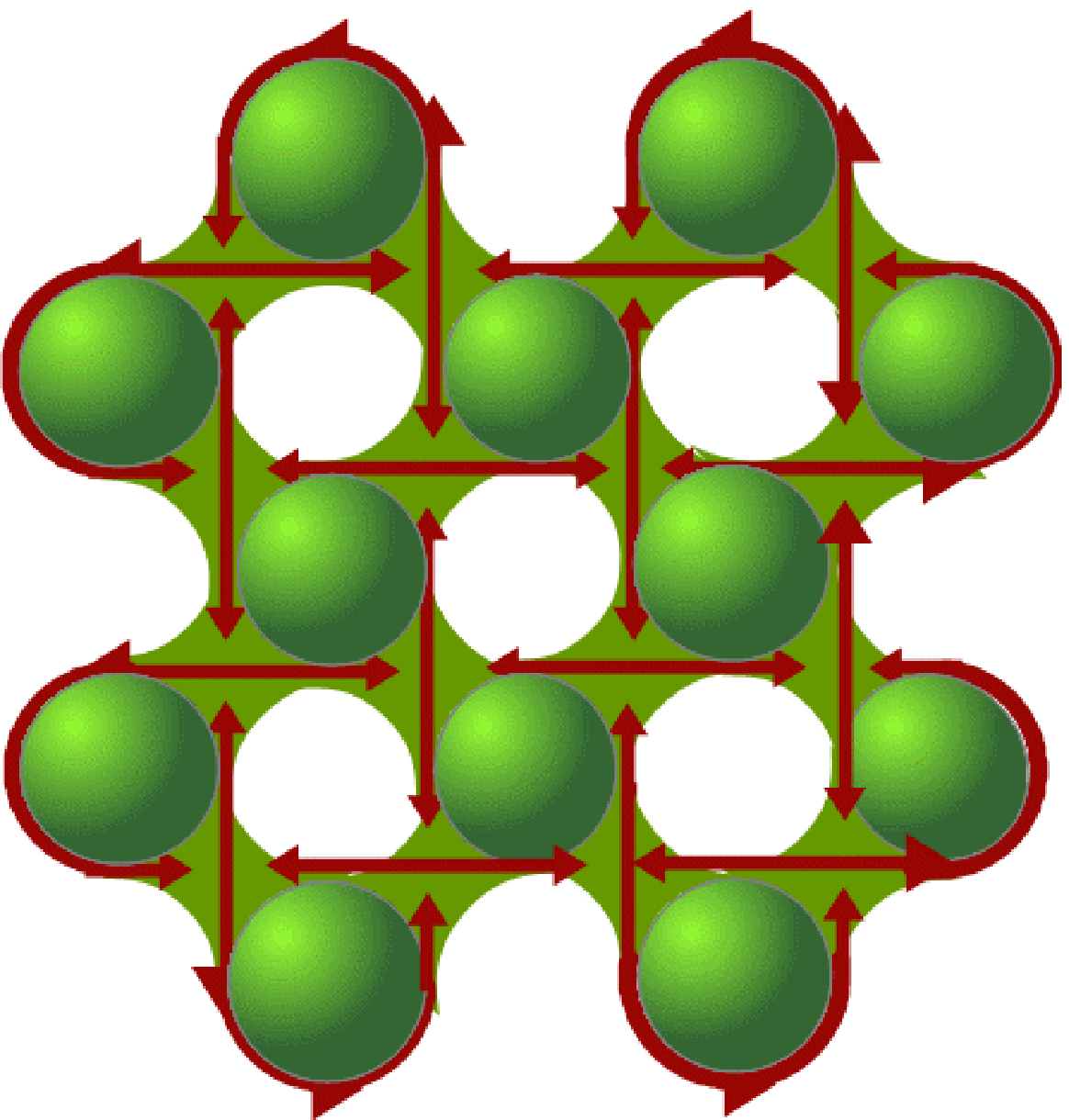}\\
\mbox{\textbf{a}} & \mbox{\textbf{b}}\\
\end{array}
$
%\end{center}\vspace{-0.3cm}
\caption{\label{vortex} (a) The vortex with opposite chirality locate on different square sublattice. The vacuum vortex lines entangle with other vacuum arcs. (b) The unoriented Seifert surface for fermions pairing state. }
\vspace{-0.2cm}
\end{figure}

The global pairing state has only two consistent patterns which
place the two local pairing patterns alternatively on the whole
lattice, any local flipping of one block spin results in
contradiction. However, if one block spin of current crossing is
braided into a vacuum state, the rest pairing states can still
exist consistently. In other words, breaking a local Cooper pair
on one lattice does not destroy the super-conducting states on
other lattice sites. For the block spin state $|+1\rangle$, there
are two braiding operators to break the Cooper pair, one is
counterclockwise braiding turns an angle of  $\pi/4$ away from
X-loop, which is denoted as $B_{t\circlearrowleft}$ (Fig. \ref{BCSpairing} (a)). It results in
a vacuum state with positive chirality following the right hand
rule, $|+0\rangle$. The other braiding operator turns an angle of
$3\pi/4$ away from X-loop, that we denote as
$B_{n\circlearrowright}$. This clockwise braiding disentangles the
crossing of pairing current into a vacuum state with negative
chirality, $|-0\rangle$. The clockwise braiding
$B_{t\circlearrowright}$ turns $|-1\rangle$ into negative vacuum
state $|-0\rangle$. While counterclockwise braiding
$B_{n\circlearrowright}$ braids $|-1\rangle$ into positive vacuum
state $|+0\rangle$. Four neighboring vacuum states with the same chirality can form a minimal loop. Electrons run through this minimal loop to go around a magnetic flux (Fig. \ref{vortex} (a)). Each minimal loop represents a vortex in superconductor. This knot square lattice for fermion pairing pattern is equivalent to the overlap of two square sub-lattices, which is denoted by the black discs and white circles correspondingly in Fig. \ref{vortex} (a). These two sublattices are the dual lattice of the unit cell, on which the magnetic flux with opposite chirality are distributed. Negative chiral vortex around a negative flux only exist at the black sites, while positive vortex only sit on white sites. Each flux site is surrounded by four unit cells. Since each braiding over an unit cell only generates two arcs with the same chirality, opposite vortex loops cannot coexist as the nearest neighbors. However, a vacuum current can separate two vortex loops to prevent them from annihilation (Fig. \ref{vortex} (a)).

Each vortex loop confines two electrons. One electron vanishes from unit cell at site $(i,j)$ and generated at site $(i-1,j-1)$, this process only draws a half circle. In the meantime, another electron must annihilates at $(i-1,j-1)$ and generates at $(i,j)$. The fermion pair current is topologically quantized. The complete vortex loop carries an integral winding number $W = \pm1$. While the half vortex loop carries a half winding number $W = \pm1/2$. A quarter arc carries a fractional winding number, $W = \pm1/4$. The complete vortex loop carries two electric charges. A quarter arc has only half charge, $e/2$. If more braiding operations are performed over the two vacuum arc within one unit cell, the vortex would entangle with supercurrent of fermion pair, or two vortex lines may also entangle each other (Fig. \ref{vortex} (a)). This nontrivial entanglement is characterized by linking number. Helicity is an effective topological quantity of entangled vortex lines,
\begin{eqnarray}\label{helicity}
H_{helicity}=\int(dx^3)\vec{J}\cdot(\vec{\nabla}\times\vec{J}),
\end{eqnarray}
where $\vec{J}$ is the familiar supercurrent of fermion pairs,
\begin{eqnarray}\label{vorhami}
J=\frac{ie\hbar}{2m}(\Delta^{\ast}\hat{D}\Delta-\Delta\hat{D}\Delta^{\ast}),
\end{eqnarray}
here $\Delta$ is the gap function for fermion pairs. Since Cooper pair is composite boson of electron pair with opposite spins, $\Delta$ is also a bosonic operator after the second quantization. $\hat{D}=\nabla+i2\pi\phi_{0}$ is the covariant derivative. The spin of electron flips by the accumulated phase factor of magnetic flux (Fig. \ref{BCSpairing} (a)), $\phi_{0} = \oint{A}dl=\int{\nabla\times{A}}dx^2=\int{B}dx^2$. The helicity Eq.(\ref{helicity}) is actually equivalent to abelian Chern-Simons action, which is topological number of many knots \cite{Duan}. If there are 2N crossing points between vortex lines over the whole vortex lattice, it takes $N$ Majorana fermions to bring entangled vacuum state to free vortex state. Similar to fractional quantum Hall state, fractional filling state also exist for two entangled vortex in superconductor.

Beside the linking number of entangled knots, Euler characteristic is also an effective topological characterization of the knot square lattice of fermion pairing model. The Euler Characteristic for an oriented, connected compact surface is $\chi(S)=2(1-g)$, where $g$ is the number of genus (holes). It is in fact the first Chern number. Seifert surface has to be constructed in order to compute the Chern number of a knot. For an arbitrary knot, Seifert's algorithm first color the projected plaquette alternatively into a checkerboard state, then lift up the plaquette with same color and view the plaquette with opposite color as hole. At each crossing point, each in-arrow must connect to its nearest neighboring out-arrow. Then the knot is decomposed into oriented loops. Filling these loops with color to generate a disk and then connect them by twisted mobius strip whose boundary projects the original crossing states in two dimensional plane, this surface is so called Seifert surface \cite{Brittenham}. To apply the Seifert algorithm on this knot square lattice, we first choose the plaquette on the white sublattice as filled surface (represented by the green spherical surface), while those on the black sublattice are holes (represented by the white blank zone in Fig. \ref{vortex} (b)). Then connect the in-arrow to out-arrow at every crossing site. It finally reduced to periodically distributed vortex loops (The green spheres in Fig. \ref{BCSpairing} (b)). We fill these loops and connect them by Mobius strip in such a way that the original current crossing are the projection of the two edges of Mobius strip (Fig. \ref{BCSpairing} (b)), which are represented by the green belt that connects two green spheres in Fig. \ref{vortex} (b). Then the knot square lattice of pairing states map into a lattice of periodically distributed Mobius strips (Fig. \ref{vortex} (b)). However this Seifert surface is still not an oriented manifold due to the opposite arrows of fermion pairing in this special model. In order to construct a closed current loop, we introduce two more currents that connect the two in-arrow to the two out-arrow in the middle waist of the Mobius strip (Fig. \ref{BCSpairing} (b)). Then it leads to an oriented current network on a unoriented Seifert surface. In fact, the Euler characteristic for this case can not distinguish an oriented surface from an unoriented one. The Seifert surface in Fig. \ref{vortex} (b) has an Euler characteristic of $\chi=2(1-5)=-8$, since it has five genus. For a square lattice with $(2N+1)\times{(2N+1)}$ unit squares, the total number of genus is $g=[(2N+1)^2+1]/2$. The corresponding Euler Characteristic is $\chi=1-(2N+1)^2$.

\begin{figure}
\begin{center}
\includegraphics[width=0.35\textwidth]{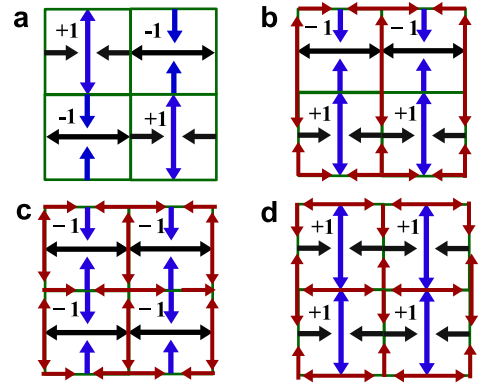}
\caption{\label{BCSpairing2} (a) The current crossing states for anti-ferromagnetic ordering, $|\psi\rangle=|\uparrow\downarrow\uparrow\downarrow\uparrow\downarrow...\rangle$. (b) The current crossing states for strip ordering, $|\psi\rangle=|\uparrow\uparrow\uparrow...\downarrow\downarrow\downarrow\rangle$. (c-d) The current crossing states for ferromagnetic ordering, $|\psi\rangle=|\downarrow\downarrow\downarrow\downarrow...\rangle$ and  $|\psi\rangle=|\uparrow\uparrow\uparrow\uparrow...\rangle$.}
\end{center}
\vspace{-0.5cm}
\end{figure}

The Euler characteristic is an effective topological number for classifying different fermion paring patterns on square lattice. The ferromagnetic ordering of current crossing is the minimal self-consistent lattice model for fermions pairings. Strip ordering and ferromagnetic ordering could show up if a local crossing flips under thermal fluctuation. In that case, two out-arrows (or in-arrows) may collide on the boundary of unit cell, on which there is no outgoing current exist. In order to construct self-consistent continuous current loops, extra current perpendicular to the two out-arrows on the boundary have to be introduced (the red arrows in Fig. \ref{BCSpairing2} (b) (c) (d)). For the ferromagnetic order state, each one of the original four unit cells is divided into four small unit cells with half lattice constant (Fig. \ref{BCSpairing2} (c) (d)). Thus the total number of unit cells increased to 16. While the current of strip ordering on the boundary between $|+1\rangle$ and $|-1\rangle$ are consistent, any extra current would results in contradiction (Fig. \ref{BCSpairing2} (b)). Thus extra current are only introduced on the boundary between $|+1\rangle$ and $|+1\rangle$ (or $|-1\rangle$ and $|-1\rangle$) (Fig. \ref{BCSpairing2} (b)). The corresponding two dimensional surface with respect to different ordering state can also be constructed following Seifert algorithm. The Seifert surface of $(2N+1)\times{(2N+1)}$ knot square lattice has a Euler characteristic $\chi=1-(2N+1)^2$ genus, while this topological number increases to $\chi=1-4(2N+1)^2$ for a homogeneous ferromagnetic ordering state. The strip ordering has an intermediate Euler characteristic number.
The ferromagnetic ordering of current crossing in real space actually can be mapped into separated current loops by Reidemeister moves. While the antiferromagnetic ordering of current crossing requires the maximal number of flipping operations on certain crossing points to map it into separated loops. The total number of flipping operations to map many entangled knots into free loops can be used to quantify the topological entanglement of a link. In this sense, the antiferromagnetic ordering of block spin for fermions pairing has the maximal topological entanglement. Thus superconductor should has maximal topological entanglement.

\begin{figure}
\begin{center}
\includegraphics[width=0.48\textwidth]{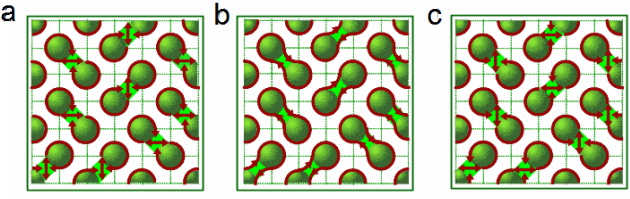}
\caption{\label{vortexdimer} (a)The dimer lattice of negative vortex pairs. (b) The dimer lattice of vortex pair coupled to vacuum. (c) The dimer lattice of positive vortex pairs.}
\end{center}
\vspace{-0.5cm}
\end{figure}

The Seifert surface provides mathematical mapping of knot square lattice into two dimensional surface with genus. If we consider the physical implementation of Seifert algorithm, the filled plaquette is physically implemented by magnetic flux in vortex loops. The connection of an in-arrow to an out-arrow is physically implementable by local braiding operation on each crossing point, which can be realized by physical magnetic field in plane. However in this knot square lattice, the filled vortex loop cannot connect to its four neighbors at the same time to construct a two dimensional surface with many holes. A vortex either exist as an isolated loop or connect only to one of its four nearest neighboring to construct a dimerized lattice of vortex pairs (Fig. \ref{vortexdimer}). Only vortex with the same chirality can form a dimer by coupling to the crossing supercurrents (Fig. \ref{vortexdimer} (a) (c)) or vacuum state (Fig. \ref{vortexdimer} (b)). Different dimer lattice have different Euler characteristic number. We make an inverse filling of the Seifert surface of dimer lattice. Then each filled vortex loop becomes a hole of continuous surface. For a lattice with $2N\times2N$ unit cells, there exist $N^2$ isolated vortex loops, and ${N^2}/2$ possible vortex dimers. The Euler characteristic of isolated vortex lattice is $\chi=2(1-N^2)$. The vacuum vortex dimer phase has an Euler characteristic $\chi=2(1-{N^2}/2)$, so does the positive and negative vortex dimer phase. Euler characteristic can not distinguish a Mobius hole from a trivial hole.

These vortex loops behave as fermion in the vortex dimer lattice, which is originated from the fermionic string arc in this knot lattice model. Each fermionic string arc can be represented by a Grassmann number. The Grassmann number obeys the following algebra, $\eta_{i}\eta_{j} = -\eta_{j}\eta_{i}$, $\eta_{i}\eta_{i}=0$. We place a Grassmann number,$\eta_{i}$ at the center of each vortex loop. These vortex loops are regularly distributed on square lattice. For a dimer lattice of the same chiral vortex loops, the total number of all different covering patterns can be computed by Kasteleyn matrix \cite{Kasteleyn}, which is equivalent to the square root of the determinant of quadratic fermion action, $S = \sum_{i<j}M_{ij}\eta_{i}\eta_{j}.$ The total number of all different covering patterns equals to the partition function of this fermion action,
\begin{eqnarray}
Z = \int{d\prod\eta_{i}{\exp{\sum_{i<j}M_{ij}\eta_{i}\eta_{j}}}}=\pm\sqrt{det[M]}
\end{eqnarray}
Note here the vacuum vortex dimer admit a self-consistent coexistence with the positive vortex dimer or negative vortex dimer. But the positive vortex dimer and negative vortex dimer can not perfectly coexist without introducing geometric frustrations (Fig. \ref{vortex} (a)). The effective action for vortex dimers on square lattice bear the same formulation as classical dimers \cite{Fendley} but with rotated wave vectors,
\begin{eqnarray}
S_{0} =\sum_{k}[i\sin{(k_{x}+k_{y})}-\sin(k_{x}-k_{y})]\eta_{k}\eta_{-k}.
\end{eqnarray}
This effective action holds for a lattice covered by pure dimers. If we consider the hybrid dimer lattice covered by positive vortex dimer and vacuum vortex dimer (or negative vortex dimer and vacuum dimer) together, then the total number of all possible covering patterns is the product of two equal partition function for pure dimers, $Z^2$. In that case, the effective action for vortex dimer lattice is
\begin{eqnarray}\label{Sh}
S_{h} =\sum_{k}[i\sin{(k_{x}+k_{y})}-\sin(k_{x}-k_{y})]\eta^{\dag}_{k}\eta_{-k}.
\end{eqnarray}
The gap closing points are periodically distributed in momentum space. This dimer counting does not take into account of internal state of dimer. If the vortex lines between two vortices were braided for for many times (Fig. \ref{vortex} (a)), the vortex dimer couples to anyon with fractional statistics, which is characterized by Laughlin wave function. The vortex loop can appear at any local site in the gapless mode of the paring model but still construct self-consistent current loops. Thus a fermion pairing lattice with vortex loops is still in super-conducting state, but the pairing energy gap closes at the vortex arcs. These vortex arcs caused the Fermi arc in the pseudogap state of unconventional superconductor.

The complete Hamiltonian for a positive vortex dimer around the $\infty$ shaped loop reads(Fig. \ref{vortexdimer} (c)),
\begin{eqnarray}
&&H_{vor} ={\sum} U \ c_{i,j\uparrow}c^{\dag}_{i,j+1}c_{i,j+1}c^{\dag}_{i-1,j+1}c_{i-1,j+1}\nonumber\\
&&c^{\dag}_{i-1,j}c_{i-1,j}c^{\dag}_{i,j\uparrow}c_{i,j\downarrow}c^{\dag}_{i,j-1}c_{i,j-1}\nonumber\\
&&c^{\dag}_{i+1,j-1}c_{i+1,j-1}c^{\dag}_{i+1,j+1}c_{i+1,j+1}c^{\dag}_{i,j\downarrow}[e^{i\int{A}dl}].
\end{eqnarray}
This Hamiltonian can be further simplified by string operator, which is a serial product of particle number operator $n_{i,j} = c^{\dag}_{i,j}c_{i,j}$ along the track,
\begin{eqnarray}
&&\hat{L}=e^{i\int{A}dl}n_{i,j+1}n_{i-1,j+1}n_{i-1,j}n_{i,j-1}n_{i+1,j-1}n_{i+1,j+1},\nonumber\\
&&H_{vor} =\sum_{i,j} - U \hat{L}\;c^{\dag}_{i,j\downarrow}c^{\dag}_{i,j\uparrow}c_{i,j\uparrow}c_{i,j\downarrow}.
\end{eqnarray}
Each vortex dimer is pinned down by a free fermion pair confined at local site $(i,j)$, on which the energy spectrum is gapped. The unit cell covered by pure vortex loop is in gapless state. The global gapless phase only exist for a lattice of isolated vortex loops and vacuum vortex dimers. In mind of the anti-commuting character of fermionic arcs, each fermonic arc can be represented by a composite Grassmann operator,
\begin{eqnarray}
\hat{\eta}_{ij}=e^{i\int{A_{ij}}dl}n_{ij},
\end{eqnarray}
The path integral of gauge potential $\exp[\int{A}dl]=\exp[i\phi]$ is only carried out along the fermionic string arcs. The string operator is the product of six Grassmann operator,
\begin{eqnarray}
\hat{\eta}_{a}&=&\eta_{i,j+1}\eta_{i-1,j+1}\eta_{i-1,j},\;\;\hat{\eta}_{b}=\eta_{i,j-1}\eta_{i+1,j-1}\eta_{i+1,j+1},\nonumber\\
\hat{L}_{ij}&=&\hat{\eta}_{a}\hat{\eta}_{b},
\end{eqnarray}
where $\hat{\eta}_{a}$ is the product of three fermion arcs around the first vortex of dimer, and $\hat{\eta}_{b}$ represent the second vortex inside the vortex dimer. The ordered Grassmann string operator switches to negative if the order of six operators is reversed. The string operator of an isolated vortex is equivalent to a Wilson loop operator, it is the product of four Grassmann operators, which is equivalent to boson. The effective Hamiltonian for a vortex dimer coupled to fermion pairing state is
\begin{eqnarray}
H_{vor} = -\sum_{i,j}U \hat{L}_{ij}\;c^{\dag}_{i,j\downarrow}c^{\dag}_{i,j\uparrow}c_{i,j\uparrow}c_{i,j\downarrow}.
\end{eqnarray}
In the mean field approximation, the corresponding Hamiltonian in momentum space becomes a dressed fermion pairing Hamiltonian after integration of Grassmann variable
\begin{eqnarray}
H_{vor}(k)=L(k)\psi_{k}^{\dag} [\Delta_{1}S_{x}+\Delta_{2}S_{y}+\Delta{N}(k) S_{z}]\psi_{k}.
\end{eqnarray}
$L(k)$ is the vortex dimer spectrum in the action Eq. (\ref{Sh}). The original fermion pairing gap closes at the gapless point of vortex dimer spectrum, $k_{x}=k_{y}$, or $k_{x}=-k_{y}$. This gapless equation induces the Fermi arc in momentum space. The gapless superconducting resonance vortex dimer state along the nodal and anti-nodal line of energy spectrum offers a theoretical explanation on pseudogap state of unconventional superconductor \cite{scalapino}. The energy current of each spin component defines the location of a point of knot in momentum space. The current knot square lattice is still in superconducting state in the presence of vortex. If the two crossing super-current between two vortex are braided more than three times in the same direction, a Majorana fermion will be generated to raise the local energy, but does not break the pairing states. The fractional statistics of this Majorana fermion can be described by Laughlin wave function.

\begin{figure}
\begin{center}
\includegraphics[width=0.45\textwidth]{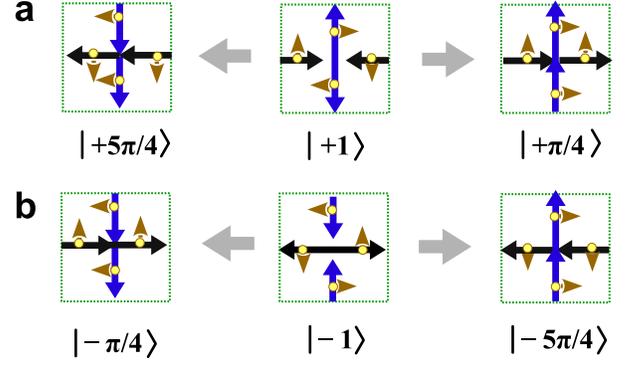}
\caption{\label{BCSpairing3} (a) The two equivalent crossing states with respect to the positive pairing pattern, $|+\pi/4\rangle,|+5\pi/4\rangle$. (b) The two equivalent crossing states with respect to negative pairing pattern current,$|-\pi/4\rangle,|-5\pi/4\rangle$.}
\end{center}
\vspace{-0.5cm}
\end{figure}

When the long range pairing pattern between two far separated crossing sites is taken into account, the self-consistent construction of fermion pairing patterns on knot square lattice includes four more degenerated patterns (Fig. \ref{BCSpairing3}), which are denoted as ($|+\pi/4\rangle,|+5\pi/4\rangle$) with respect to $|+1\rangle$ and ($|-\pi/4\rangle,|-5\pi/4\rangle$) with respect to $|-1\rangle$. The generation of a positive spin with positive momentum is equivalent to the annihilation of a negative spin with negative momentum. This equivalent correspondence obeys the same Feynmann diagram rule for particle and anti-particle in conventional quantum field theory. Within this enlarged crossing states space, one fermion could travel across many unit lattice spaces to meet at the local pairing sites. In that case, the fermion pairing Hamiltonian carries long range pairing interaction that shares the same formulation as Eq. (\ref{pairhami}). The fermion-antifermion pair form a conserved continuous current without vertical convection current.

\section{Anyons of quantum knot lattice model on honeycomb lattice}

\begin{figure}
\begin{center}
\includegraphics[width=0.45\textwidth]{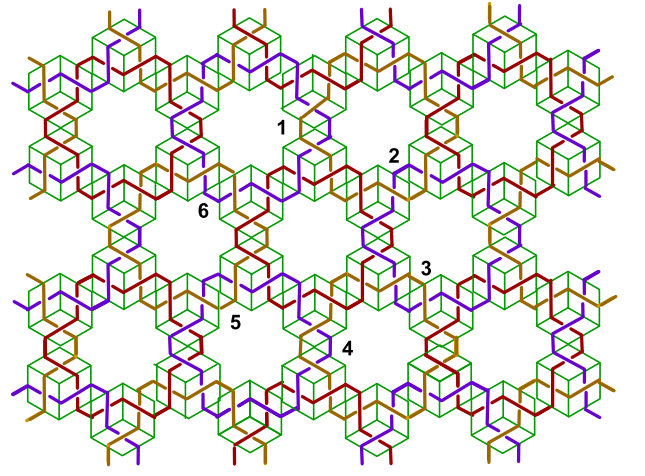}
\caption{\label{kitaev} One layer of the multilayer knot patterns for effective mapping of Ising model and Kitaev model on honeycomb lattice.}
\end{center}
\vspace{-0.5cm}
\end{figure}

\begin{figure}
\begin{center}
\includegraphics[width=0.39\textwidth]{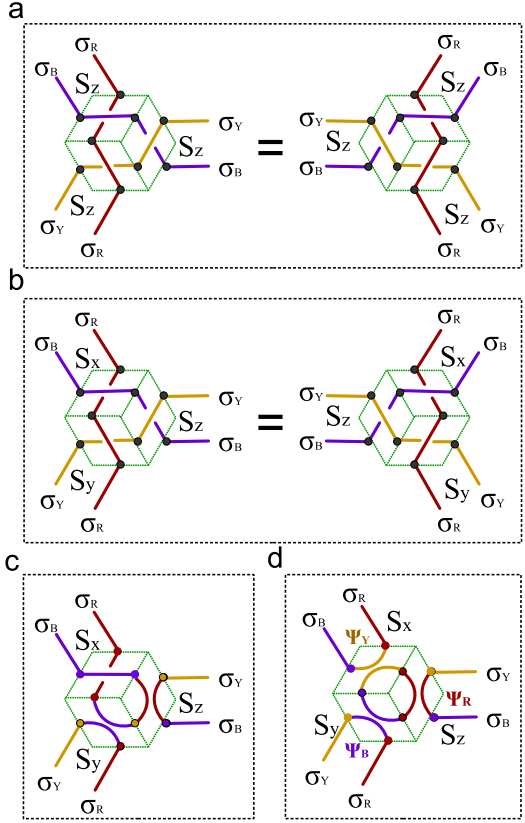}
\caption{\label{yang} (a) Three color anyon currents entangled each other to implement the local tribein for Ising model on honeycomb lattice. The three braiding satisfy Yang-Baxter equation. (b) Three color anyon currents entangled each other to implement coupling style in Kitaev honeycomb model. (c) The fusion rule of the three color anyons into two color Majorana fermions in Hilbert section of $S^{z}$, and the fusion of two Majorana fermions. (d) The fusion rule of the three color anyons into three color Majorana fermions in Hilbert section of $S^{z}$.}
\end{center}
\vspace{-0.5cm}
\end{figure}

The multi-knot lattice model can be extened to honeycomb lattice, triangular lattice, and so on. In order to implement spinor for quantum spins, we have to use double current along each bond as a geometric representation of spinor. The complete Hamiltonian for the multi-knot lattice model of spins includes long range spin-spin coupling terms,
\begin{eqnarray}
\hat{H}_{L}=\sum_{j=1}^{m}J_{\alpha}{S}^{\alpha}_{i}S^{\alpha}_{i+j}, \;\; \alpha=x,y,z.
\end{eqnarray}
The conventional Ising model, Kitaev model as well as other spin-spin coupled quantum models are the special case of knot lattice model for the nearest neighbor coupling interactions. The non-commutative character of spin operators induced non-trivial quantum physics beyond classical Ising spin.

\subsection{The knot lattice model of transverse field Ising chain model}

\begin{figure}
\begin{center}
\includegraphics[width=0.32\textwidth]{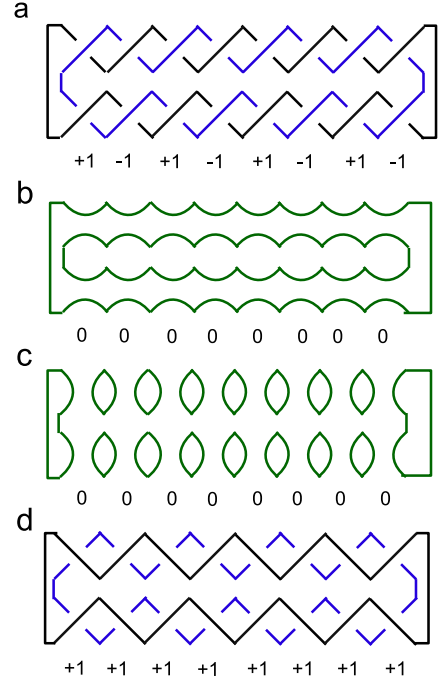}
\caption{\label{transising} The two entangled loops as eigenstate representation of one dimensional transverse Ising model.}
\end{center}
\vspace{-0.5cm}
\end{figure}

When the two-state Ising model or block spin-1 Ising model contains more than two different spin component operator, the Ising spin $S^z$ component is non-commutative with the other spin components, which induced a more complicate geometric implementation of the eigenstate of quantum spin models. We first take one dimension Ising spin chain model as an example to explore the knot lattice configuration for entangled quantum states. The short range transverse field Ising chain model includes both $ S^{z}$ and $ S^{x}$,
\begin{eqnarray}\label{Ht}
H_{t}=-\sum_{\langle{ij}\rangle}(J_{z} S^{z}_{i}S^{z}_{j} + h_{i} S_{i}^{x}).
\end{eqnarray}
Replacing $S_{i}^{x}$ with $S_{i}^{y}$ leads to an equivalent quantum model under duality transformation. The knot representation of this model is two loops periodically entangled with each other (Fig. \ref{transising}). For a vanished transverse field $h_{i}=0$, the spin chain reduced to classical Ising model. For the antiferromagnetic coupling $J_{z}>0$, the ground state is two loops with the maximal winding numbers, one loop wraps around the other at every lattice site (Fig. \ref{transising} (a)). For ferromagnetic coupling $J_{z}<0$, its ground state is two separated loops, i.e., One loop is above the other everywhere on the whole chain of lattice (Fig. \ref{transising} (d)). A weak transverse field acts as a perturbation on the ground state by flipping the crossing states from over-crossing to under crossing (or vice versa) upon certain number of lattice sites instead of the whole lattice. While if coupling strength between the nearest neighboring spin drops to zero, $J_{z}=0$, the eigenstate of the spin chain is determined by the eigenstate of spin operator $S^{x}$ alone, $(|\uparrow\rangle\pm|\downarrow\rangle)$, i.e., $S^{x}(|\uparrow\rangle\pm|\downarrow\rangle)=\pm1(|\downarrow\rangle\pm|\uparrow\rangle)$. The eigenstate knot lattice of transverse Ising model is a bilayer knot lattice, the bottom layer carrying $|\uparrow\rangle$ plus (or minus for a positive $h_{i}$) the upper layer carrying $|\downarrow\rangle$. If $S_{i}^{x}$ is replaced by $S_{i}^{y}$ in Hamiltonian Eq. (\ref{Ht}), a phase factor $e^{i\pi/2}$ must be added to the upper knot lattice layer with $|\downarrow\rangle$ (Fig. \ref{spinx} (d)) in order to fulfill the eigen-equation of  $S_{i}^{y}$,  i.e., $S^{y}(|\uparrow\rangle\pm i|\downarrow\rangle)=\pm 1(|\uparrow\rangle\pm i|\downarrow\rangle)$. The transverse Ising model admits a quantum phase transition when the neighboring coupling strength equals to the transverse field strength $J_{z} = h$. This critical point is derived by exact solution of this Ising chain model, which maps spin coupling into fermion pairing Hamiltonian \cite{sachdev} by Jordan-Wigner transformation Eq. (\ref{spincmap}),
\begin{eqnarray}\label{Htpair}
H_{t}(k)&=&\sum_{\langle{k}\rangle}(2[h-J_{z}\cos(k)]c^{\dag}_{k}c_{k}\nonumber\\
&-&iJ_{z}\sin(k)[c^{\dag}_{-k}c^{\dag}_{k}+c_{-k}c_{k}]-h).
\end{eqnarray}
This pairing Hamiltonian can decompose into similar coupling type between pseudo-spin and complex fermions as Eq. (\ref{pseudospinfermi}),
\begin{eqnarray}
H_{t}(k)=\psi_{k}^{\dag} [\epsilon_{k}\textbf{I} - {\Delta} S_{x}]\psi_{k}-h.
\end{eqnarray}
here $\epsilon_{k}=2[h-J_{z}\cos(k)]$, $\Delta=iJ_{z}\sin(k)$. $\psi_{k}=[c_{k},c_{k}^{\dag}]$. The critical point of this conventional transverse Ising model lies at the gapless point, where the exact spectrum vanishes, $E = -2\sqrt{(J^2+h^{2}-2Jh\cos{k})}=0$. For the ferromagnetic coupling at critical point, $J_{z} = h<0$. The Ising coupling drives the ground state of each layer into a state with two separated loops (Fig. \ref{transising} (d)). The transverse field prefers a superposition state of two layers of knot chain with opposite global crossing states and a plus sign between them. Thus the ground state this quantum Ising model is composed of four separated loops. In order to reach the gapless state, an odd number of spins must be flipped on the two layers simultaneously. This inevitably increased the winding number of the four loops. Another operation is to add $\exp[i\pi]$ phase factor to certain lattice sites in one layer, but keep the other layer the same. This operation increases the energy by unit of transverse field. While adding $\exp[i\pi]$ phase upon a spin state is equivalent to flipping its state to the opposite, i.e., from $(|\uparrow\rangle+\exp[i\pi]|\downarrow\rangle)$ = $(|\uparrow\rangle-|\downarrow\rangle)$ to $(|\uparrow\rangle+|\uparrow\rangle)$. The decreased energy in the output of $S^{x}$ is transferred to the increased part in the output energy of $S^{z}$. The neighboring coupling and transverse field compete each other to reach balance point of energy output, which is the critical point of quantum phase transition. The energy carrier here is actually kink excitation. The total number of winding number is the total number of kinks in the loop chain. The maximal number of kinks on a chain of $2N$ spin is $N$. The maximum winding number is also $N$. The total winding number at the critical point is zero. The total wining number above the critical point has opposite sign to that below the critical point. The total winding number here is the linking number of the four entangled knots. In this sense, this quantum phase transition is also a topological phase transition. Since the effective spin component has only one wave vector, it cannot draw a two dimensional loop in momentum space. However if auxiliary parameters and the next-nearest neighboring interaction are introduced in an extended transverse Ising model, the closed loops in auxiliary parameter can still characterize quantum phase transition \cite{GZhang}.

The spin flipping in this one dimensional chain can be implemented by the same braiding operations for Ising spin in two dimensional lattice (Fig. \ref{spinx} (a)). One spin flipping is realized by at least two braiding in the same direction. Single braiding could result in vacuum states, that is not the eigenstate of two-states Ising spin. Thus we extend the two-state Ising spin to block spin-1 operator, $S_{i}^{\alpha}$ has three eigenstates ,$|\pm1\rangle$ and $|0\rangle$. For a spin chain oriented along X-direction, there are two independent braiding, one is $\hat{B}_{x}$, which generates one up and one down Majorana fermion arc (the green arcs in Fig. \ref{transising}). The other is $\hat{B}_{y}$, generating a left and a right Majorana fermion arc. These braiding operators are implementable by Magnetic field. If the spin chain is acted only by $\hat{B}_{y}$, the resulting vacuum state is compose of $N$ unit circles for $N$ lattice sites. This is the dimerized insulator state (Fig. \ref{transising} (b)). For the other global braiding operation of $\hat{B}_{x}$, the generated vacuum state is two separated loops (Fig. \ref{transising} (c)). Particles can run through the whole lattice but split into two bands. If there are only two $\hat{B}_{y}$ braiding on the left and the right edge correspondingly, and $\hat{B}_{x}$ braiding over the whole inner section, the two separated loops can fuse into one. This leads to the gapless vacuum state. Different vacuum states can be distinguished by topological numbers.

Similar to the two dimensional block spin-1 Ising model, this block spin-1 Ising chain also have non-abelian anyons in the eigenenergy level. The superposition of vacuum state and Majorana fermion state is naturally embedded in the Hilbert space. The eigenstate in vacuum state is a natural quantum qubit for topological quantum computation \cite{Kitaev1}. Repeating the braiding operations drives the Majorana fermion to jump from bottom layer to upper layer (or vice versa). The topological correlation for this block spin-1 chain is easier to calculate than the two-dimensional lattice model. The Jones polynomial for a given quantum state here is exactly computable by Skein recursion relation, which view the vacuum state as the superposition of $|\uparrow\rangle$ and $|\downarrow\rangle$. Thus every quantum state is associated with a Jones polynomial. However different state may share the same Jones Polynomial, we call it topological degeneracy, which fulfill the requirement for topological quantum computation.

\subsection{The knot lattice of spin 1/2 Ising model on honeycomb lattice}

Since there are three currents intersect at one lattice site, we abandon the magnetic monopoles
and positrons to avoid local frustrations. Here we introduce three
color anyons, the red ($\sigma_{_{R}}$), yellow ($\sigma_{_{Y}}$)
and blue anyons ($\sigma_{_{B}}$), a pair of anyons is running in
two currents that is oriented by one arm of the local tribein
which has three arms separated by $2\pi/3$. One possible
implementation of $\sigma_{_{R}}$ is an electron attached by a
flux phase factor $e^{i\int{dA}}$, so does $\sigma_{_{Y}}$ and
$\sigma_{_{B}}$. An over-crossing occurs on each arm of the
tribein which is enveloped by an hexagonal unit cell (Fig.
\ref{kitaev}). Three sequential braiding at the three
intersecting squares obey Yang-Baxter equation (Fig. \ref{yang}
(a) (b)),
\begin{eqnarray}\label{yangbaxter}
R^{\sigma_{_Y}\sigma_{_B}}R^{\sigma_{_R}\sigma_{_B}}R^{\sigma_{_R}\sigma_{_Y}} = R^{\sigma_{_R}\sigma_{_Y}}R^{\sigma_{_R}\sigma_{_B}}R^{\sigma_{_Y}\sigma_{_B}}.
\end{eqnarray}
Here $R^{\sigma_{_p}\sigma_{_q}}$ are the braiding operators on the over crossing points. Thus a rotated tribein or its mirror configuration is equivalent to the original coupling tribein. We first input Ising coupling between neighboring unit cells,
\begin{eqnarray}\label{Hhoney}
H_{h}=\sum_{i}J_{z}(S^{z}_{i,\textbf{e}_x}S^{z}_{i+\textbf{e}_x}+S^{z}_{i,\textbf{e}_y}S^{z}_{i+\textbf{e}_{_y}}
+S^{z}_{i,\textbf{e}_z}S^{z}_{i+\textbf{e}_z}).
\end{eqnarray}
The local crossing state of each arm in $i$th unit cell is only
acted by $S_{i}^{z}$. The conventional Ising spin on lattice is
represented by single  $\uparrow_{i}$ or $\downarrow_{i}$. While
here every unit cell has three internal spin components that has
the freedom to point up or down. In order to implement classical
Ising model, we only choose two block spin configurations,
$|\Uparrow\rangle_{i}$ = $|\uparrow\uparrow\uparrow\rangle_{i}$
and $|\Downarrow\rangle_{i}$ =
$|\downarrow\downarrow\downarrow\rangle_{i}$. The knot
configuration in Fig. \ref{yang} (a) corresponds to
$|\uparrow\uparrow\uparrow\rangle_{i}$. Flipping all of the three
crossing states of Fig. \ref{yang} (a) results in
$|\downarrow\downarrow\downarrow\rangle_{i}$, which indicates the
generation of statistical phase $\exp[{i\pi}]$ . Each spin flip is
performed by braiding one pair of anyons twice. Three spin flips
requires 6 times of braiding anyon on the three arms. Thus each
anyon carries a statistical phase factor, $\exp[{{i\pi}/6}]$.

The coupling interaction between crossing points within the
hexagon unit cell can also be introduced to construct an Ising
model on Kagome lattice \cite{paddison}. The three frustrated spins within the
triangle lattice either generate three kinks or one kink. The
block up-spin/down-spin above represents three kinks running in
clockwise/counterclockwise direction. The single kink state exist
for the case that one current is above the other two which always
have one current lifted in opposite direction. These frustrated
states are generated by spin flipping upon
$|\Downarrow\rangle_{i}$ or $|\Uparrow\rangle_{i}$,
\begin{eqnarray}\label{blockup}
\;|\uparrow\downarrow\uparrow\rangle_{i},
\;|\downarrow\downarrow\uparrow\rangle_{i},
\;|\downarrow\uparrow\uparrow\rangle_{i}\;,
\;|\downarrow\uparrow\downarrow\rangle_{i},
\;|\uparrow\uparrow\downarrow\rangle_{i},
\;|\uparrow\downarrow\downarrow\rangle_{i}\;.
\end{eqnarray}
They are in fact frustrated triple spin states on triangle lattice, obeying the following Hamiltonian,
\begin{eqnarray}\label{Hfru}
H_{f}&=&\sum_{i}J_{z}(S^{z}_{i,\textbf{e}_x}S^{z}_{i+\textbf{e}_x}+S^{z}_{i,\textbf{e}_y}S^{z}_{i+\textbf{e}_{_y}}
+S^{z}_{i,\textbf{e}_z}S^{z}_{i+\textbf{e}_z}\nonumber\\
&+&S^{z}_{i,\textbf{e}_x}S^{z}_{i,\textbf{e}_y}+S^{z}_{i,\textbf{e}_y}S^{z}_{i,\textbf{e}_z}+S^{z}_{i,\textbf{e}_z}S^{z}_{i,\textbf{e}_x}).
\end{eqnarray}
The coupling term within the unit cell has an equivalent formulation by the sum of three arm spins,
\begin{eqnarray}\label{Hfru2}
H_{f}&=&\sum_{i}J_{z}(S^{z}_{i,\textbf{e}_x}S^{z}_{i+\textbf{e}_x}+S^{z}_{i,\textbf{e}_y}S^{z}_{i+\textbf{e}_{_y}}
+S^{z}_{i,\textbf{e}_z}S^{z}_{i+\textbf{e}_z})\nonumber\\
&+&\sum_{i}\frac{J_{z}}{2}[(S^{z}_{i,\textbf{e}_x}+S^{z}_{i,\textbf{e}_y}+S^{z}_{i,\textbf{e}_z})^2-3].
\end{eqnarray}
The ground state requires $(S^{z}_{i,\textbf{e}_x}+S^{z}_{i,\textbf{e}_y}+S^{z}_{i,\textbf{e}_z})=\pm1$, since the sum of the three arm spins can not reach zero. The frustrated spin configurations Eq. (\ref{blockup}) is the ground state. The first excited stat of single hexagonal unit cell is $ E = 3{J_z}$ corresponds the three kink states. $ E = -{J_z}$ corresponds the 6 degenerated single kink states (Eq. (\ref{blockup})). Only flipping one spin could transform the ground state with single kink to the first excited states with three kinks. One spin flip contributes a $\exp[{i\pi}]$ phase factor to the block spin. Here a spin flipping is performed by braiding two anyons twice. Thus each anyon carries an abelian phase factor $\exp[{i\pi/2}]$. Each braiding costs an energy unit $2{J_z}$.

For single triangular unit cell, braiding anyon within the ground
state shows other statistical phases. For instance, in order to
map $\;|\uparrow\downarrow\uparrow\rangle_{i}$ to
$|\downarrow\downarrow\uparrow\rangle_{i}$, only the spin in
$\textbf{e}_x$ is flipped by braiding anyons twice. The
the collective wave function acquires a phase $\exp[{i2\pi}]$.
Thus the anyon in $\textbf{e}_x$ carries a phase $\exp[{i\pi}]$.
While mapping state $\;|\uparrow\downarrow\uparrow\rangle_{i}$ to
$\;|\downarrow\uparrow\downarrow\rangle_{i}$ requires three spin
flips which is carried out by 6 braiding on local tribein. Thus
each anyon carries a statistical phase $\exp[{i\pi/3}]$. The
statistical factors of anyons are not uniformly distributed along
different mapping paths from one eigenstate to another. Thus we
call the anyons within these frustrated states as non-abelian
anyons.

These highly degenerated frustrate states have an exact one-to-one
mapping to multi-layer knot lattice configurations similar to what
we showed before. Thus each state could be labeled by a
topological linking number of those entangled loops. This
topological linking number is in fact equivalent to the total
magnetization of spins. One special topological character of these
frustrated state is there exist at least one current that can be
disentangled from the other two without cutting at each local
triangle. Repeating this operation for all of the other 5
triangles around the hexagon plaquette, we can always find a free
loop current around the local hexagon plaquette. Anyon keeps
running in this loop around the hexagon plaquette without losing
energy or charges, but generates a magnetic flux passing through
the center of hexagon plaquette. As showed in last section,
electrons running in knot lattice model bears quantum Hall effect.
For this two-state Ising spin on Kagome lattice, it only
implements the quantum Hall effect with even number of magnetic
fluxes. The two-state Ising spin $S_{i}^{z}=\pm1$ on this kagome
lattice must be replaced by three states Ising spin
$S_{i}^{z}=\pm1,0$ in order to implement the fractional quantum
Hall effect.

\subsection{The knot lattice of spin 1/2 Kitaev honeycomb lattice model}

This knot lattice model can also be constructed to explore the
non-abelian anyon in the gapless phase of Kitaev honeycomb model
\cite{Kitaev2}. The knot patterns for constructing honeycomb
lattice is showed in Fig. \ref{kitaev}, here the local tribein is
acted by different spin operators on each arm,
\begin{equation}\label{Htri}
H_{k}=\sum_{i}(J_{x}S^{x}_{i}S^{x}_{i+\textbf{e}_x}+J_{y}S^{y}_{i}S^{y}_{i+\textbf{e}_{_y}}
+J_{z}S^{z}_{i}S^{z}_{i+\textbf{e}_z}).
\end{equation}
Different ferromagnetic coupling strengths are assigned in each
direction of local tribein, i.e., $J_{l}<0, (l = x,y,z)$. Here
$(S^{x},S^{x},S^{x})$ are spin-1/2 operators. In the knot lattice
model (Fig. \ref{kitaev}), we introduce three color anyons, the
red ($\sigma_{_R}$), yellow ($\sigma_{_Y}$) and blue anyons
($\sigma_{_B}$), to implement this anisotropic coupling type. The knot pattern generated by the red and blue anyon is only acted by spin operator $S^{x}$. $S^{y}$ acts on the red- and yellow
anyon. While $S^{z}$ acts on yellow and blue anyon (Fig.
\ref{yang} (a)). The Yang-Baxter equation still holds for this
knot lattice (Fig. \ref{yang} (a)). Extending this anisotropic
coupling tribein over the whole honeycomb lattice, it naturally
leads to an equivalent model to Kitaev model on honeycomb lattice,
except here every local coupling tribein has been rotated for
consistence (Fig. \ref{kitaev}).

The three color anyons carry topological color charges. A running
red anyon generates yellow and blue field to drive the motion of
yellow anyon and blue anyon correspondingly. In gapped phase of
this knot lattice model, these anyons obey abelian anyon fusion
rules (Fig. \ref{yang} (b) (c)),
\begin{eqnarray}\label{RYB}
\sigma_{_R}\times\sigma_{_Y} = \psi_{_B}, \;\;\; \sigma_{_Y}\times\sigma_{_B} = \psi_{_R}, \;\;\; \sigma_{_B}\times\sigma_{_R} = \psi_{_Y}.
\end{eqnarray}
When the three tribein all fuse into Majorana fermions, a triplet of Majorana fermion is left in the middle, $\psi_{R}\psi_{Y}\psi_{B}$, which is a fermionic cluster (Fig. \ref{yang} (c)). The knot lattice model has three conserved plaquette operators around three adjacent hexagon plaquette (Fig. \ref{kitaev}),
\begin{eqnarray}\label{plaquette}
W_{_R} &=& S^{z}_{1}S^{z}_{2}S^{z}_{3}S^{z}_{4}S^{z}_{5}S^{z}_{6},\;\;\;
W_{_B} = S^{y}_{1}S^{y}_{2}S^{y}_{3}S^{y}_{4}S^{y}_{5}S^{y}_{6},\;\;\;\nonumber\\
W_{_Y} &=& S^{x}_{1}S^{x}_{2}S^{x}_{3}S^{x}_{4}S^{x}_{5}S^{x}_{6}.
\end{eqnarray}
here $(1,2,3,4,5,6)$ indicate the six lattice sites around each
hexagon. $W_{_R}$ draws a closed loop around the red loops in Fig.
\ref{kitaev}, so does the blue loops and yellow loops for $W_{_B}$
and $W_{_Y}$ on other hexagon plaquette. The plaquette operator is
in fact Wilson loop, which can be viewed as generalized magnetic
flux. A conserved color anyon runs in each Wilson loop. These
Wilson loop operators commute with Kitaev Hamiltonian,
$[W_{p},H_k]=0$, (p = R, B, Y). Thus Wilson loop and the knot
lattice share the same ground state. The minimum energy state of
the knot lattice model is reached at vortex-free state, i.e.,
$W_{p}=+1$. A vortex is generated in the plaquette if $W_{p}=-1$.

\begin{figure}[htbp]
\centering
\par
\begin{center}
$
\begin{array}{c@{\hspace{0.03in}}c}
%\multicolumn{1}{1}{\mbox{}} & \multicolumn{1}{1}{\mbox{}} \\
\includegraphics[width=0.24\textwidth]{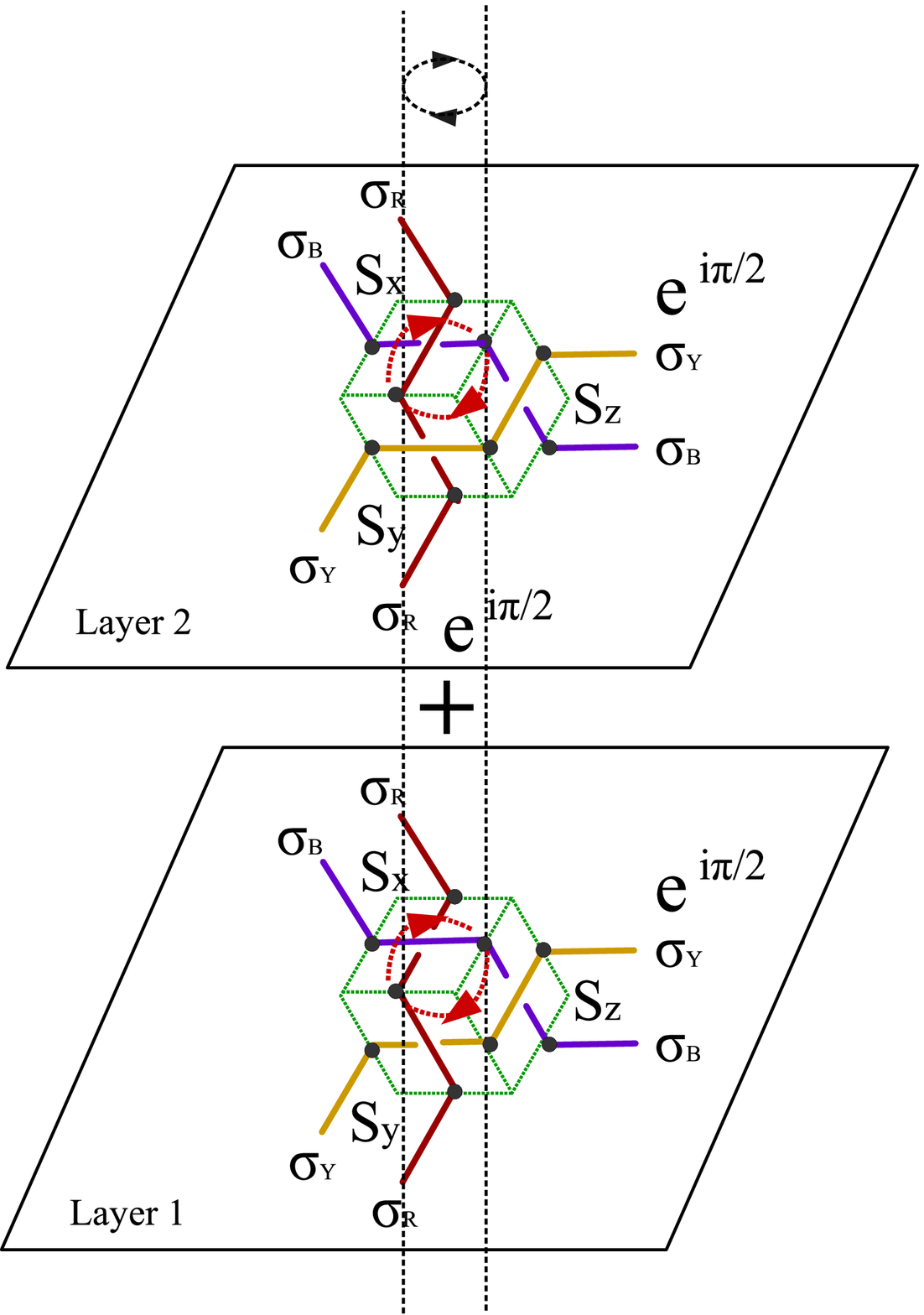}&\includegraphics[width=0.24\textwidth]{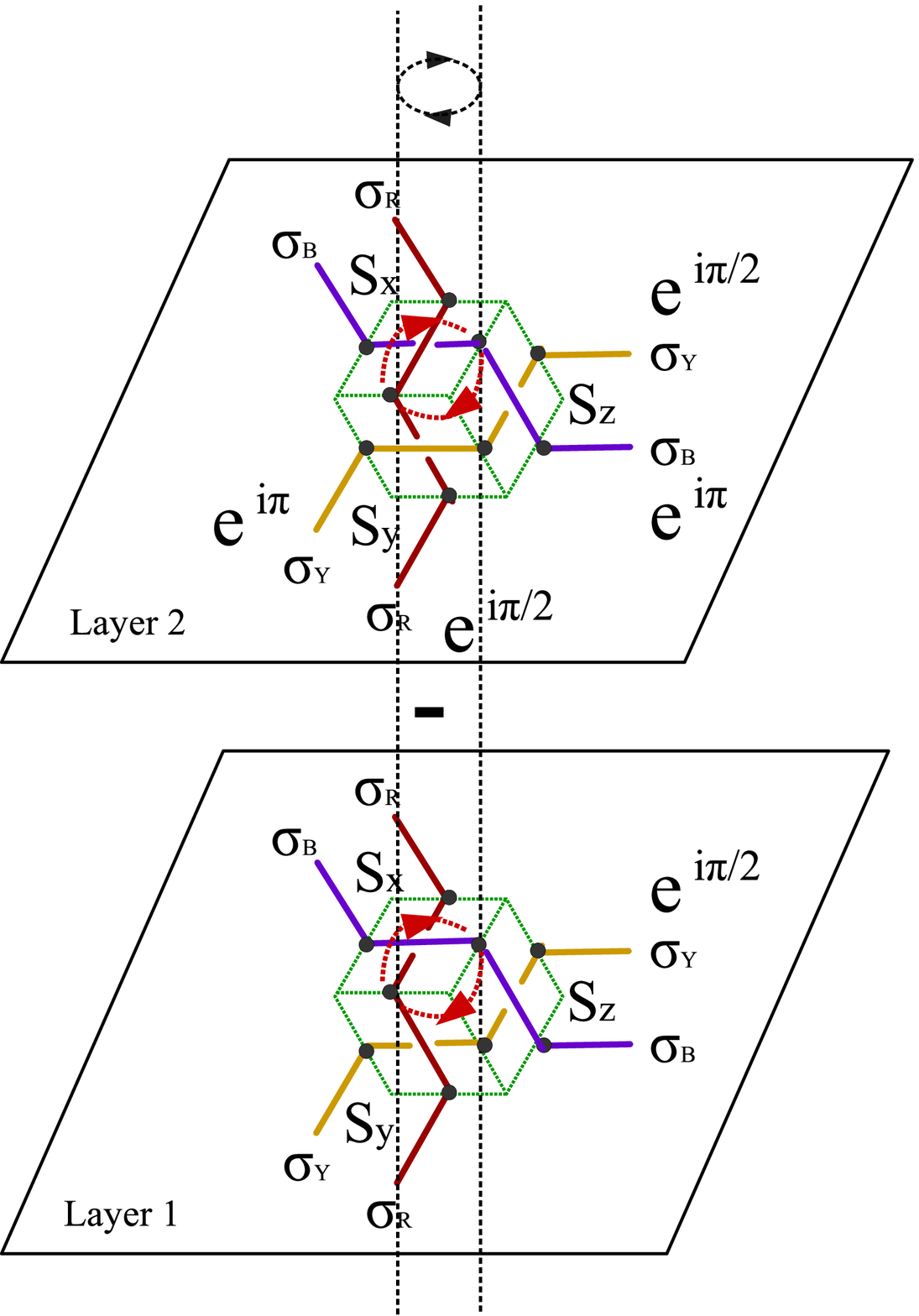}\\
\mbox{\textbf{a}} & \mbox{\textbf{b}}\\
\end{array}
$
\end{center}\vspace{-0.3cm}
\caption{\label{bilayer} (a) The knot configuration of positive eigen-energy state of spin operator.. (b) The knot configuration of negative eigen-energy state of spin operator.}
\vspace{-0.2cm}
\end{figure}

The knot lattice pattern of the ground state in each plaquette is
consist of $(2^{6}-1)$ layers of link lattice, since each Wilson
loop operator is constructed actually by 6 effective spin
operators. The eigenstate of Wilson loop operator is superposition
state of 6 spins values, which is either $|\downarrow\rangle$ or
$|\uparrow\rangle$. The ground state of this Wilson loop is a spin
liquid state, which does not show magnetically ordered state. We
denote the spin configuration at the ground state of Wilson loop
operator as,
\begin{eqnarray}\label{wilsonground}
|g\rangle &=& \sum |s_{1},s_{2},s_{3},s_{4},s_{5},s_{6}\rangle, \nonumber\\
W_{p} |g\rangle &=& \prod_{q=1}^{6} (\pm 1)_q |g\rangle = + 1 |g\rangle,  \;\;\; p = R, B, Y.
\end{eqnarray}
For example, the eigenstate of Wilson loop operator, $W_{R}$, is
the superposition state of every possible spin configuration with
even number of $|\downarrow\rangle$,
\begin{eqnarray}\label{blockz}
|g\rangle &=& | \downarrow_{1}\uparrow_{2}\downarrow_{3}\uparrow_{4}\uparrow_{5}\uparrow_{6}\rangle+| \uparrow_{1}\downarrow_{2}\downarrow_{3}\uparrow_{4}\uparrow_{5}\uparrow_{6}\rangle+\cdots, \nonumber\\
W_{_R} |g\rangle  &=&  + 1 |g\rangle.
\end{eqnarray}
Flipping one spin at any one of the six sites generates a
plaquette vortex. For the other two plaquette operators, $W_{B}$
and $W_{Y}$, the spin configuration of ground state is consists of
a bilayer knot lattice, so that the spin operator fulfills the
eigen-equation at every lattice site,
\begin{eqnarray}\label{sxsysz}
S_{i}^{x}(|\uparrow\rangle_{i}^{x} + |\downarrow\rangle_{i}^{x}) &=& +1(|\downarrow\rangle_{i}^{x} + |\uparrow\rangle_{i}^{x}),\nonumber\\
S_{i}^{y}(|\uparrow\rangle^{y} + i |\downarrow\rangle_{i}^{y}) &=& +1(|\uparrow\rangle_{i}^{y} +i |\downarrow\rangle_{i}^{y}),\nonumber\\
S_{i}^{z}(|\uparrow\rangle_{i}^{z} + |\uparrow\rangle_{i}^{z}) &=& +1(|\uparrow\rangle_{i}^{z} +|\uparrow\rangle_{i}^{z}).
\end{eqnarray}
One exemplar bilayer knot pattern that fulfils the equations above
is the sum of two layers (Fig. \ref{bilayer}). The $e^{i\pi/2}$
phase factor on the $S_{i}^{y}$ sector is assigned by the its
eigenstate construction Eq. (\ref{sxsysz}). While the $e^{i\pi/2}$
phase factor on $S_{i}^{z}$ sector is assigned by the commutator
relation of spin operators, $S^{a}S^{b}=i\epsilon_{abc}S^{c}$. For
the negative eigenvalue of $S_{i}^{p}$,
$S_{i}^{p}|\downarrow\rangle_{i}=-1|\downarrow\rangle_{i}$, we
introduce a minus between the two layers. The $S^{x}$ and $S^{y}$
sector state spontaneously flipped a sign. To avoid the vanishing
of eigenstate of $S_{i}^{z}$, we add a $e^{i\pi}$ phase upon the
the sector of $S_{i}^{z}$ in the second layer. At the same time,
to satisfy the equation
$S^{a}S^{b}|g\rangle=i\epsilon_{abc}S^{c}|g\rangle$, the product
of the eigenvalues of $S^{x}$ and $S^{y}$ must be negative to be
consistent with the eigenvalue of $S_{i}^{z}$. Thus we add a
$e^{i\pi}$ phase shift to the sector of $S^{y}$. Repeating this
knot lattice construction process for the six spins of the
plaquette, we can derive the superposition state of 62 layers of
knot plaquette as the eigenstate of the Wilson loop operator.

Braiding the color anyons in the knot lattice with respect to the
eigenstate of Kitaev model follows the same fusion Eq.
(\ref{braidings2}) as that for Ising model. Unlike the bilayer
configuration of Ising ground state, the ground state of one
plaquette in Kitaev honeycomb model has 62 layers. Braiding two
anyons across the 62 layers synchronously in clockwise (or
counterclockwise) direction would generate either a vacuum state
or a Majorana fermion state in each layer following the basic
fusion rules,
\begin{eqnarray}\label{BRIpsi}
&&\sigma_{_B}\times\sigma_{_R} = I + \psi_{_Y},\;\;\sigma_{_R}\times\sigma_{_Y} = I + \psi_{_B},\;\;\nonumber\\
&&\sigma_{_Y}\times\sigma_{_B} = I + \psi_{_R}.
\end{eqnarray}
The anyon braiding in $S^{x}$ and $S^{y}$ sector carries phase
factors due to commutator relations of spin operators and
eigenstate construction.

\begin{figure}
\begin{center}
\includegraphics[width=0.4\textwidth]{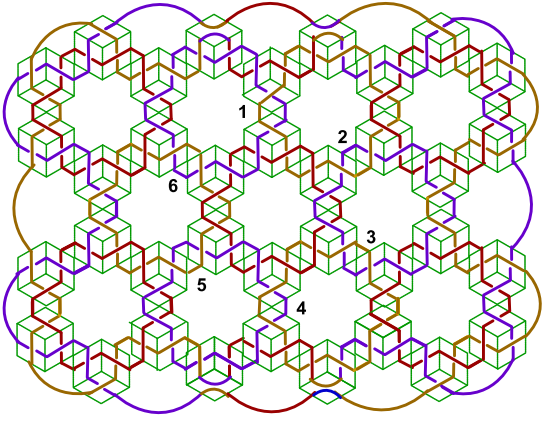}
\caption{\label{edgecurrent} The consistent existence of edge current on the upper and bottom boundary of Kitaev model on honeycomb lattice.}
\end{center}
\vspace{-0.5cm}
\end{figure}

The knot lattice in real space also map out knot configuration in
momentum space, since the energy of system is independent of
representation parameters. The fermions operator on knot lattice
have periodical distribution in real space, thus it has a natural
mapping into momentum space by Fourier transformation. Then the
huge number of knot lattice layers in which electronic wave
propagates could reduce to a statistical distribution of
oscillating frequency in wave vector space. This distribution is
summarized by the spectrum of interacting system. The spin 1/2
Kitaev Hamiltonian was mapped into quadratic fermion Hamiltonian,
which is equivalent to complex fermion pairing Hamiltonian
\cite{Kitaev2}\cite{feng}\cite{hdchen}. In Kitaev's Majorana
fermion pairing representation of one spin operator,
[$S^{\alpha}=ib^{\alpha}c$, ($\alpha=x,y,z$)], the spin-spin
coupling Hamiltonian is mapped into a fermion coupling Hamiltonian
in momentum space \cite{Kitaev2},
\begin{eqnarray}\label{kitaevkk}
H_{kk}=\frac{1}{2}\sum_{k,\alpha,\beta}iA_{\alpha\beta}(k)(\psi_{-k,\alpha}\psi_{k,\beta}),
\end{eqnarray}
here $\alpha,\beta $ represent the local tribein $[\textbf{e}_x,
\textbf{e}_y, \textbf{e}_z]$ inside each unit cell. The spectrum
matrix has an off-diagonal representation,
\begin{eqnarray}\label{Akk}
A(k)=\left[\begin{array}{cc}
    0 & f({k}) \\
    -f({k})^{\ast} & 0
  \end{array}\right],
\end{eqnarray}
where the spectrum function
$f(k)=2(J_{x}e^{i\textbf{k}\cdot\textbf{r}_{1}}+J_{y}e^{i\textbf{k}\cdot\textbf{r}_{2}}+J_{z})$.
$\textbf{r}_{i}$ are the translational invariant unit vectors on
the hexagonal lattice, which points to the the next nearest
neighboring unit cell,
$\textbf{r}_{1}=(\frac{1}{2},\frac{\sqrt{3}}{2})$,
$\textbf{r}_{2}=(-\frac{1}{2},\frac{\sqrt{3}}{2})$. The Energy
spectrum is $E(k)=\sqrt{f({k})^{\ast}f({k})}$. The Kitaev
Hamiltonian in momentum space can also decompose into fermion-spin
coupling formulation,
\begin{eqnarray}\label{Hkkspin}
H_{k}(k)=\frac{1}{2}\sum_{k,\alpha,\beta}i(I_{y}S_{y}+I_{x}S_{x})_{\alpha,\beta}(\psi_{-k,\alpha}\psi_{k,\beta}),
\end{eqnarray}
Here $S_{x}$ and $S_{y}$ are conventional Pauli matrices. The
currents along the direction of x-spin and y-spin components are
\begin{eqnarray}\label{HkkIxIyIz}
I_{y}&=&J_{x}\cos(\frac{k_{x}}{2}+\frac{\sqrt{3}k_{y}}{2})+J_{y}\cos(-\frac{k_{x}}{2}+\frac{\sqrt{3}k_{y}}{2})+J_{z},\nonumber\\
I_{x}&=&J_{x}\sin(\frac{k_{x}}{2}+\frac{\sqrt{3}k_{y}}{2})+J_{y}\sin(-\frac{k_{x}}{2}+\frac{\sqrt{3}k_{y}}{2}).
\end{eqnarray}
These two current depicts the Fourier knot in momentum space, which usually express the variable $k_{x}$ by spatial frequency and a free parameter, i.e., ($k_{x}=\omega_{x}k$, $k_{y}=\omega_{y}k$). These two currents are actually the projection of the energy spectrum to X-spin and Y-spin, $E(k) = \sqrt{I_{x}^2+I_{y}^2}$. In the exact solutions by Jordan-Wigner transformation, $I_{y} = \epsilon_{k}$ represents the kinetic energy of free fermion.  $I_{x} = \Delta$ is the energy gap for excites a fermion pair. The energy spectrum $ E(k) = \sqrt{\epsilon_{k}^2+\Delta^2}$ has a similar form as p-wave BCS pairing model. If we consider the evolution of a time dependent knot lattice state with eigenenergy $E(k)$, its final state oscillates as a matter wave, $|\psi_{f}\rangle= A\exp[\frac{i}{\hbar}Et]|\psi_{i}\rangle.$ Higher $E(k)$ indicates a faster oscillation with higher energy. In the mean time, the energy current also oscillates in space. $\omega_{x}$ counts the number of oscillations in unit space along X-direction. Kitaev model in momentum space is an effective coupling model between fermion and pseudospin vector. When the pseudospin vector rotates in momentum space, the fermion current in the color channels also fluctuates following time dependent Heisenberg equation. In the gapless phase $E=0$, the Majorana fermion in each unit hexagon cell has a constant existence probability even though it still oscillates from site to site. The flipping probability of current crossing state of the multi-layer knot lattice has a static value. The gapless spectrum exists when the three coupling coefficients satisfy the following inequalities \cite{Kitaev2},
\begin{eqnarray}\label{gaplessJ}
&&|J_{x}| < |J_{y}|+|J_{z}|,\;\;\;|J_{y}| < |J_{z}|+|J_{x}|,\nonumber\\
&&|J_{z}| < |J_{x}|+|J_{y}|.
\end{eqnarray}
In this case, the three coupling bonds of local tribein reaches a balance, then noncommutative characteristic of three component spin operators dominates the system. The gapless phase is an ordered phase of spin liquid which shows up after the quantum entropy generated by oscillating Majorana fermions is suppressed. The gapless phase in Kitaev model corresponds to a vanishing spectrum of fermions pairing system, which is in fact the locations of vortex in momentum space.

\begin{figure}
\begin{center}
\includegraphics[width=0.42\textwidth]{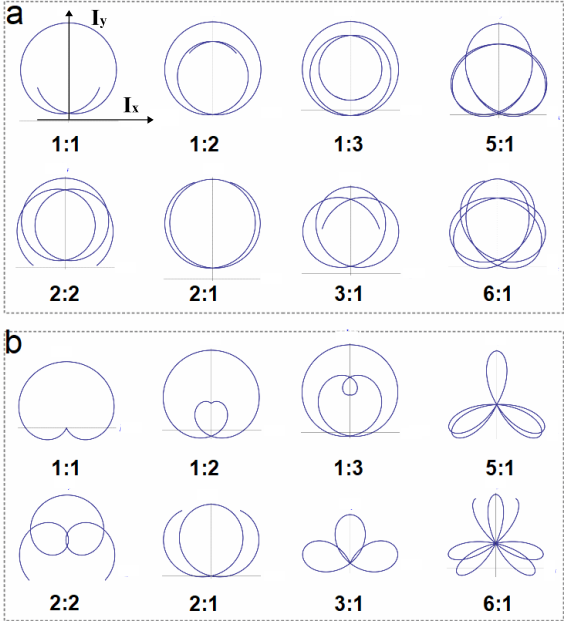}
\caption{\label{knotserial} (a) The Fourier knot serial in momentum space for the gapless phase of Kitaev model with respect to different frequency ratios, $\omega_{x}:\omega_{y}$. Parameter setting is [$J_{x}=40$, $J_{y}=10$, $J_{x}=40$]. (b) The Fourier knot serial with respect to different frequency ratios in the gapped phase of Kitaev model. Parameter setting is [$J_{x}=10$, $J_{y}=10$, $J_{x}=40$].}
\end{center}
\vspace{-0.5cm}
\end{figure}

\begin{figure}
\begin{center}
\includegraphics[width=0.45\textwidth]{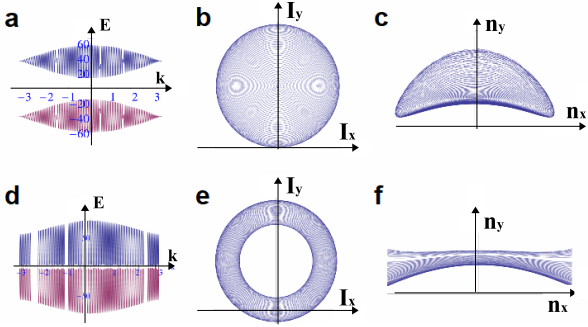}
\caption{\label{kitaevknot} (a) The energy spectrum of the gapped phase, with [$J_{x} = 10$, $J_{y} = 10$, $J_{x} = 40$], and frequency ratio, $\omega_{x}:\omega_{y} = 1:100$. (b) The Fourier knot of current in momentum space in gapped phase. (c) The knot of normalized current in momentum space of gapped phase with the same parameters as (a,b). (d) The energy spectrum of the gapless phase, with [$J_{x} = 40$, $J_{y} = 10$, $J_{x} = 40$], and frequency ratio, $\omega_{x}:\omega_{y} = 1:100$. (e) The Fourier knot of current in momentum space of gapless phase. (f) The knot of normalized current in gapless phase with the same parameters as (d,e).}
\end{center}
\vspace{-0.5cm}
\end{figure}

The Fourier knots of current and its corresponding normalized current in momentum space have different topology for different phases (Fig. \ref{knotserial}, Fig. \ref{kitaevknot}). In the gapless phase (with parameter setting ($J_{x}=40$, $J_{y}=10$, $J_{x}=40$)), the serial knot for different frequency ratios draw different knots around a forbidden hole in the middle. For the equal frequency, $1:1$, the knot is a simple circle (Fig. \ref{knotserial} (a) $1:1$). It draws a trefoil knot for  $1:3$ and an overlapping trefoil knot for $6:1$. For other ratios, $\omega_{x}:\omega_{y}=1:\omega_{y}$, the knot wraps $\omega_{y}$ rounds with a winding number $W =\omega_{y}$ (Fig. \ref{knotserial} (a)). For other ratios, $\omega_{x}:\omega_{y} = \omega_{x} : 1$, the winding number is $W =\omega_{x}$. For arbitrary frequency ratios, the winding number is $W =\omega_{x}+\omega_{y}-1$. As the frequency goes higher to $\omega_{x}:\omega_{y}=1:100$, the knot in momentum space approaches to a toroid plate (Fig. \ref{kitaevknot} (e)). The corresponding normalized currents, ($n_{x}=I_{x}/E(k)$,\;\; $n_{y}= I_{y}/E(k)$), depicts an open band (Fig. \ref{kitaevknot} (f)). In the meantime, the oscillating two bands touch each other periodically along $k=0$ (Fig. \ref{kitaevknot} (d)). For the gapped phase, the energy current draws serial knots which pass the interior of a circular area without forbidden hole (Fig. \ref{knotserial} (b)). It winds $\omega_{y}$ closed loops for frequency ratio $\omega_{x}:\omega_{y}=1:\omega_{y}$. Trefoil knot also appears for  frequency ratio $\omega_{x}:\omega_{y}=3:1$ (Fig. \ref{knotserial} (b)). The energy spectrum shows two separated bands with many oscillations enveloped in one wave package (Fig. \ref{kitaevknot} (a)). The energy current knot turns into a filled disc at the high frequency ratio of $\omega_{x}:\omega_{y}=1:100$ (Fig. \ref{kitaevknot} (c)). The corresponding normalized current also depicts a complicate knot which winds into a closed half-moon shape (Fig. \ref{kitaevknot} (c)). Thus the energy current of gapless phase has an Euler characteristic $\chi=0$, while the gapped phase has a nonzero Euler characteristic, $\chi=2$. This knot lattice implementation of Kitaev model has a straightforward extension into knot lattice in three dimensions by anyonic loop model \cite{yu}.

The block spin-1 Kitaev honeycomb may also demonstrate fractional quantum hall effect. Most crossing lines become turning arcs that transform one type of anyon into another. The conserved plaquette operators still exist for this block spin-1 Kitaev Hamiltonian due to the same commutator relation of spin operators. The three plaquette operators, [$W_{p}$, (p = R, B, Y)], bear the same formulation as Eq. (\ref{plaquette}). The ground state of this block spin-1 plaquette operator is the same as that of spin-1/2 Kitaev Hamiltonian. However the plaquette excitations are completely different from that of spin-1/2 Kitaev model. The first excited state is determined by $W_{p}=0$, (p = R, B, Y). If any one of the six unit cell of hexagon plaquette joins in the vacuum state of zero crossing state, then it generates one plaquette excitation that we called vacuum excitation. There exist 62 possible vacuum excitations in total. The topological operation of these vacuum excitation is to disentangle two neighboring loops. The vacuum excitation is generated by braiding two anyons once (Fig. \ref{yang} (d)). Braiding anyon twice upon the ground along the same direction leads the plaquette to the second excited states. The three states of $S=\pm,0$ are eigenstate of $S^{z}$ sector. While in the $S^{x}$ or $S^{x}$ sector, even number of braiding plus Majorana fermion flipping does not generate the eigenstate of $S^{x}$. In that case, the bilayer knot of eigenstates for $S^{x}$ or $S^{y}$ have to be acted synchronously.

\subsection{The gapless edge current of non-abelian anyon in a finite knot lattice}

\begin{figure}[htbp]
\centering
\par
\begin{center}
$
\begin{array}{c@{\hspace{0.06in}}c}
%\multicolumn{1}{1}{\mbox{}} & \multicolumn{1}{1}{\mbox{}} \\
\includegraphics[width=0.23\textwidth]{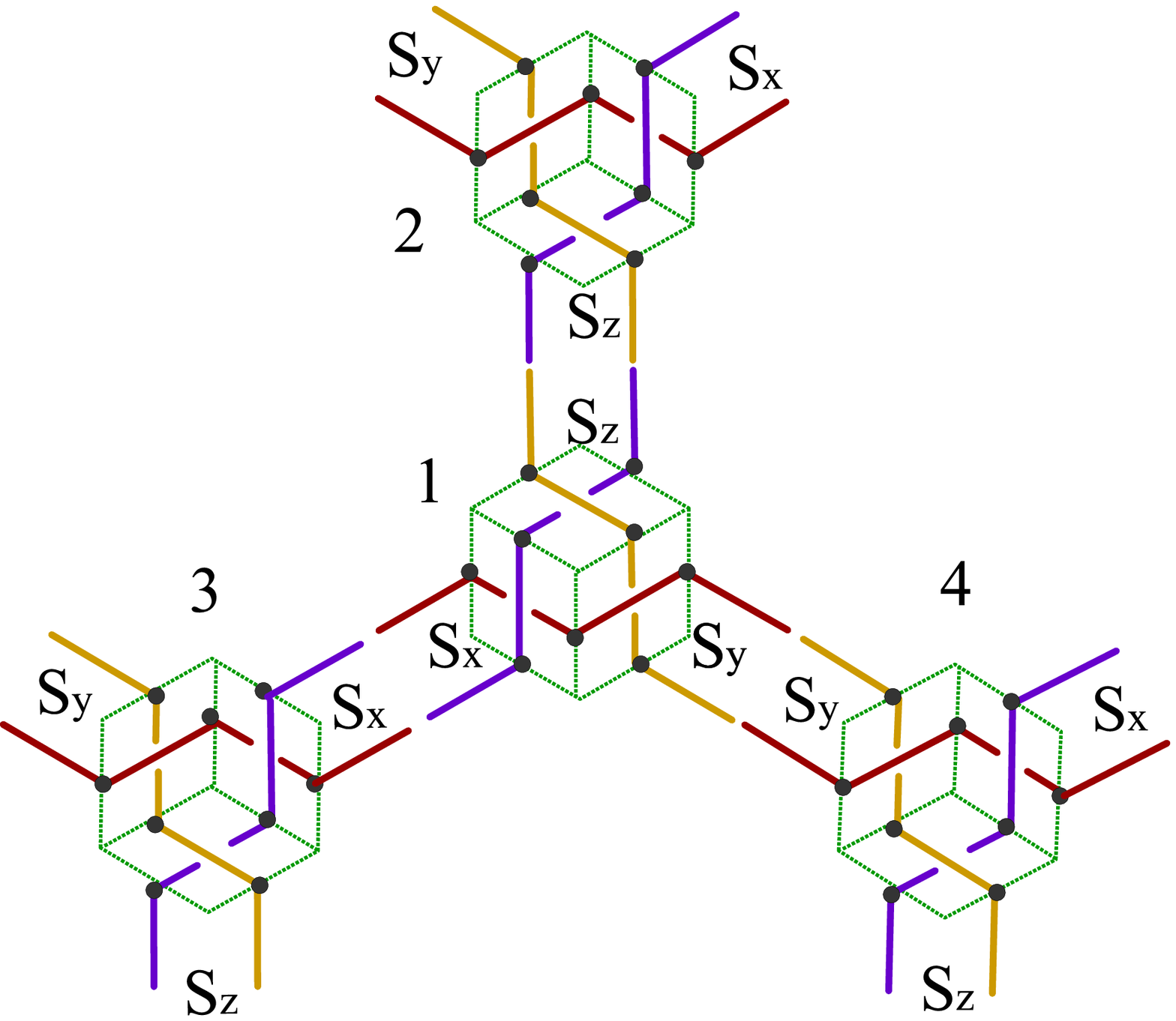}&\includegraphics[width=0.22\textwidth]{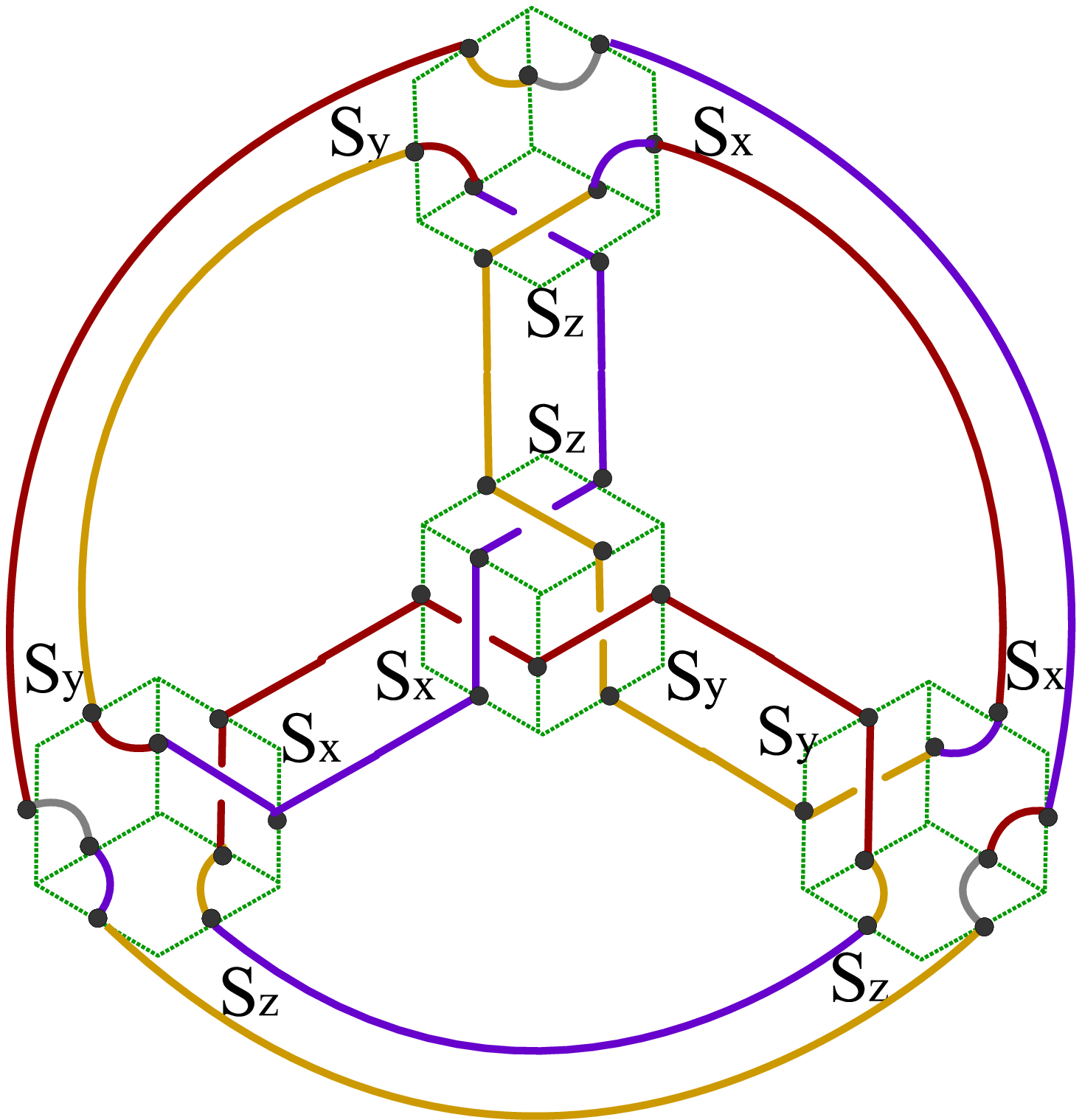}\\
\mbox{\textbf{a}} & \mbox{\textbf{b}}\\
\end{array}
$
\end{center}\vspace{-0.3cm}
\caption{\label{tetra} (a) The knot lattice of four spin's tribein mismatch each other on each bond. (b) The knot configuration of edge states with exotic fusion rules. The grey arcs indicates fermionic vacuum string.}
\vspace{-0.2cm}
\end{figure}

For a finite lattice without periodic boundary condition, the
boundary coupling types have to be carefully arranged in order to
map the Kitaev model consistently. Because there are always three
color strings branches on the upper boundary or bottom boundary,
there is no way to connect the strings with the same color
without over crossing or under crossing other current. The only
consistent connection pattern is fusing color anyon into colored
Majorana fermions (Fig. \ref{edgecurrent}). Two edge currents of
mixed anyons and Majorana fermions exist both on the upper
boundary and bottom boundary (Fig. \ref{edgecurrent}). However, if
the lattice structure is not perfect hexagonal lattice, instead it
contains certain plaquette formed by odd number of edges, then the
anyons in the edge current does not obey abelian fusion rules.
There must exist some fermionic vacuum state and unpaired Majorana
fermions.  For instance, the minimal extension of spin 1/2 Kitaev
model on a lattice with odd plaquette is tetrahedron lattice which
is constructed by four triangles \cite{tieyan}. The Hamiltonian
simply reads (Fig. \ref{tetra} (a)),
\begin{eqnarray}\label{kita4}
H_{t}&=&J_{x}S^{x}_{1,\textbf{e}_x}S^{x}_{3,\textbf{e}_x}
+J_{y}S^{y}_{1,\textbf{e}_y}S^{y}_{4,\textbf{e}_y}
+J_{z}S^{z}_{1,\textbf{e}_z}S^{z}_{2,\textbf{e}_z}\nonumber\\
&+&J_{x}S^{x}_{2,\textbf{e}_y}S^{x}_{4,\textbf{e}_y}+J_{y}S^{y}_{2,\textbf{e}_y}S^{y}_{3,\textbf{e}_y}
+J_{z}S^{z}_{3,\textbf{e}_y}S^{z}_{4,\textbf{e}_y}.
\end{eqnarray}
The tribein of each lattice cannot directly connect to
its neighbor sites without introducing extra crossings (Fig.
\ref{tetra} (a)). If extra over-crossings are introduced for
consistence, the tetrahedron lattice model finally turns into
honeycomb lattice model again. If vacuum and exotic Majorana state
are introduced on the edge bonds, the Kitaev-type tetrahedron
model can also be constructed consistently (Fig. \ref{tetra} (b)).
However the anyon on the edges obeys exotic fusion rules
\begin{eqnarray}\label{fusionedge}
&&\sigma_{_R}\times\sigma_{_B} = \psi_{_{Y}} + \psi_{_{B}},\;\;\sigma_{_R}\times\sigma_{_B} = I + \psi_{_{B}},\;\;\nonumber\\
&&\sigma_{_B}\times\sigma_{_Y} = \psi_{_{Y}} + \psi_{_{B}},\;\;\sigma_{_B}\times\sigma_{_Y} = I + \psi_{_{Y}},\;\;\nonumber\\
&&\sigma_{_R}\times\sigma_{_Y} = I + \psi_{_{R}}.
\end{eqnarray}
Thus abelian anyons run in the inner land current channels and
obeys the conventional fusion rule Eq. (\ref{RYB}). Non-abelian
anyons runs on the edge and generate Majorana fermions and
fermionic vacuum (grey bonds in Fig. \ref{tetra} (b))
simultaneously. There are three unpaired Majorana fermions running
in the edge loop. Odd loops breaks the time-reversal symmetry
under the transformation $T S_{i}^{\alpha} T^{-1} =
-S_{i}^{\alpha}$. Thus the three running Majorana fermion runs in
one direction, either in clockwise direction or in
counterclockwise direction.

Each spin operator in Kitaev honeycomb model is expressed by a pair of Majorana fermion \cite{Kitaev2}, $S^{\alpha}_{j}=ib^{\alpha}_{j}c_{j}.$ In the vacuum state of this knot lattice model, $b^{x}_{j}$ represents yellow arc, and $b^{y}_{j}$ for blue arc, $b^{z}_{j}$ for red arc. $c_{j}$ corresponds to the fermionic arc forming the closed loop around the $j$th lattice site. The effective Majorana fermion Hamiltonian of the gapless current reads (Fig. \ref{tetra} (b)),
\begin{eqnarray}\label{teHamimajo}
H_{t}^{e}=J_{x}\hat{o}_{2}b^{z}_{4}c^{y}_{2}c^{y}_{4}+J_{y}\hat{o}_{3}b^{x}_{2}c^{z}_{3}c^{z}_{2}+J_{z}b^{y}_{3}\hat{o}_{4}c^{x}_{3}c^{x}_{4},
\end{eqnarray}
Here $\hat{o}_{j}$ is fermionic vacuum operator, which has an effective representation by Clifford algebra. The product of three bond operators, $u_{ij}=c^{y}_{i}c^{y}_{j}$, construct a Wilson loop operator, $W =W_{a}W_{b}W_{c} u_{24}u_{23}u_{34}$, with ($W_{a},W_{b},W_{c}$) as the rest part of the three independent loops.
This Wilson loop operator commutes with the bulk Hamiltonian. The edge Hamiltonian reduces to a single fermion operator under a fixed gauge, $u_{ij}=+1$,
\begin{eqnarray}\label{teHamiIB}
H_{t}^{e}=J_{x}\hat{o}_{2}b^{z}_{4}+J_{y}\hat{o}_{3}b^{x}_{2}+J_{z}\hat{o}_{4}b^{y}_{3}.
\end{eqnarray}
Thus the coupling type between spins on the edge is translated one step back along the coupling parameter sequence. The fermionic vacuum exist as a quasi-excitation coupled to an unpaired Majorana fermion.

\begin{figure}
\begin{center}
\includegraphics[width=0.3\textwidth]{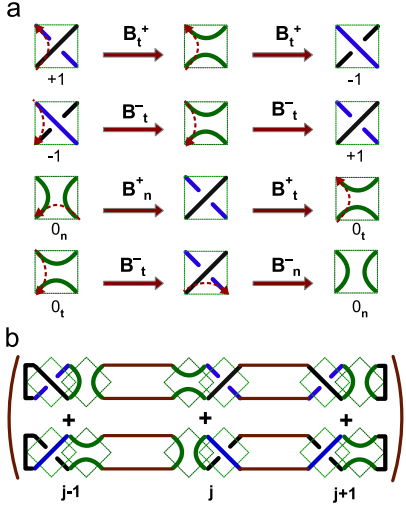}
\caption{\label{edgemode} (a) Two sequential braiding map one spin state to the other state, and one vacuum states to the other. (b) The eigenstate of knot lattice for the gapless edge current Hamiltonian.}
\end{center}
\vspace{-0.5cm}
\end{figure}

To generate chiral anyon current on the boundary of honeycomb lattice, we can introduce a pentagon or triangular plaquette on the upper boundary or bottom boundary. For an extension of Kitaev model on lattice with odd plaquette \cite{yao}\cite{cpsun}, this chiral edge current was born to exist on the boundaries. The most general Hamiltonian for edge current of non-abelian anyons chain model reads,
\begin{eqnarray}\label{HamitetraOB}
H_{t}^{c}=\sum_{ \langle{j}\rangle}J_{x}\hat{o}_{3j}b_{3j+1}+J_{y}\hat{o}_{3j+1}b_{3j+2}
+J_{z}\hat{o}_{3j+2}b_{3j+3}.
\end{eqnarray}
Here $b$ is a general Majorana fermion. $\hat{o}$ is a fermionic vacuum operator. One usual construction of Majorana fermion by conventional fermion operator is
\begin{eqnarray}
b_{j}=(c_{j}+c^{\dag}_{j}), \;\;b_{l}=(c_{l}-c^{\dag}_{l})/i.
\end{eqnarray}
The ground state of this Majorana fermion is $|g\rangle=(|0\rangle-c^{\dag}_{j}|0\rangle)/\sqrt{2}$, i.e., $b_{j}|g\rangle=-1|g\rangle.$ While the vacuum fermion has no conventional construction. The only self-consistent construction of vacuum fermion and Majorana fermion simultaneously is based on knot lattice chain model (Fig. \ref{transising}). The creation operator can be defined as product of two clockwise braiding operators (Fig. \ref{edgemode}). While the annihilation operator is the product of two counterclockwise braiding operators. The vacuum operator is also product of braiding operators (Fig. \ref{edgemode} (a)),
\begin{eqnarray}\label{c=BBo=B}
c_{j} &=& B^2_{n\circlearrowright,j}, \;\;c^{\dag}_{j} = B^2_{n\circlearrowleft,j}, \;\;
\hat{o}=B_{n\circlearrowright,j}B_{t\circlearrowright,j},\nonumber\\
c_{j} &=& B^2_{t\circlearrowleft,j}, \;\;c^{\dag}_{j} = B^2_{t\circlearrowright,j}, \;\;
\hat{o}=B_{n\circlearrowleft,j}B_{t\circlearrowleft,j}.
\end{eqnarray}
The creation operator, annihilation operator and vacuum operator have two different representations by braiding operators in different direction (Fig. \ref{edgemode} (a)). In order to match the conventional symbol of quantum operators, we replace the symbol ($\circlearrowright$ and $\circlearrowleft$) with ($-$ and $+$),
\begin{eqnarray}\label{c=BBo=B+-}
c_{j} &=& (B^{-}_{n,j})^2, \;\;c^{\dag}_{j} = (B^{+}_{n,j})^2, \;\;
\hat{o}_{j}=B^{-}_{n,j}B^{-}_{t,j},\nonumber\\
c_{j} &=& (B^{+}_{t+j})^2, \;\;c^{\dag}_{j} = (B^{-}_{t,j})^2, \;\;
\hat{o}_{j}=B^{+}_{n,j}B^{+}_{t,j}.
\end{eqnarray}
The indices ${n}$ and $t$ indicate normal and tangential direction along which the braiding is performed. The fermion chain model of non-abelian anyon are expressed by pure braiding operators,
\begin{eqnarray}\label{HamitetraBk}
&&H_{t}^{c}=\sum_{ \langle{j}\rangle}J_{x}\hat{o}_{3j+1}[(B^{-}_{n,3j+1})^2+(B^{+}_{n,3j+1})^2]\nonumber\\
&&+J_{y}\hat{o}_{3j+2}[(B^{-}_{n,3j+2})^2+(B^{+}_{n,3j+2})^2]\nonumber\\
&&+J_{z}\hat{o}_{3j+3}[(B^{-}_{n,3j+3})^2+(B^{+}_{n,3j+3})^2],
\end{eqnarray}
where fermionic vacuum operator is
\begin{eqnarray}\label{o=bb-bb}
\hat{o}_{j}=[B^{-}_{n,j}B^{-}_{t,j}-B^{+}_{n,j}B^{+}_{t,j}].
\end{eqnarray}
The non-commutable characteristic of braiding operators is consistent with non-abelian anyon statistics. The eigenstate of this non-abelian anyon Hamiltonian can be constructed by the sum of two knot chains (Fig. \ref{edgemode} (b)), i.e., $H_{t}^{c}|{\psi}\rangle=\pm|{\psi}\rangle$. For an isolated crossing state, single braiding operator is neither fermion nor boson. Since exchanging two pairs of double braiding operator contributes $\exp[i\pi]$, each braiding is equivalent to an anyon which bears an effective statistical factor $\exp[i\pi/2]$. However if there exist many connected neighboring zero energy states and only a few crossing states, the statistical phase factor from collective spin up to spin down is not $\exp[i\pi]$ any more. In that case, the statistical phase is determined by the filling factor,
\begin{eqnarray}
\nu=\frac{L_{link}}{N(B)},
\end{eqnarray}
$N(B)$ is total number of braiding to transform one collective state to another one. This number has a straightforward counting from a Hamiltonian in real space. An equivalent counting can also be performed in momentum space under Fourier transformation,
\begin{eqnarray}
B^{-}_{n,3j}& =& \frac{1}{\sqrt{N}}\sum_{k_{x}}e^{-ik_{x}{j3a}}B^{-}_{n}(k_{x}),\nonumber\\
B^{+}_{n,3j}& =& \frac{1}{\sqrt{N}}\sum_{k_{x}}e^{ik_{x}{j3a}}B^{+}_{n}(k_{x}).
\end{eqnarray}
The anyon fermion pair model is mapped into a coupling model of four braiding operators. The braiding operator in momentum space braids different energy levels. The spin currents with opposite spin in quantum spin Hall effect cannot cross each other in energy spectrum, this defines an over-crossing state or under-crossing state. The level crossing avoidance point defines different vacuum state. Higher excited knot states of energy levels in momentum space also exist for non-abelian anyon models.

\begin{figure}
\begin{center}
\includegraphics[width=0.35\textwidth]{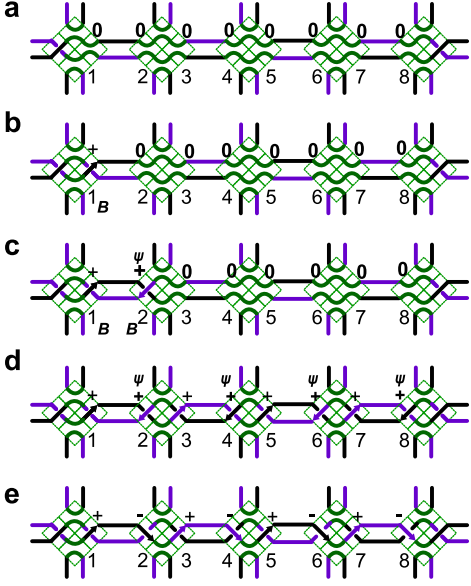}
\caption{\label{kanestrip} (a) one vacuum strip knot covering four neighboring unit cells. (b-c) The strip knot under one and two braiding. (d) The strip knot under after eight braiding. (e) The disentangled strip knot after four flipping operations by Majorana fermion $\psi$. }
\end{center}
\vspace{-0.5cm}
\end{figure}

The gapless edge current is actually one dimensional Majorana fermion chain including vacuum state. One dimensional quantum  chain of ordinary fermions with long range hopping terms is a more general case of non-abelian anyon.
\begin{eqnarray}\label{longHami}
\hat{H}_{L}=\sum_{i}[\sum_{j=1,\alpha}^{m}e^{i\phi}t_{j}{c}^{\dag}_{i\alpha}c_{i+j,\alpha}].
\end{eqnarray}
Here $m$ is the distance number that counts the number of lattice site between two lattice sites that is connected by continuous long range channels. In the presence of the long range hopping Hamiltonian Eq. (\ref{longHami}), long loops covering many sites would show up in the eigenstates. For the exemplar strip ordering phase in Fig. \ref{kanestrip}, the long range hopping Hamiltonian reads,
\begin{eqnarray}\label{longHami4}
\hat{H}_{L}=\sum_{j=1,\alpha}^{4}e^{i\phi}t_{j}{c}^{\dag}_{i\alpha}c_{i+j,\alpha}.
\end{eqnarray}
Suppose there exist a string of gapless modes covering four neighboring unit cells (Fig. \ref{kanestrip} (a)). Under one braiding operation at first sites (Fig. \ref{kanestrip} (a)), it generates one  $|+1\rangle$ state, this knot maps to a trivial circle by Reidemeister moves. The output state under two braiding is $|+1,-1\rangle$ (Fig. \ref{kanestrip} (b)), this knot has a linking number $L_{link}=1$, which results in a filling factor $\nu = 1/2$. The output knot under three same braiding is $|+1,+1,+1\rangle$, it take one flipping on $2$nd or two flips on $1$st and $3$rd sites to map into trivial knot state, thus the filling factor here is $1/3$ or $2/3$. The strip knot has eight alternative crossings (Fig. \ref{kanestrip} (d)) after eight braiding, which has a linking number of $L_{link}=4$. In order to map the maximal linked state into trivial circle, it takes four flipping operations at the $2$nd, $4$th, $6$th, and $8$th crossing sites. Then the ferromagnetic phase of this strip loop could ba map into trivial circle by Reidmeister move. This defines a collective vacuum state (Fig. \ref{kanestrip} (e)). Each flipping operation is equivalent to a Majorana fermion operator. The four flipping costs eight braiding operators. Thus the filling factors of this collective vacuum state is $\nu=4/8=1/2$. This is a local filling factor. It turns into a global filling factor for a periodical distribution of strip knots over a two dimensional knot lattice (Fig. \ref{kaneknot}). If the strip vacuum state covers $2n$ unit cells, the filling factor could reach $(n\pm1)/2n$. In this sense, the fractional filling state is essentially a topological finite size effect.

\subsection{Knot lattice implementation of quantum lattice model with quantum Hall states}

\begin{figure}
\begin{center}
\includegraphics[width=0.4\textwidth]{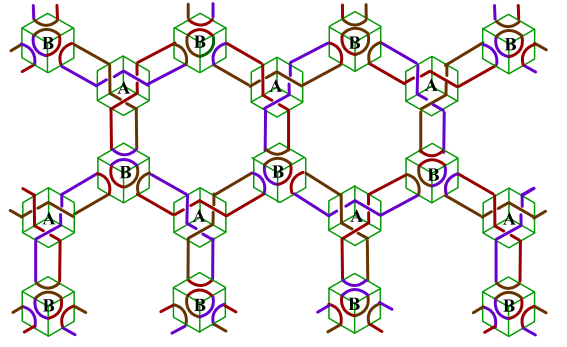}
\caption{\label{haldanelattice} One of many layers of honeycomb knot lattice implementation of Haldane model with the next nearest neighboring hopping terms for integral quantum Hall effect.}
\end{center}
\vspace{-0.5cm}
\end{figure}

Haldane model on honeycomb lattice demonstrates quantum Hall effect without Landau levels \cite{haldane}. The honeycomb knot lattice can also construct the eigenstate of Haldane model (Fig. \ref{kitaev}, Fig. \ref{haldanelattice}). The honeycomb lattice is composed of two triangular sub-lattice. In Haldane model, besides the hopping of fermions between nearest neighbors, the hopping of fermions among the next nearest neighbor sites are also introduced \cite{haldane},
\begin{eqnarray}\label{Haldane}
H_{hal} &=& \sum_{ \langle{ij}\rangle\alpha}t_{1}{c}^{\dag}_{i\alpha}c_{j,\alpha}
+ \sum_{\langle\langle{ij}\rangle\rangle\alpha}e^{i\phi}t_{2}{c}^{\dag}_{i\alpha}c_{j,\alpha}+h.c.\nonumber\\
&+& M \left(\sum_{ {i\in{A,\alpha}}}{c}^{\dag}_{i\alpha}c_{i,\alpha}-\sum_{ {i\in{B,\alpha}}}{c}^{\dag}_{i\alpha}c_{i,\alpha}\right).
\end{eqnarray}
The phase factor is $\phi=2\pi(2\phi_{a}+\phi_{b})/\phi_{0}$. If the next nearest neighbor sites are directly connected by continuous current channels (For instance, knot lattice in Fig. \ref{haldanelattice}), the next nearest neighbor hopping terms in Hamiltonian Eq. (\ref{Haldane}) can be directly performed on knot lattice to flip the local crossing state. However every lattice site must be an over-crossing state or under crossing state, in order to keep continuous channel connecting any next nearest neighbors. If there exist some block spin $|0\rangle$ states on certain lattice sites, there would not exist a minimal channel to connect certain next nearest neighbors (Fig. \ref{yang} (c)). Thus a superposition state of many honeycomb knot lattice with only crossing states can construct the eigenstate of Haldane model. Since the braiding operator is only allowed to perform on two states,$|+1\rangle$ or $|-1\rangle$, the dominator of filling factor can only be even number. Opposite writhing current (Fig. \ref{selflink}) can implement the two opposite magnetic field on renormalized sublattice. The Hall resistance in integral quantum Hall effect is quantized by first Chern number of Berry phase. The first Chern number is equivalent to Euler characteristic on a discrete lattice. In order to map the Chern number to Euler characteristic, the energy current in momentum space must be trianglized into grid. How to derive an exact mathematical mapping is still an unsolved problem in mathematics.

As showed in last section, fractional filling factor in quantum Hall effect is essentially quantized by the linking number of crossing current. The block vacuum state in block spin-1 knot lattice is inevitable to produce fractional quantum Hall states. As we already seen in knot square lattice model, more braiding within one unit square inevitably creates more lattice sites within the unit square. This means fractional Hall effect only exist for long range hopping dynamics to derive higher linking number. Since fractional filling factor only depends on topological braiding operations, it is independent of specific lattice structure. Each lattice site must admits a block spin 1 state, i.e., two crossing states ($|+1\rangle$, $|-1\rangle$) and two vacuum states ($|0_{t}\rangle$, $|0_{n}\rangle$). A global vacuum state is a set of minimal loop around the center of each unit cell, there is no next nearest neighboring hopping. Global vacuum state does not match Haldane model with next nearest neighbors hopping terms. If all of the vacuum sates are placed on $B$ lattice sites and none on $A$ lattices sites, the next nearest neighboring hopping still survives to construct Haldane model (Fig. \ref{haldanelattice}). It takes three braiding to bring the trefoil knot around $A$ into vacuum states, the corresponding writhing number decreases from three to zero. Three crossings were eliminated during this braiding process. Thus the filling factor of the vacuum excitation is one.

\begin{figure}
\begin{center}
\includegraphics[width=0.4\textwidth]{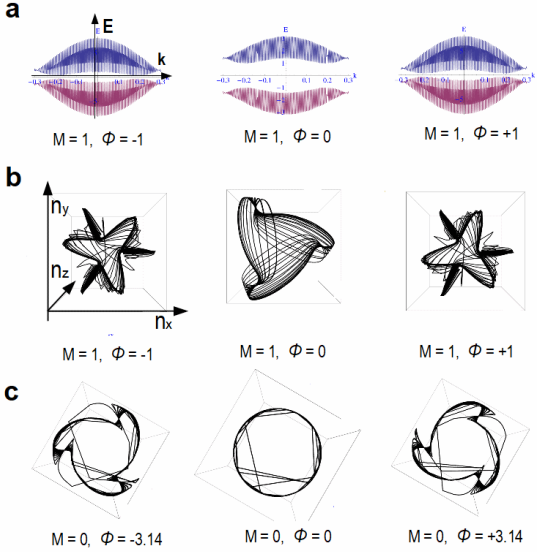}
\caption{\label{haldanenknot} (a)(b) The energy spectrum and knot of Haldane model in momentum space with respect to different phases $\phi$ and on-site energy $M=1$. (c) The knot of Haldane model in momentum space on-site energy $M=0$. The frequency ratio for the figures above is $\omega_{x}:\omega_{y}$=1:100 and $t_{2}=-2$. }
\end{center}
\vspace{-0.5cm}
\end{figure}

The energy spectrum of Haldane model in momentum space \cite{haldane} can be decomposed as the function of Fourier knot drew by the three energy current with respect to the three spin components,
\begin{eqnarray}\label{HaldaneIxyz}
I_{x}&=&\cos(-\sqrt{3}\pi\omega_{x}k+\pi\omega_{y}k)+\cos(\sqrt{3}\pi\omega_{x}k+\pi\omega_{y}k)\nonumber\\
&+&\cos(-2\pi\omega_{y}k),\nonumber\\
I_{y}&=&\sin(-\sqrt{3}\pi\omega_{x}k+\pi\omega_{y}k)+\sin(\sqrt{3}\pi\omega_{x}k+\pi\omega_{y}k)\nonumber\\
&+&\sin(-2\pi\omega_{y}k),\nonumber\\
I_{z}&=&[M-2t_{2}\sin(\phi)[\sin(-\sqrt{3}\pi\omega_{x}k+3\pi\omega_{y}k)\nonumber\\
&+&\sin(\sqrt{3}\pi\omega_{x}k+3\pi\omega_{y}k)].
\end{eqnarray}
The Hall conductivity $\sigma^{xy}$ is quantized at the Fermi level, $\nu({e^2}/h)$. The integer here $\nu=+1$ ($\nu=-1$) for $\phi>0$($\phi<0$) \cite{haldane}. This integer measures the sign of the winding number of knot in momentums space. The circular trefoil knot in momentum space for $\phi=-3.14$ and $\phi=-1$ turn into the opposite direction of that for positive phases, i.e., $\phi=3.14$ and $\phi=1$ (Fig. \ref{haldanenknot} (b-c)). The phase $\phi$ only modifies the initial phase of the oscillating energy spectrum wave, but does not change the gap characteristic between two bands, except here the energy gap reaches the maximal value at zero phase (Fig. \ref{haldanenknot} (a)). The trefoil knot in momentum space here is the dual mapping of trefoil knot in real space (Fig. \ref{haldanelattice}).

\begin{figure}
\begin{center}
\includegraphics[width=0.4\textwidth]{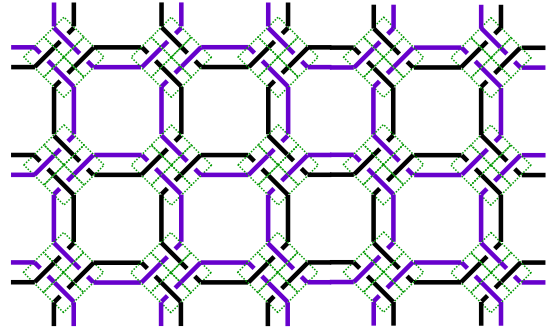}
\caption{\label{kaneknot} The square knot lattice of double currents for implementation of Haldane model and Kane-Mele model.}
\end{center}
\vspace{-0.5cm}
\end{figure}

The Haldane model is a topological band model with Chern number $C=1$. A topological flat band model with Chern number $C=2$ could be constructed on square lattice with the third nearest neighboring hopping \cite{yanggu}. This square knot lattice is a natural implementation of Kane-Mele model \cite{KaneMele}, which induces spin-orbital coupling into Haldane model \cite{haldane}. Each unit cell here is composed of four unit squares with internal crossing currents. Fermions with opposite spin run in separate channels (the purple channel and black channel in Fig. \ref{kaneknot}). For those channels turning left (right) to reach the next nearest neighbors, $\mu_{ij}=+1$($\mu_{ij}=-1$). The eigenstate of Kane-Mele model can be constructed by multi-layer knot configurations without block vacuum state. In order to implement higher order of fractional quantum Hall states, the Kane-Mele model must be extended to include long range hopping terms. A multi-layer generalization of Haldane model with the third nearest neighboring hopping is topological flat band model with arbitrary Chern numbers \cite{yanggu}. The same physics hold for the Haldane model on honeycomb lattice with the third nearest neighbors hopping,
\begin{eqnarray}\label{FQHkane}
H_{3} &=& H_{hal} + \sum_{j}[\sum_{m=1,\alpha}^{3}t_{3}{c}^{\dag}_{j\alpha}
c_{j+c_{m},\alpha}].
\end{eqnarray}
Three layers of knot lattice construction of Haldane model on Honeycomb lattice generates a topological band model with Chern number $C=3$ \cite{yanggu} as well as Chern number $C=2$ \cite{yuzhai}. Note inter-layer hopping does not exist in this construction. A straightforward construction of topological flat band model with Chern number $C=N$ is consists of N layers Haldane model with the third nearest neighboring hopping \cite{yanggu}.

\begin{figure}
\begin{center}
\includegraphics[width=0.35\textwidth]{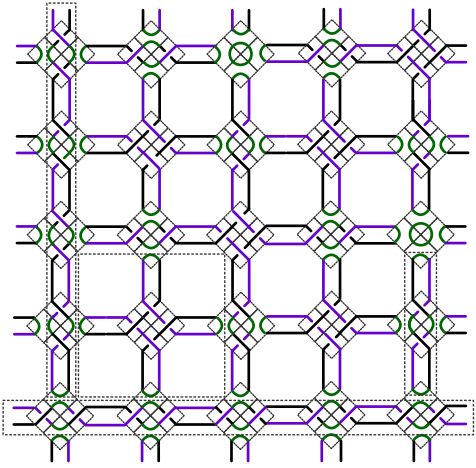}
\caption{\label{kane} Examples of elementary knot patters for different long range hopping terms for a topological band model of fractional quantum Hall states.}
\end{center}
\vspace{-0.5cm}
\end{figure}

\begin{figure}
\begin{center}
\includegraphics[width=0.45\textwidth]{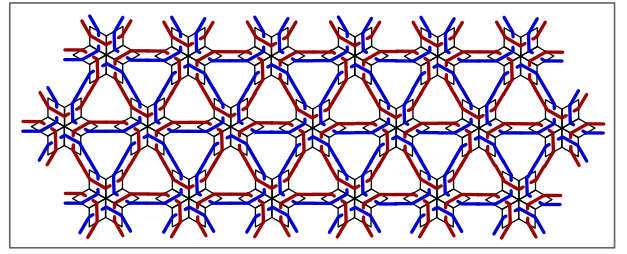}
\caption{\label{triangle} The knot triangular lattice for implementing topological flat band model with Chern number $C=3$.}
\end{center}
\vspace{-0.5cm}
\end{figure}

One exemplar knot lattice for fractional quantum Hall state of $1/3$ filling is topological flat band model with Chern number $C=2$. The local knot pattern on two dimensional square lattice visualize hopping distance of particles. For instance, the next nearest neighboring hopping on square lattice run through a small knot (Fig. \ref{kane}). The next-next-nearest-neighboring hopping runs along the big cross channel. The Fourth neighboring hopping terms run along two entangled loops covering many unit cells either along the horizontal line or vertical line (Fig. \ref{kane}). The vacuum state within one unit cell acts as a cut off of a long knot pattern. To construct knot lattice with respect to fractional quantum Hall state, a combination of these basic knot elements are arranged periodically on the whole lattice. Similar to the one dimensional strip knot, the initial state of the big crossing-knot could be set to vacuum state or ferromagnetic state. The filling factor for a given knot state is determined by its linking number and the number of braiding to reach that state from initial state. Fractional quantum Hall is a natural existence in this knot square lattice since electron runs in the channel that always turns from X-direction into Y-direction or vice versa. The long range hopping terms in Hamiltonian also expands an energy spectrum of free particle into extra parameter space. Each energy level in the conventional model containing only nearest neighbor hopping terms now expands to many hyperfine levels. The conventional fully filled energy levels now become partially filled, which results in fractional filling states. Fractional filling states bears a topological origin, it does not depends on specific lattice structure. A triangular knot lattice can also be constructed for topological flat band model with Chern number $C=3$ (Fig. \ref{triangle}).

The minimal vacuum loop excitation in square knot lattice model is different from that of honeycomb knot lattice. There are four Majorana arcs surrounding one lattice site to form the minimal loop in vacuum excitation of the square knot lattice (Fig. \ref{kaneknot}). Thus each vacuum loop excitation behaves as Boson. While the minimal vacuum loop around each center of hexagon unit cell is formed by three Majorana fermion arcs, thus the vacuum loop behaves as fermion (Fig. \ref{yangbaxter} (d)). Every pair of Majorana fermion arcs indicates the existence of one magnetic flux. Similar to block spin-1 Ising model, if the crossing state is braided for many times, it would generate extra crossing points within one unit cell. Then Majorana fermion operator as a crossing state flipper have to be introduced to bring the local crossing back to Hilbert space.

A more direct implementation of fractional quantum Hall effect on honeycomb knot lattice is to introduce an external magnetic field by placing one Dirac magnetic monopole at the center of a sphere lattice, or one magnetic monopole ring around the center line of a torus lattice. Then the input electrons carrying different gauge phases flow into different knot channels. Its Hall resistance bears the fundamental character of fractional quantum Hall effect. The partition function of the knot lattice is still a topological invariant. However every hexagon unit cell was attached by a complex coefficient. The total linking number of the knot lattice is now a complex number as well. We could start with the vacuum state without crossing at any lattice site and use Skein relation and Jones polynomial repeatedly to compute the partition function.

\section{Conclusion}

The entangled multiknot lattice model is essentially a new geometric representation of quantum many-body system with long range coupling interaction. It naturally encoded abelian anyon, non-abelian anyon, integral and fractional quantum Hall states, and so on. Under the exact one-to-one mapping between the over crossing or under crossing states of two current lines and the eigenstate of quantum operators, an arbitrary eigenstate of quantum many body model can be exactly represented by knot lattice patterns. Then each quantum state bears a topological character due to the topological field theory of many knots. In the case of short range coupling interactions, this multiknot lattice model spontaneously reduces to conventional quantum spin model. The eigenstate is a collective wave function of many entangled loop currents. The integral and fractional filling factors for the collective eigenstate of this knot lattice model is originated from the topological linking number of many knots, which can be computed by Chern-Simons field theory. The fractional statistics of anyons is explicitly demonstrated by braiding operation on local crossing states. The anyon physics revealed by this knot lattice model has a widespread existence in conventional fermion and spin models, such as Ising model, quantum Hall system, topological insulator model \cite{qi}\cite{Hasan}, BCS fermion pairing model, kitaev model, Haldane model and Kane-mele model, and so on. It offers knot representation on Ising anyon, an exact computation of integeral and fractional Hall conductance by linking number com, a Lissajous knot representation in the momentum space of topological insulator, and a new explanation to the pseudogap state by vortex dimer state. This knot model also predict two edge currents moving in the same direction on odd width of lattice, and fractional filling states in vortex dimer state as well as multi-vortex state in unconventional superconductors. Different quantum phases can be quantified by different topological linking numbers. For instance, the linking number shows different values in the disorder phase of Ising model from that of magnetically ordered phase. The linking number is computable by Jones-polynomial and non-abelian Chern-Simons field theory \cite{Witten} as well as abelian Chern-Simons action \cite{Duan} which includes self-linking number, writhing number and twisting numbers.

The multi-knot lattice model has spontaneous extension to three dimensional lattice models, such as Ising model, extended Kitaev type model \cite{yu} and so on. Since Chern-Simons action is also a topological action of three dimensional knot \cite{tieyanknot}. A periodical lattice of three dimensional knot can be extended to study four dimensional quantum Hall effect \cite{JPHu}. The multiknot lattice also has many theoretical extension in Chern-Simons field theory and super string theory \cite{Marino}\cite{Dilao} \cite{Babaev}\cite{Faddeev}\cite{siJHEP}. Beside these theoretical extensions, it is also more realistic to experimentally implement the knot lattice model by a network of optical fibers, in which spinning photons demonstrates optical spin Hall effect \cite{Bliokh}. This optical network may provide a different approach to topological quantum computation \cite{Nayak} by constructing braiding on graphene \cite{XWan}\cite{ZXHu}.

\textbf{Acknowledgment} This work is supported by Fundamental Research Foundation for the Central Universities and "National Natural Science Foundation of China"(Grant No. 11304062).\\
$*$ E-mail: tieyansi@hit.edu.cn.

\end{document}